\documentclass[titlepage,12pt,openright,twoside]{report}

\usepackage{amsmath}
\usepackage{amsfonts}
\usepackage{fancyhdr}
\usepackage{epsf}
\usepackage[top=1.2in,bottom=1.2in,left=1.2in,right=1.0in,twoside]{geometry}
\makeatletter
\def\cleardoublepage{\clearpage\if@twoside \ifodd\c@page\else
    \hbox{}
    \vspace*{\fill}
    \vspace{\fill}
    \thispagestyle{empty}
    \newpage
    \if@twocolumn\hbox{}\newpage\fi\fi\fi}
\makeatother

\newcommand{\nin}{\noindent} 
 
\newcommand{\be}{\begin{equation}} 
\newcommand{\ee}{\end{equation}} 
\newcommand{\ba}{\begin{eqnarray}} 
\newcommand{\ea}{\end{eqnarray}} 
 
\newcommand{\br}{{\bf r}} 
\newcommand{\brp}{{\bf r}^{\prime}}

\newcommand{\brb}{\overline {\bf r}} 
 
\newcommand{\brbp}{{\overline {\bf r}}^{\prime}}

\newcommand{\bq}{{\bf q}} 
\newcommand{\bqp}{{\bf q}^{\prime}}

\newcommand{\bp}{{\bf p}} 
\newcommand{\bpb}{{\overline {\bf p}}} 
\newcommand{\bpbp}{{\overline {\bf p}}^{\prime}} 

\newcommand{\tp}{t^{\prime}} 
\newcommand{\tb}{\overline{t}}
\newcommand{\tbp}{\overline{t}^{\prime}}
\newcommand{\bt}{\overline{t}}
\newcommand{\btp}{\overline{t}^{\prime}}

\newcommand{\tst}{{\tilde s}}

\newcommand{\dd}{{\rm d}}
\newcommand{\Hc}{{\cal H}}

\newcommand{\Dc}{{\cal D}}

\newcommand{\Tr}{{\rm Tr}}
\newcommand{\Pc}{{\cal P}}
\newcommand{\ii}{\mathrm{i}}

\begin{document}


\thispagestyle{empty}

\begin{center}
\vspace*{\stretch{1}}

\Large
\textbf{The Loschmidt echo in classically chaotic systems: \\
Quantum chaos, irreversibility and decoherence}\\
\vspace{2cm}

\textmd{Fernando Mart\'{\i}n Cucchietti}\\
\vspace{3cm}

\normalsize

Presentado ante la Facultad de Matem\'{a}tica, Astronom\'{\i}a y F\'{\i}sica \\
como parte de los requerimientos para acceder al grado de \\
Doctor en F\'{\i}sica. \\

\vspace{2cm}

Universidad Nacional de C\'{o}rdoba\\ 
Junio de 2004 \\

\vspace*{\stretch{2}}
\end{center}

\normalsize

\vspace{2cm}

\begin{minipage}[c]{0.4\textwidth}
\centering
\rule[0.2cm]{6cm}{0.1mm}\\
Lic. Fernando M. Cucchietti \\
Autor\\
\end{minipage}%

\vspace{2cm}

\begin{minipage}[c]{0.4\textwidth}
\centering
\rule[0.2cm]{6cm}{0.1mm}\\
Dr. Horacio M. Pastawski \\Director\\
\end{minipage}%

\cleardoublepage

\thispagestyle{empty}
\newenvironment{dedication}
	{\cleardoublepage \thispagestyle{empty} \vspace*{\stretch{1}} \begin{flushright} \em}
	{\end{flushright} \vspace*{\stretch{3}} \clearpage}
		\begin{dedication}
		A Soledad
		\end{dedication}
		\thispagestyle{empty} \cleardoublepage
		
\pagenumbering{roman} \setcounter{page}{1}
\chapter*{Abstract\markboth{Abstract}{Abstract}}

The Loschmidt echo (LE) is a measure of the sensitivity of quantum mechanics to perturbations in the evolution operator. It is defined as the overlap of two wave functions evolved from the same initial state but with slightly different Hamiltonians. Thus, it also serves as a quantification of irreversibility in quantum mechanics. 

\smallskip

In this thesis the LE is studied in systems that have a classical counterpart with dynamical instability, that is, classically chaotic. An analytical treatment that makes use of the semiclassical approximation is presented. It is shown that, under certain regime of the parameters, the LE decays exponentially. Furthermore, for strong enough perturbations, the decay rate is given by the Lyapunov exponent of the classical system. Some particularly interesting examples are given.

\smallskip

The analytical results are supported by thorough numerical studies. In addition, some regimes not accessible to the theory are explored, showing that the LE and its Lyapunov regime present the same form of universality ascribed to classical chaos. In a sense, this is evidence that the LE is a robust temporal signature of chaos in the quantum realm. 

\smallskip

Finally, the relation between the LE and the quantum to classical transition is explored, in particular with the theory of decoherence. Using two different approaches, a semiclassical approximation to Wigner functions and a master equation for the LE, it is shown that the decoherence rate and the decay rate of the LE are equal. The relationship between these quantities results mutually beneficial, in terms of the broader resources of decoherence theory and of the possible experimental realization of the LE.

\newpage

\chapter*{Acknowledgements\markboth{Acknowledgementes}{Acknowledgements}}

Working on my thesis dissertation turned out to be different than what I naively expected at the beginning. In particular, I never suspected that the learning component would be so large, not only in scientific but also in personal matters. Many people contributed, in some way or another, to this process: this is my small attempt to honor those people.

\medskip

I owe special thanks to my advisor Horacio M. Pastawski for generously sharing not only his knowledge, but also (and more important) his passion for physics. More than five years working together is, in practice, more than can be reflected in these words; it is actually in my future work that the imprint of this relationship will be most noticeable.  

\medskip

I am quite grateful to Patricia R. Levstein for her dedication and patience. It is always a pleasure to work with her; indeed, if it were not for my growing interest in the theoretical aspect of the Loschmidt echo I would have never abandoned the joy of working by her side at the spectrometer. 

\medskip

Equally important to me and to this thesis is the great experience I had with all my collaborators. They are (in alphabetical order) Karina Chattah, Diego Dalvit, Rodolfo Jalabert, Caio Lewenkopf, Eduardo Mucciolo, Juan Pablo Paz, Jesus Raya, Oscar Vallejos, Diego Wisniacki, and Wojciech Zurek. Without exception, they all have  unreservedly taught me more than I can thank for. Even more, I owe a good part of my professional self confidence to the respectful treatment as a colleague I always received from them. 

\medskip

An important part of obtaining a doctorate degree is remaining mentally healthy during the process. Friends and family provided me a great environment for this purpose. I would like to thank specially my parents Tito and Cristina, my sister Vanina, my grandma Elva and my family in law Mery, Pety, Maru y Laura for their unconditional support throughout these years, for their love and for always trusting in me. 

\medskip

The friends I would like to give thanks to can be divided roughly in two groups. First, the people from office 324 in C\'{o}rdoba: Gonzalo Alvarez, Fernando Bonetto, Ernesto Danieli, Luis Fo\'{a} Torres, Pablo Gleiser, Marcelo Montemurro and Silvina Segu\'{\i}, for sharing this formation stage with me and being such great people. Second, the people from Los Alamos: Diego Dalvit, Juan Pablo Paz and Augusto Roncaglia, and all my  friends in Santa Fe, for turning this far--from--home time an incredible experience. Also very important, I am grateful to Luis Teodoro and Augusto Roncaglia for reading the manuscript and providing helpful suggestions.

\medskip

Thanks to the people in LANAIS and in FaMAF that, despite the difficulties of being in a country like Argentina, always strove to give me a comfortable working environment. I feel specially indebted to SeCYT for the economical support during four (not so calm) years.

\medskip

Perhaps for being the most important, I left this thanks for the end. As I said above, my journey through this dissertation was not without hard work and even surprises. I really believe that I would have never done so successfully and happily without my wife Soledad by my side. Year after year her constant support, enthusiasm, sacrifices, understanding, and above all, love, have given me not only the strength but also the motivation to go on in this enterprise. Thank you Sol, I can only aspire to be such a great partner for you in your endeavors. This thesis is partly yours.

\renewcommand{\contentsname}{Table of Contents}

\tableofcontents 

\newpage

\setcounter{page}{1}
\pagenumbering{arabic} 
\setcounter{chapter}{0}

\chapter{Introduction}
\label{chap:Introduction}

\begin{quote}
{\em Can nature possibly be as absurd as it seems to us in these atomic experiments?}

Werner Karl Heisenberg
\end{quote}

In 1872 Boltzmann published his first formulation of his now famous $H$ theorem, in which he provided a proof of irreversibility (growth of entropy) from mechanics. His derivation used statistical techniques which had recently been developed by Maxwell. The early misunderstanding of some of the implications of these tools led Boltzmann to use a strong deterministic language in his conclusions. His subsequent work would go back to the understanding and development of further proofs of his $H$ theorem over the next two decades, at the same time advancing some of the most profound concepts of statistical mechanics. 

However, his early flaws were readily picked up by the Austrian physicist Josef Loschmidt, who in 1876 \footnote{A comment on the same lines was made two years previously by William Thomson, later known as Lord Kelvin} enunciated a theorem that showed the impossibility of deriving the second law from mechanics. His argument was based in that the microscopic laws of mechanics are invariant under time reversal. For every mechanically possible motion that leads towards equilibrium (and growth of entropy), there is another one, equally possible, that leads away from it. This evolution, reducing the entropy and thus violating the second law of thermodynamics, is set in motion by taking the final state of the previous evolution as the new initial state and then reversing all the individual molecular velocities.

Although Boltzmann was not mentioned directly by Loschmidt, he was greatly concerned with this ``reversibility paradox'' (as it became known later). Ultimately, it brought Boltzmann to a proper statistical understanding of the second law and his $H$ theorem, realizing the existence of statistical fluctuations, and leading him to his final expressions for entropy using the probability of states compatible with the values of the thermodynamic variables.

The statistical arguments, however, do not make Loschmidt's argument untrue, they only prove that such occurrences are extremely improbable. In the spirit of Maxwell's daemon {\em gedanken} experiment, let us imagine a supernatural being that has the power of reversing all the velocities of the particles trapped in a box. The fact remains that such a creature has the intrinsic power of decreasing the entropy of the system under his will. We call this being a {\em Loschmidt daemon}. Furthermore, an external observer, measuring some variable of the particles in the box before and after the action of the daemon, would see a recurrence in his measurements which we call a {\em Loschmidt echo}.

Several decades passed until it was possible to give a measure of the powers the daemon needed to perform this time reversal effectively. This came with the advent of chaos theory, observed empirically for the first time in 1960 by the meteorologist Edward Lorenz. In the foundations of this theory is the observation that some systems have equations of motion that are hypersensitive to their initial conditions (an example of this is the weather). In this sense, it is known that any prediction of future states of the system will rapidly differ from its actual evolution. It should be noticed that this does not mean that the motion is not deterministic, it is only extremely difficult to predict (where again the weather provides an everyday example). Further mathematical development of the theory showed that these systems, despite their unpredictability, share many broad features that characterize them. An important feature for this work is the so called Lyapunov exponent, which is the rate at which two very close initial states diverge exponentially in time. The Lyapunov exponent is a property of the Hamiltonian of the system and does not depend on the distance between initial conditions, which is only a prefactor of the divergence in time.
The explanation of irreversibility provided by chaos is based on the notions of mixing and coarse graining. The former is the property of chaotic systems of generating a uniform distribution in phase space over the proper energy shell for any initial distribution. The latter, on the other hand, refers explicitly to the sensitivity to initial conditions: the coarseness of our instruments prevents us to prepare specific initial states that will evolve diminishing their entropy.

The devastating consequence of these conclusions for the Loschmidt daemon are the following: to achieve his feat, it must possess an exponentially increasing precision (with the complexity of the system) of the time reversal operation. In the thermodynamic limit, the hypersensitivity of the classical equations of motion implies that the Loschmidt daemon needs to be perfect: otherwise, his attempts to reduce the entropy will quickly fail and go back to the usual thermodynamic prescription. It is, in a way, Boltzmann's concept of ``molecular disorder'' or {\em stossszhalansatz}  (in an extremely more developed fashion) that  settles the century old paradox.
However, the Loschmidt daemon has apparently an exit door: becoming quantum. 

It is fairly simple to demonstrate that changes in the initial conditions do not grow with quantum evolution, the main reason being that the evolution operators are unitary. Suppose we have an initial state $\left| \psi (0) \right>$ and another one very close to it denoted $\left| \psi_\delta(0) \right>$, such that the initial distance between them is measured by the overlap $\delta(0) = | \left< \psi_\delta (0)|  \psi (0) \right> |^2$. This distance in time can be expressed using the quantum evolution operator $U(t)$,
\ba
\delta(t)&=& | \left< \psi_\delta (t)|  \psi (t) \right> |^2 \nonumber \\
&=& | \left< \psi_\delta (0)| U^\dagger(t) U(t)|  \psi (0) \right> |^2 = \delta(0), \nonumber
\ea
because $U$ is a unitary operator. Conclusive numerical evidence of this insensitivity to initial conditions even when the underlying Hamiltonian is classically chaotic was presented in \cite{Casati86}. This property of quantum mechanics could fairly imply that a time reversal as proposed above has better chances of being successful in this context. 

The relevant question now is: how to define the action of the Loschmidt daemon in quantum mechanics? A simple way is given by the observation that in the Schr\"{o}dinger equation, a change in the sign of the Hamiltonian could be absorbed as a change in the sign of time, and therefore is equivalent to a time reversal. The Loschmidt daemon's powers are then restricted to ``flipping'' the sign of the Hamiltonian, something that could be in fact less demanding than changing velocities of particles.
So much simpler, actually, that an experimental realization was possible in the setting of Nuclear Magnetic Resonance experiments. As early as 1950, Hahn \cite{Hahn50} noticed that a $\pi$ pulse in the $X-Y$ plane in a sample under a strong magnetic field in the $Z$ direction would be equivalent to changing the sign of the local magnetic field of the spins. This in turn inverted the decay of the total magnetization (given by inhomogeneities of the external field for each spin), and produced the first realization of a Loschmidt echo. The fact that the sequence only changed the sign of the spin-field term of the Hamiltonian, leaving aside interactions and other terms, made the magnitude of the echo decay with the time waited to perform the operation, therefore leading to an imperfect time reversal.

Further improvements were performed by Rhim, Kessemeir, Pines and Waugh \cite{Rhim71A,Rhim70}. They were able to change the sign of the dipolar interaction between spins through a pulse sequence that changed the axis of quantization of the spins in the sample. Theirs was the first implementation of a many body Loschmidt echo (albeit later called Magic echo in the NMR community). However it was still far from a perfect echo, since its magnitude also decreased indicating some failure in the time reversal. Furthermore, since the only available information was the total magnetization of the sample, no further insight on the microscopic nature of this decay was possible. Although the majority of the technological components needed were available at the time, it took more than two decades of conceptual progress to combine them to produce a more controlled Loschmidt daemon.

In the meantime, theoretical studies focused more on what meant chaos in quantum mechanics. Being unable to provide a dynamical definition, researchers found that the spectral properties of systems with a classically chaotic analog presented particular features that distinguished them from integrable systems. For instance, Casati et al. \cite{Casati80} and later Bohigas et. al. \cite{Bohigas84} observed that the distribution of level spacings in a classically chaotic system was the same as the distribution obtained from random matrices with the appropriate symmetries (for instance time reversal or spin symmetry), which are more amenable to analytical studies. In particular, chaotic systems present a Wigner-Dyson distribution of the spacings, which has a marked zero for degenerate levels. Integrable systems on the other hand show a Poissonian (exponential) distribution, indicating that degeneracies abound. Heller \cite{Heller84}, on another line, showed that the spatial density of the wave function in chaotic billiards showed {\em scars}, marked lines that corresponded to the position of classical periodic orbits. Another example is the finding by Szafer et. al. \cite{Szafer93}, who showed that the energies of a chaotic system as a function of an external parameter displayed repulsion and spectral rigidity (also found in random matrices). They also were able to compute the value of the parameter after which perturbation theory broke down and the level velocity correlations decayed to zero. There exists a multitude of studies of spectral properties of chaotic systems, but we shall focus on just two that directly treat dynamics.

One of the first results that connected the classical chaotic motion to a quantum observable was also due by Heller \cite{Heller90}, who showed that given an initial wave packet along a periodical orbit of a chaotic system, one should observe recurrences in the autocorrelation function that were attenuated exponentially with the a rate equal to the Lyapunov exponent. 

The second work is closely related to the discussion that interests us, the Loschmidt daemon. Peres in \cite{Peres84} proposed that, given the paradox between quantum mechanics and classical chaos, one should look for sensitivity not in the initial conditions but rather in the Hamiltonian that governs the evolution. In specific terms, he proposed to study the overlap of the same initial wave function $\left| \psi_0 \right>$ evolved with two slightly different Hamiltonians,
\be 
M(t)=\left\vert m(t)\right\vert ^{2}=\left\vert \left\langle \psi 
_{0}\right\vert e^{\mathrm{i}(\mathcal{H}_{0}+\Sigma)t/\hbar}\ e^{-\mathrm{i}%
\mathcal{H}_{0}t/\hbar}\left\vert \psi_{0}\right\rangle \right\vert ^{2}.
\label{LE}%
\ee 
By virtue of the structure of quantum operators, this equation also describes the magnitude of an imperfect Loschmidt echo: take an initial state $\left| \psi_0 \right>$, evolve it with a given Hamiltonian $\Hc_0$ for a time $t$, perform a {\em faulty} sign change in the Hamiltonian represented by the addition of a unitary term $\Sigma$, and compute what is the overlap with the desired (initial) state.

Peres' work \cite{Peres84} contains two main results. Frst, using perturbation theory he showed that for short times or very small $\Sigma$, $M$ decayed quadratically. Second, and perhaps more important and profound, 
he provided numerical evidence that the long time behavior of $M$ for classically chaotic or integrable systems was clearly distinguishable. While the former decayed rapidly to a small constant given by the inverse of the size of the Hilbert space, $M$ in integrable systems showed strong oscillations and recurrences and did not saturate at such a small value. Later on, in Ref. \cite{PeresBook}, Peres would indicate that in numerical computations for chaotic systems $M$ appeared to decay exponentially until the saturation value was reached.

Peres' seminal paper sparked a wave of related work \cite{Schack92,Schack93,Benet93,Schack96,Ballentine96,Schack96b}.  Of importance to us is Ref.~\cite{Schack96b}. There, using an information theoretical approach, Schack and Caves were able to show that in classical dynamics perturbing the evolution had the same effect as that of changing the initial conditions: linear increase of entropy with the Lyapunov exponent. 

To explain the following analytical breakthrough, it is better refer back to its experimental motivation. After the Magic echo of the 70's, it took another 20 years to combine it with a technique called cross-polarization to probe the inner dynamics of the spin network. The setup is the following: using a rare spin species as a local probe and the cross-polarization technique, magnetization can be injected and after some time measured by the probe in only one spin of a large network \cite{Zhang1}. Therefore, one can access microscopic information about the dynamics directly. In between the injection and the measurement, the Magic echo sequence can be applied. The result, dubbed a Polarization echo \cite{Zhang1}, allowed to track the behavior under time reversal of a {\em local} excitation of the spin system, unlike the Magic echo which only provides information of the magnetization on a global level. 

The group of Levstein and Pastawski developed this matter further, and aimed to shed light on the problem of irreversibility. Among their main results, the following are those that are most pertinent to this discussion. For spin systems weakly coupled to the environment, the decay of the Loschmidt echo was found to have a Gaussian shape \cite{Levstein98,Usaj98}. The width $\tau_\phi$ of this Gaussian was observed to depend mainly on the dipolar interaction constant between the spins \cite{Usaj98}. Even more, they were able to show that $\tau_\phi$ depends only weakly on the strength $\omega$ of the RF field used to perform the pulse sequence \cite{Pastawski00} (where it is argued that the most important terms of $\Sigma$ are proportional to $\omega^{-1}$). The long detour through the experiments that led to these striking results, along with their interesting conceptual framework and analysis is reviewed in \cite{PastawskiBook}. 

The main conclusion to be extracted from the last two findings is that the typical decay time of the Loschmidt echo in an isolated many body spin system depends only on properties of the Hamiltonian of the system, and is independent of the perturbation $\Sigma$. The similarity of this effect to that discussed above for chaotic systems is striking, and strongly suggests that the many body system presents an hypersensitivity to perturbations in much the same way as classical chaotic systems. 

Sadly, analytical tools to treat this problem in many body systems do not exist, or at least are not sufficiently simple or capable of providing a solution. An alternative problem amenable to analytical treatment, and with enough elements to at least mimic the most prominent behavior, is a single particle in a classically chaotic system. In this sense, we assume that the chaotic Hamiltonian supplies enough complexity in the dynamics to make up for the intricacies of the many body situation. Furthermore, it is at least reasonable to assume that if hypersensitivity to perturbations were to be found in quantum mechanics, one would expect to observe it in chaotic systems such that the classical behavior is consistently recovered.

These are the ideas and assumptions behind the work of Jalabert and Pastawski \cite{Jalabert01} that is the basis for this thesis. I will briefly mention their main result since it will be derived later in full detail. Studying the Loschmidt echo for a single particle in a classically chaotic Hamiltonian, Jalabert and Pastawski showed that there exists a regime of the parameters where $M(t)$ decays exponentially with a rate given by the Lyapunov exponent of the classical system. This striking result triggered a large amount of analytical and numerical work in different groups on many aspects of the theory, a process I had the privilege of participating actively since its very beginnings. 

This thesis is an account of the work I did in this period of great excitement over the field. The results of the investigations I took part of are not presented in chronological order for pedagogical reasons\footnote{This choice, however, takes away the opportunity to witness the emotional roller-coaster of scientific research.}.
The organization of this work is the following: In chapter 2, I introduce a generalization of the original calculation of \cite{Jalabert01}, as well as an ample discussion on the implications of the theoretical results. Ensuing, some particular examples are given that serve to gain insight on the (somewhat intricate) semiclassical calculations. In chapter 3, I first present some numerical results that support the theory. Afterwards, the main topic of the chapter is addressed, namely the universality of the Lyapunov decay of the LE. Briefly, this universality is understood as that commented above for classical chaos. The subtle issue of an apparently classical behavior emerging from a quantum object is approached in chapter 4. For this purpose, a relation between the LE and the theory of the quantum to classical transition in open systems is demonstrated. This relation proves to be useful in providing fresh perspectives and insight of the previously obtained results.

At the end of each chapter a summary of the main results is given. This leaves for the conclusions some remarks on the general character of the problem, comments on work by other groups not mentioned in the body of the thesis, and finally some considerations on future investigations.

\chapter{The semiclassic approximation to the Loschmidt echo} 
\label{chap:Semiclassical}

\begin{quote}
{\em You can never solve a problem on the level on which it was created.}

Albert Einstein.
\end{quote}
 
We left the previous chapter with the purpose of tackling the  problem of a complex many body Hamiltonian with many degrees of freedom applying a rather crude first approach: a single body in a chaotic system. An even simpler approximation might be to consider a random Hamiltonian, however by doing this we would also strip the problem from plenty of structure (we will see this in the following). Chaos is our attempt to introduce complexity while at the same time retaining some analytical tractability, hoping that at least some physical insight will be gained. 
A powerful tool that lets one take into account the classically chaotic 
properties of motion in the description of quantum dynamics is the so called 
{\em semiclassical approximation}. In this chapter we will use it to 
analyze the problem of the Loschmidt echo (LE), and show that it 
succesfully describes some of its relevant regimes. Afterwards, we
will consider some particular examples that help develop intuition on the subject,
apart from being useful to compare with numerical tests. 
Finally (and in the spirit of gaining intuition) we will consider an interesting analytically solvable example: an inverted harmonic oscillator that, although presents dynamical instability, is not chaotic.
 
\section{General Approach} 
\label{sec:GeneralLE}
 
This section contains calculations of the Loschmidt echo for a generic chaotic Hamiltonian ${\cal H}_0$ and a perturbation $\Sigma$ that has a random time and spatial dependence.
This latter restriction will be relaxed later in one of the examples of the next section.
 
\subsection{Semiclassical Evolution} 
\label{sec:SCEvolution}
 
Let us consider as an initial state a Gaussian wave packet of width $\sigma$ 
and initial mean momentum $\bp_0$, 
\be 
\psi (\overline{{\bf r}},t\!=\!0)=\left( \frac{1}{\pi \sigma ^{2}}\right) 
^{d/4}\exp \left[ \frac{{\rm i}}{\hbar }{\bf p}_{0}\cdot (\overline{{\bf 
r}}-%
{\bf r}_{0})-\frac{1}{2\sigma ^{2}}(\overline{{\bf r}}-{\bf r}_{0})^{2}%
\right] .  \label{InitialPacket} 
\ee
Such an initial state is a 
typical choice in semiclassical approximations because it is a 
good representation of a classical state and, not less important, 
it usually simplifies analytical calculations. The generality of 
results based on this approximation can be questioned, 
since it is not true that the behavior of the LE is the same for a general initial state. 
While not formally proved, the general feeling is that any localized state (in space, 
momentum, etc.) will show classical properties appearing in the 
decay of the LE. On the other hand, it has been shown that this is 
not the case for other choices like eigenstates \cite{WisniackiCohen} of the system or 
random states \cite{Wang02}. We adhere however to the usual prescription, noting 
that the results obtained will be as general as the decomposition 
of the initial state into a superposition of Gaussians \cite{Jacquod02,Wang02}. 
 
The time evolution of an initial state $\psi(\brb,0)=\left<\brb|\psi(t=0)\right>$ is given by 
\be 
\psi (\br,t)=\int d\br \ K(\br,\brb;t) \ \psi (\brb,0) \ , 
\label{DefinitionPropagator} 
\ee 
with the quantum propagator 
\be 
K(\br,\brb;t)=\left\langle {\bf r}\right| e^{-{\rm i} 
{\cal H}t/\hbar }\left| \brb \right\rangle . 
\label{QuantumPropagator} 
\ee 
 
It is usually at this point that the semiclassical approximation is made (only a brief 
summary of it follows asmany good textbooks exist on the subject \cite{Gutzwiller,Brack}). 
This consists of an expansion of the full quantum propagator 
in a sum of propagators {\em but only} over all possible classical trajectories\footnote{Trajectories not included, but allowed by quantum mechanics, are for instance those with tunneling through energy forbidden regions.}
$s(\br,\brb,t)$ going from $\brb$ to $\br$ in time $t$, 
\ba 
K(\br,\brb;t) &\simeq&\sum_{s(\brb,\br,t)}K_{s}(\br,\brb;t) \ ,  \nonumber \\ 
K_{s}(\br,\brb;t) &=&\left( \frac{1}{2\pi \ii \hbar }\right) 
^{d/2}C_{s}^{1/2} \ \exp{\left[\frac{{\rm i}}{\hbar}S_{s}(\br,\brb;t) 
-{\rm i}\frac{\pi }{2}\mu _{s}\right]} \ . 
\label{SemiclassicalPropagator} 
\ea 
The approximation is valid in the limit of large energies for which the de Broglie wavelength $(\lambda _{dB}=2\pi /k_{dB}=2\pi \hbar /p_{0})$ is the minimal length scale. 
$S_{s}(\br,\brb;t)=\int_{0}^{t} \dd \bt {\cal L}_{s} (q_{s}(\tb ),\dot{q}_{s}(\tb);\tb)$ is the action over the trajectory $s$, and ${\cal L}$ the Lagrangian. The Jacobian $C_{s}=\left| \det B_{s}\right| $ accounts for the conservation of classical probabilities, with the matrix 
\be 
\left( B_{s}\right) _{ij}=-\frac{\partial ^{2}S_{s}}{\partial {\bf r}%
_{i}\partial \overline{{\bf r}}_{j}} \ , 
\label{MatrixB} 
\ee 
obtained from the derivatives of the action with respect to the various 
components of the initial and final positions. $\mu _{s}$ is the Maslov 
index, counting the number of conjugate points of the trajectory $s$, but it will be disregarded since it does not play any role in the LE. 
 
Let us consider fairly concentrated initial wave--packets, which will let us expand the action around trajectory $s$ to first order 
\be 
S_{s}(\br,\brb;t)\simeq S_{\hat s}(\br,\br_{0};t)- 
\bpb_{\hat s}\cdot (\brb-\br_{0}) \ ,
\label{ActionExpansion} 
\ee 
where $\left. \nabla _{\brb_{\rm i}}S_{s}\right|_{\brb=\br_{0}} = 
-\bpb_{s,{\rm i}}$ and $\bpb_{s,{\rm i}}$ is the i-th component of the 
initial momentum of trajectory $s$.   
We are lead to work with trajectories ${\hat s}$ that join $\br_{0}$ 
to $\br$ in a time $t$, which are slightly modified with respect to the 
original trajectories $s(\brb,\br,t)$. We can therefore write 
 
\begin{eqnarray} 
\psi(\br,t) &=&\sum_{s(\br_{0},{\bf r},t)}K_{s}(\br,\br_{0};t) 
\int d\brb \ \exp{\left[-\frac{{\rm i}}{\hbar}\bpb_{s} 
\cdot(\brb-\br_{0})\right]} \ \psi (\brb,0) \nonumber \\ 
&=&\left( 4\pi \sigma ^{2}\right)^{d/4}\sum_{s(\br_{0},\br,t)} 
K_{s}(\br,\br_{0};t) \ \exp{\left[-\frac{\sigma^{2}}{2\hbar^{2}} 
\left(\bpb_{s}-{\bf p}_{0}\right) ^{2}\right]} \ , 
\label{SemiclassicPacket} 
\end{eqnarray} 
 
\nin where we have neglected second order terms of $S$ in $(\brb-\br_{0})$ 
since we assume that the initial wave packet is much larger 
than the de Broglie wavelength ($\sigma \gg \lambda _{dB}$). 
Eq.~(\ref{SemiclassicPacket}) shows that only trajectories with initial 
momentum $\bpb_{s}$ closer than $\hbar /\sigma $ to ${\bf p}_{0}$ are 
relevant for the propagation of the wave-packet, and it is the expression 
for the wave function at time $t$ that will let us obtain a tractable 
form for the Loschmidt echo (even though further approximations are still 
needed). 
 
\subsection{Semiclassical Loschmidt Echo} 
\label{sec:SCLE}
 
Combining Eqs.~(\ref{LE}) and (\ref{SemiclassicPacket}) one readily obtains the 
semiclassical expression for the amplitude of the Loschmidt echo, 
 
\ba
m(t)&=&\left(\frac{\sigma^{2}}{\pi \hbar^{2}}\right)^{d/2}\int \dd \br 
\sum_{s,\tilde{s}}C_{s}^{1/2}C_{\tilde{s}}^{1/2} \ 
\exp{\left[\frac{{\rm i}}{\hbar}(S_{s}-S_{\tilde{s}})- 
\frac{{\rm i}\pi}{2}(\mu _{s}-\mu _{\tilde{s}})\right]}  \nonumber \\
&\times& \exp{\left[-\frac{\sigma^{2}}{2\hbar^{2}} 
\left(\left(\bpb_{s}-\bp_{0}\right)^{2}+ 
\left(\bpb_{\tilde{s}}-\bp_{0}\right)^{2}\right) \right]} \, 
\label{EchoAmplitude} 
\ea 

\nin where $s$ ($\tilde{s}$) are trajectories traversed with the unperturbed (perturbed) Hamiltonian $\Hc_0$ ($\Hc_0+\Sigma$).
Let us first evaluate this equation for the zero perturbation 
($\Sigma =0$) case. Here we need to restrict the sum to the terms 
with $s=\tilde{s}$ (the ones we leave aside are terms with a 
highly oscillating phase and are corrections of smaller order). 
Thus we obtain 
 
\be
m_{\Sigma=0}(t)=\left(\frac{\sigma^{2}}{\pi \hbar ^{2}}\right)^{d/2}
\int \dd\br \sum_{s({\bf r}_{0},{\bf r},t)}C_{s} \ 
\exp{\left[-\frac{\sigma^{2}}{\hbar^{2}}\left(\bpb_{s}- 
\bp_{0}\right)^{2}\right]}  =1_s \ . 
\label{MSigma0} 
\ee
 
\nin where we have performed the change from the final position 
variable $\br$ to the initial momentum $\bpb_{s}$ using the 
Jacobian $C$, and then simply carried out a Gaussian integration 
over the variable $\bpb_{s}.$ Notice the subindex $s$ to the unity is a remainder that the result is $1$ to first order ($s=\tilde{s}$) and that small corrections to $1$ could exist. $1_s$ is therefore the ``semiclassical'' unity \cite{Vallejos01}.
 
To proceed analytically in the $\Sigma \neq 0$ case, we need to 
perform a rather controversial approximation. We will assume that 
the perturbation is sufficiently weak so that it does not change 
appreciably the classical trajectories associated with ${\cal 
H}_0$, at least in the time interval of interest. In terms of 
Eq.~(\ref{EchoAmplitude}), this means we will only keep terms where $s \sim \tst$. 
Clearly, in a chaotic system this is a no-hope situation, 
where individual trajectories are per definition exponentially 
sensitive to perturbations. Thus, the time regime of validity of 
the approximation is logarithmically short. However, it was observed in numerical 
tests that this so called {\em classical perturbation 
approximation} holds for times much longer than expected. Even 
though one can argue that terms where $s \neq \tst$ cancel out 
because of rapid oscillations or averaging, a more subtle cause 
for this robustness has been pointed out \cite{Cerruti02,Heller03}. Despite the sensitivity 
of individual points in phase space, the whole manifold of trajectories in chaotic systems 
displays a rather strong structural stability. In terms of such an approximation this  
means that instead of claiming that trajectories $\tst $ are weakly affected by the perturbation, one can always resort to a ``replacement" trajectory $s'$ that moves close to $s$ \cite{Heller03}.
 
Within the classical perturbation approximation then Eq.~(\ref{EchoAmplitude}) can be cast as 
\be 
m(t)\simeq\left( 
\frac{\sigma^{2}}{\pi\hbar^{2}}\right)  ^{d/2}\int 
\mathrm{d}\mathbf{r}\ \sum_{s}\ C_{s}\ \exp\left[ 
\frac{\mathrm{i}}{\hbar }\Delta S_{s}\right]  \ \exp{\left[ 
-\frac{\sigma^{2}}{\hbar^{2}}\left[ \left( 
{\overline{\mathbf{p}}}_{s}-\mathbf{p}_{0}\right)  ^{2}\right] 
\right]  }\ . 
\label{EchoAmplitudeAprox}%
\ee 
Where $\Delta S_{s}$ is the modification of the action, associated with the trajectory $s$, by the effect of the perturbation $\Sigma$. It can be obtained as 
\be 
 \Delta S_{s}= - 
\int_{0}^{t}d\overline{t}\ \Sigma_{s}(\mathbf{q}(\overline 
{t}),\mathbf{\dot{q}(}\overline{t}),\overline{t}) \ , 
\label{DeltaS}%
\ee 
when the perturbation is in the potential part of the 
Hamiltonian: if it is in the kinetic term there is an irrelevant 
change of sign. 
 
Using expression (\ref{EchoAmplitudeAprox}) we can write the LE as 
\ba
M(t)&=&\left(  \frac{\sigma^{2}}{\pi\hbar^{2}}\right) 
^{d}\int\mathrm{d} 
\mathbf{r}\int\mathrm{d}\mathbf{r}^{\prime}\sum_{s(\mathbf{r}_{0} 
,\mathbf{r},t)}\sum_{s^{\prime}(\mathbf{r}_{0},\mathbf{r}^{\prime},t)} 
C_{s}C_{s^{\prime}}\exp\left[  \frac{\mathrm{i}}{\hbar}\left( 
\Delta S_{s}-\Delta S_{s^{\prime}}\right)  \right]  \nonumber \\ 
& \times &\exp\left[ 
-\frac{\sigma^{2} }{\hbar^{2}}\left[  \left( 
\mathbf{p}_{s}-\mathbf{p}_{0}\right) ^{2}+\left( 
\mathbf{p}_{s^{\prime}}-\mathbf{p}_{0}\right) ^{2}\right]  \right].
\label{MComplete} 
\ea 
 
As in Ref.~\cite{Jalabert01}, the LE can be decomposed as 
\be 
M(t)=M^{\mathrm{nd}}(t)+M^{\mathrm{d}}(t) \ , 
\label{DiagNoDiag}%
\ee 
where the first term (non-diagonal) contains trajectories $s$ and 
${s}^{\prime}$ exploring different regions of phase space, while 
in the second (diagonal) ${s}^{\prime}$ remains close to $s$. Such 
a distinction is essential when considering the effect of the 
perturbation over the different contributions. One could object 
that the separation is rather arbitrary and not complete, in the 
sense that it has not been precisely defined yet and that it does 
not contemplate cases between the two categories (which are 
likely to exist due to the chaotic nature of the system). A 
mathematical definition for the separation will be given later 
in the treatment of the diagonal contribution, and this will help 
dividing more precisely the terms in the two categories. In any 
case, numerical experiments will show that such a separation is 
sufficient to describe the most prominent behavior of $M(t)$. 

\subsection{Non diagonal terms}
\label{sec:NDTerms}

To proceed, one needs to enter information about the perturbation. In this
and in the following section we will introduce the calculation of the LE for a
quite general form of the perturbation, requiring knowledge of only statistical
properties of the perturbation correlators. 

Let us first consider a perturbation in the potential term of the Hamiltonian,
that depends randomly on the position and in time,
$\Sigma=\Sigma(\br,t)$. In particular, the potential needs to be continuous and have a finite range $\xi$ in order to allow the application of the semiclassical tool (this is given by $\xi
k_{dB}\gg1$). Other restrictions will be specified below when necessary. The correlation function of the above potential is given by%

\begin{equation}
C_{\Sigma}(|\bq-\bqp|,t-\tp) = 
\overline{\Sigma^2} C_{S}(|\bq-\bqp|)C_{T}(t-\tp)=
\langle \Sigma(\bq,t)\Sigma(\bqp,\tp)\rangle \,
\label{Correlator}
\end{equation}

\nin where we have assumed that the time correlation $C_T$ is independent of the
spatial one $C_S$. $\overline{\Sigma^2}^{1/2}$ is the typical strength of the
perturbation. We require that at least $C_S$  or $C_T$  decay sufficiently fast,
so that
\be
\int_0^{\infty} \dd r \ C_S(r)=\xi<\infty 
\ \mbox{or} \ \int_0^\infty \dd t \ C_T(t)=\tau_0< \infty,
\label{FastCorrelators}
\ee
which for chaotic systems is a sensible approximation due to the random-like behavior of observables in these systems. Above $\xi$ is the
typical correlation distance of $C_S$, and $\tau_0$ is the typical decay time of
$C_T$.
The finite range of the potential is a crucial ingredient in order
to bridge the gap between the physics of disordered and dynamical systems
\cite{Jalabert00,Mirlin03} and to obtain the Lyapunov regime
\cite{Jalabert01}. 
Moreover, taking a finite $\xi$ or $\tau_0$ is not only helpful
for computational or conceptual purposes, but it constitutes a sensible
approximation for an uncontrolled error in the reversal procedure
$\mathcal{H}_{0}\rightarrow-\mathcal{H}_{0}+\Sigma$ as well as an approximate
description for an external environment, which is likely to extend over a certain
typical length instead of being local.

As discussed above, in the leading order of $\hbar$ and for
sufficiently weak perturbations, we can neglect the changes in the classical
dynamics associated with the external source. One simply modifies the contributions to
the semiclassical expansion of the LE associated with a trajectory $s$ (or in
generally to any quantity that can be expressed in terms of the propagators)
by adding the extra phase $\Delta S$ of Eq.~(\ref{DeltaS}). Let us assume 
that the velocity along the trajectories remains unchanged with respect to its initial value 
$v_{0}=p_{0}/m=L_{s}/t$.

For trajectories of length $L_{s}\gg \xi$, $L_{s}\gg v_0 \tau_0$, the contributions to $\Delta S$ from
segments separated more than $\xi$ or $v_0 \tau_0$ are uncorrelated. The stochastic
accumulation of action along the path can be therefore interpreted as
determined by a random-walk process, resulting in a Gaussian distribution of
$\Delta S_{s}(L_{s})$. This approximation has also been verified numerically in
Ref.~\cite{Heller03}, but it is worth noticing that it could depend on the shape of the perturbation and the chaoticity of the system \cite{Casati04}. In this sense the integration over trajectories represents an average for the non--diagonal terms $M^{\mathrm{nd}}$, and we can write as the product of the averages,
\be
\left< M^{\mathrm{nd}} (t)\right>= \left| \left(  \frac{\sigma^{2}}{\pi\hbar^{2}}\right) 
^{d/2} \int\mathrm{d} 
\mathbf{r} \sum_{s(\mathbf{r}_{0} 
,\mathbf{r},t)}
C_{s} \left\langle \exp\left[  \frac{\mathrm{i}}{\hbar}\left( 
\Delta S_{s}\right)  \right] \right\rangle  \exp\left[ 
-\frac{\sigma^{2} }{\hbar^{2}} \left( 
\mathbf{p}_{s}-\mathbf{p}_{0}\right) ^{2} \right]  \right|^2.
\label{MNonDiagonal} 
\ee
Using the above considerations on the statistical properties of $\Delta S_{s}$, the ensemble average over the propagator (\ref{SemiclassicalPropagator}) [or over independent trajectories in Eq.~(\ref{EchoAmplitudeAprox})] of the phase differences can be written as

\begin{equation}
\langle\exp\left[  \frac{\mathrm{i}}{\hbar}\Delta S_{s}\right]  \rangle
=\exp\left[  -\frac{\langle\Delta S_{s}^{2}\rangle}{2\hbar^{2}}\right]  \ ,
\label{AverageDeltaS}%
\end{equation}

\noindent and therefore $M^{\mathrm{nd}}$ is entirely specified by the variance%

\begin{equation}
\langle\Delta S_{s}^{2} \rangle= \overline{\Sigma^2}
\int_{0}^{t} \mathrm{d}\tb  \int_{0}^t \mathrm{d}\tbp
C_{S}(|\bq_s(\tb)-\bqp_s(\tbp)|)C_{T}(\tb-\tbp).
\label{VarianceS}
\end{equation}

Since the length $L_{s}$ of the trajectory is supposed to be much larger than
the decay distance of the correlators, the integral over $\tau=\tb-\tbp$ can be taken from 
$-\infty$ to $+\infty$, while the integral on $\hat{t}=(\tb+\tbp)/2$ gives a factor 
of $t$. Two regimes are readily solved, the first one when the time dependence of the
perturbation is slow compared to the spatial change, $\xi \ll v_0/\tau_0$, and
one obtains

\be
\left\langle \Delta S_{s}(t)^{2}\right\rangle \simeq\overline{\Sigma^2}\int_{0}^{t}d\overline
{t}\int_{-\infty}^{\infty}d\tau C_{S}\left[  \left|  \mathbf{q}_{s}%
(\overline{t}-\tau/2)-\mathbf{q}_{s}(\overline{t}+\tau/2)\right|  \right]
=\frac{v_0 t\hbar^{2}}{\widetilde{\ell_S}},
\label{SpatialDeltaS}
\ee
where $C_{T}(\tau)$ is assumed constant and the mean free path of the perturbation is defined as
\be
\frac{1}{\widetilde{\ell_S}}=\frac{\xi \overline{\Sigma^2}}{v_0^2 \hbar^2}.
\label{SpatialFGR}
\ee

On the other extreme, when $\tau_{0}\ll \xi/v,$ we have the opposite regime and 
the decay of $M^{nd}$ will be given by
\be
\left\langle \Delta S_{s}(t)^{2}\right\rangle \simeq\overline{\Sigma^2}\int_{0}^{t}d\overline
{t}\int_{-\infty}^{\infty}d\tau C_{T}(\tau)=
\frac{v_0 t\hbar^{2}}{\widetilde{\ell_T}},
\label{TemporalDeltaS}
\ee
with 
\be
\frac{1}{\widetilde{\ell_T}}=\frac{ \tau_0 \overline{\Sigma^2}}{v_0 \hbar^2}.
\label{TemporalFGR}
\ee

Replacing Eqs.~(\ref{SpatialDeltaS}) or (\ref{TemporalDeltaS}) into Eq.~(\ref{MNonDiagonal}) and using the Jacobian $C_s$ to perform the Gaussian integral,
\be
M^{\mathrm{nd}}=\exp{\left( -\frac{v_0 t}{\widetilde{\ell}} \right) }.
\label{MNDFinal}
\ee

\noindent The \textquotedblleft elastic mean free path" $\widetilde{\ell}$ and the
mean free time $\widetilde{\tau}=\tilde{\ell}/v_{0}^{{}}$ associated with the
perturbation determines the rate of decay of $M^{\mathrm{nd}}$ and will constitute a measure of the strength of the coupling. This is not to be confused with any typical time or distance that might exist in the unperturbed Hamiltonian. $\widetilde{\ell}$ is to be regarded as the distance over which the perturbation has a sensible effect on the action accumulated in the trajectory.

Taking averages over the perturbation is technically convenient, but not
crucial. These results would also arrive from considering a
single perturbation and a large number of trajectories exploring
different regions of phase space.

The intermediate regime, when the temporal and spatial scales of the
perturbation coincide, is only accesible through numerical simulations or further
assumptions on the form of the correlators. We will take the latter path in
the next section.
However, before that, let us take a brief detour to explore the association of $M^{\mathrm{nd}}$ with the well known Fermi Golden Rule (FGR).

\subsubsection{Random Matrix approach: the Fermi Golden rule}
\label{sec:RMtheory}

Straying momentarily from the semiclassical treatment, we study here the non diagonal terms with tools from random matrix theory (RMT), and show how the LE is related to spectral features of the system. In particular, the non diagonal terms just discussed can be shown to arise from a Fermi Golden rule treatment. 

The computation of $M^{\rm nd}(t)$ by the statistical approach is actually a standard
random-matrix result (see, for instance, Ref. \cite{Agassi75} or
Appendix B of Ref. \cite{Lutz99}). The connection between those terms and the FGR was first pointed out in Ref. \cite{Jacquod01}. For instructional purposes, let us describe the derivation. The connection to the random matrix theory is made by
the Bohigas' conjecture \cite{Bohigas84}, which states that Hamiltonians with a classically chaotic equivalent have the same spectral properties of random matrices with certain distribution of its components. Consequently, the matrix elements
\be 
\label{V_nn'}
\Sigma_{nn^\prime} = \langle n | \Sigma({\bf r}) | n^\prime
\rangle 
\ee 
with respect to the eigenstates of $\Hc_0$ are Gaussian distributed,
regardless of the form of $\Sigma({\bf r})$ (and as long as the distribution is not long tailed). Noticing that the average
\ba
\left< M^{\rm nd}\right>&=&\left| \left< m(t) \right> \right|^2 \nonumber \\
&=& \left| \left< \psi_0 \right| \left< e^{-\ii \Hc t/\hbar} \right> \left| \psi_0 \right> \right|^2,
\label{AverageMND}
\ea
it is clear that we need to calculate the average of the quantum propagator
\be 
U(t) = e^{-\ii \Hc t/\hbar}\theta(t). 
\ee 
This task is usually carried out in the energy representation by
introducing the Green function operator
\be
G(E) = \frac{1}{E + \ii \eta - \Hc} \;, \quad \mbox{with}
\quad \eta \rightarrow 0^+ \;.
\ee
The formal expansion of $G$ in powers of $\Sigma$ and the rules for
averaging over products of Gaussian distributed matrix elements give
\be 
\overline{G} = G_0 \frac{1}{1 - \overline{\Sigma G_0\Sigma}G_0}\,,
\ee 
where $G_0 = (E + \ii \eta - \Hc_0)^{-1}$. The matrix representation
of $\overline{G}$ is particularly simple. In the eigenbasis of $\Hc_0$
it becomes
\be 
\overline{G}_{nn^\prime}(E) = 
\frac{\delta_{nn^\prime}} {E + \ii \eta - E_n - \gamma_n (E)},
\ee
where $E_n$ is the $n$-th eigenvalue of $\Hc_0$ and 
\ba 
\gamma_n (E) = \sum_{n^\prime} \overline {\Sigma^2_{nn^\prime}}
(G_0)_{n^\prime} \equiv \Delta_n(E) - \frac{i}{2} \Gamma_n(E),
\ea
with
\ba 
\Delta_n(E) & = & \mbox{PV} \, \sum_n
\frac{\overline{\Sigma^2_{nn^\prime}}} {E - E_n} \nonumber \\ \Gamma_n(E)
& = & 2\pi \sum_n \overline{\Sigma^2_{nn^\prime}} \delta(E - E_n)\,.
\ea
Where PV stands for principal value. The real part $\Delta_N(E)$
only causes a small shift to the eigenenergy $E_n$ and will thus be
neglected. Whenever the average matrix elements
$\overline{\Sigma^2_{nn^\prime}}$ show a smooth dependence on the indices
$n$, it is customary to replace $\Gamma_n$ by its average value,
\be
\label{GammaFGR}
\Gamma = 2\pi \overline{\Sigma^2}/\Delta,
\ee
where $\Delta$ is the mean level spacing of the unperturbed
spectrum. In most practical cases, $\Gamma$ and $\Delta$ can be
viewed as local energy averaged quantities. Hence, the average
propagator in the time representation becomes
\be
\label{avKRMT}
\overline{U}_{nn^\prime}(t) = \delta_{nn^\prime}
\exp\!\left(-\ii \frac{E_n t}{\hbar} - \frac{\Gamma t}{2\hbar}
\right) \theta(t) \,.
\ee
It is worth stressing that $\Gamma$ arises from a
nonperturbative scheme; nonetheless, it is usually associated with the
Fermi golden rule due to its structure.

Now we need to use the average propagator obtained in Eq.~(\ref{avKRMT}) in the expression of Eq.~(\ref{AverageMND}). This step also gives us a more precise meaning to the smooth energy dependence of $\Gamma(E)$: In this construction the latter has
to change little in the energy window corresponding to the energy uncertainty of $\psi({\bf r}, t)$, which is determined by $\sigma$. Thus, the RMT final expression for $M^{nd}(t)$ is
\be
\label{RMTFGRdecay}
M^{\rm nd}_{\rm RMT}(t) = \exp(-\Gamma t/\hbar),
\ee
with $\Gamma$ given by Eq. (\ref{GammaFGR}). Equation
(\ref{RMTFGRdecay}) does not hold for very short times, since we
neglected the smooth energy variations of $\Gamma_n$ and
$\Delta_n$. It is beyond the scope of RMT to remedy this situation,
since for that purpose nonuniversal features of the model have to be
accounted for.

Despite sharing the same formal structure as Eq.~(\ref{MNDFinal}), we should also demonstrate that both the semiclassical and the RMT exponents are the same. This was done in Ref. \cite{Cucchietti02Smooth} for the specific case of a two dimensional billiard with the quenched disorder perturbation used by Jalabert and Pastawski \cite{Jalabert01}, and in principle could be shown for other models. A general proof however is still not available.

We will however focus on the connection of Eq.~(\ref{RMTFGRdecay}) with the spectral properties of the system. Let us first notice that the structure of the average RMT propagator (\ref{avKRMT}) tells us that the decay of $M^{\rm nd}$ for a general initial state is the same as for any eigenstate $\left| n \right>$ of $\Hc_0$. In this case it is easy to write an expression for $M^{\rm nd}$, 
\ba
M^{\rm nd}(t)&=&\left| \left< n \right| U^\dagger_\Sigma(t) U_0(t) \left| n \right> \right|^2 \nonumber \\
&=& \left| e^{-\ii E_n t/\hbar} \left< n \right| U^\dagger_\Sigma(t) \left| n \right> \right|^2 ,
\label{RMT1}
\ea
which is equal to the survival or return probability $P_{n}(t)$ of state $\left| n\right>$ under the action of Hamiltonian $\Hc=\Hc_0 + \Sigma$. Expanding in the basis $\left| \phi \right>$ of $\Hc$,  it is straightforward to obtain
\ba
M^{\rm nd}(t)&=&P_n(t)=\left| \sum_\phi \left| \left< n | \phi \right> \right|^2 e^{-\ii E_\phi t/\hbar} \right|^2 \nonumber \\
&=& \left| \int \ \eta (E) \ e^{-\ii E t/\hbar} \ \dd E \ \right|^2 , \label{RMT2}
\ea
where we have defined the local density of states (LDOS) 
\be
\eta(E)=\sum_\phi \left|\left< n |\phi\right>\right|^2 \delta(E-E_\phi),
\label{LDOS}
\ee
also known as the strength function \cite{Casati93,Casati96}. The LDOS tells us how much the original eigenstates expand into the basis of the perturbation. The derivation of Eq.~(\ref{RMTFGRdecay}) and the relation (\ref{RMT2}) thus serve to demonstrate that for random matrices $\eta(E)$ has a Lorentzian shape,
\be
\eta(E)=\frac{1}{\pi}\frac{\Gamma}{\Gamma^2+E^2}.
\label{LDOS2}
\ee
=
\subsection{Diagonal terms: The Lyapunov regime}
\label{sec:DTerms}

The remaining term in Eq.~(\ref{DiagNoDiag}), $M^{\dd}$ comes from the contribution of
trajectories $s$ and $s^\prime$ [from Eq.~(\ref{MComplete})]
that remain close in such a way that their action differences
$\Delta S_s$ are not uncorrelated. In a more precise sense, we will define such a
set of trajectories as those around which we can expand the perturbation as
\be
\Sigma(\bq,t)=\Sigma(\bq_0,t_0)+\nabla \Sigma\left[  \bq_0,t_0\right]
\cdot\left[  \mathbf{q}-\mathbf{q}_{0}\right]  + \frac{\partial \Sigma}{\partial t}(\bq_0,t_0) \ (t-t_0),
\label{PertExpansion}
\ee
where $\bq_0$ lies on the trajectory $s$ and $\bq$ in $s^\prime$.
Using this, the action difference
\begin{equation}
\Delta S_{s}(t)-\Delta S_{s^{\prime}}(t)=\int_{0}^{t}dt^{\prime}
\Sigma(\mathbf{q}_{s}(t^{\prime}),t^{\prime})-\Sigma(\mathbf{q}_{s^{\prime}}(t^{\prime
}),t^{\prime}), 
\label{DiagActionDiferences}%
\end{equation}
can be written as
\be
\Delta S_{s}(t)-\Delta S_{s^{\prime}}(t) \simeq \int_{0}^{t}dt^{\prime}
\nabla \Sigma\left[  \mathbf{q}_{s}(t^{\prime}),t^{\prime}\right]  \cdot\left[
\mathbf{q}_{s}(t^{\prime})-\mathbf{q}_{s^{\prime}}(t^{\prime})\right]
\label{DiagPhaseChange}
\ee
where the term with the time derivative becomes null because both coordinates
are evaluated at the same time, see Eq. (\ref{DiagActionDiferences}).

Taking in consideration these terms, the average of Eq.~\ref{MComplete} gives
\ba
M(t) &\simeq&\left(  \frac{\sigma^{2}}{\pi\hbar^{2}}\right)^{d}
\int \! d\mathbf{r}\int\! d\mathbf{r}^{\prime}\!\!
\sum_{{s({\bf r},{\bf r}_0, t)}\atop{s^\prime({\bf r^\prime},{\bf r}_0, t)}}
C_{s}C_{s^{\prime}}
  \left\langle \exp\left[  \frac{i}{\hbar}\left(  \Delta S_{s}
(t)-\Delta S_{s^{\prime}}(t)\right)  \right]  \right\rangle \nonumber \\
& & \exp\left[  -\frac{\sigma^{2}}{\hbar^{2}}\left(  \left(
\overline{\mathbf{p}}_{s}-\mathbf{p}_{0}\right)  ^{2}+\left(  \overline
{\mathbf{p}}_{s}-\mathbf{p}_{0}\right)  ^{2}\right)  \right]  .
\label{MDiagonal}
\ea
which, assuming again a Gaussian distribution for the fluctuations 
of the phase difference [Eq.~(\ref{DiagPhaseChange})], leads us to consider the force correlator of 
the perturbation
\ba
\lefteqn{
\left\langle \exp\left[  \frac{i}{\hbar}\left(  \Delta S_{s}
(t)-\Delta S_{s^{\prime}}(t)\right)  \right]  \right\rangle = 
}\nonumber \\
& &\exp \left[ -\frac{1}{\hbar^2}\int_0^t \dd \tb \int_0^t \dd \tbp
C_\nabla (| \bq(\bt)-\bqp(\btp)|,\tb-\tbp)
\left( \bq(\bt)-\bqp(\btp) \right) ^2 \right],
\label{AveragePhase}
\ea
where 
\be
C_\nabla (| \bq(\bt)-\bqp(\btp) |,\tb-\tbp) =
\left\langle \nabla \Sigma(\bq(\bt),\bt) \cdot \nabla \Sigma(\bqp(\btp),\btp)
\right\rangle.
\label{ForceCorrelator}
\ee

The difference between the intermediate points of both trajectories can be
expressed using the matrix $B$ of Eq.~(\ref{MatrixB}):
\be
\bq_s(\bt)-\bqp_{s^{\prime}}(\btp)=
B^{-1}(\bt)\left( \bpb_s - \bpbp \right) =
B^{-1}(\bt)B(t)\left( \br-\brp \right) .
\ee
In a chaotic system, $B^{-1}(t)$ is dominated by the largest eigenvalue
$e^{\lambda t}$. Therefore we make the simplification
$B^{-1}(\bt)B(t)=\exp{\left[\lambda(\bt-t)\right]}I$, with $I$ the unit matrix
and $\lambda$ the Lyapunov exponent. By doing so we have discarded marginally stable regions with anomalous time behavior, in a sense using the hypothesis of strong
chaos.

In order to continue we need further approximations of the force correlator.
As we will see in
the sequel, we can lose some generality here because the effect of the
correlator [Eq.~(\ref{ForceCorrelator})]
appears only in the prefactor of $M^d(t)$ and not in its exponent, and thus it is
not relevant to the shape of the decay.
We restrict ourselves only the cases where Eq.~(\ref{ForceCorrelator}) can be written as
\be
C_\nabla (\left| \bq-\bqp \right|,\bt-\btp)=(\nabla_\bq \cdot \nabla_\bqp)
\left\langle \Sigma(\bq,\bt) \Sigma(\bqp,\btp)\right\rangle.
\label{AproxForceCorrelator0}
\ee
Therefore, using Eq.~(\ref{Correlator}),
\be
C_\nabla (\left| \bq-\bqp \right|,\bt-\btp)=\overline{\Sigma^2}C_T(\bt-\btp)(\nabla_\bq \cdot \nabla_\bqp) C_S(|\bq-\bqp|).
\label{AproxForceCorrelator}
\ee
Notice that the correlator
\be
(\nabla_\bq \cdot \nabla_{\bqp}) C_S(q\equiv|\bq-\bqp|)=
\frac{1-d}{q}\frac{\partial C_S(q)}{\partial q}-
\frac{\partial^2 C_S(q)}{\partial q^2},
\ee
and we require that it decays sufficiently fast. Using the above expressions we obtain
\be
\left\langle \exp\left[  \frac{i}{\hbar}\left(  \Delta S_{s}
(t)-\Delta S_{s^{\prime}}(t)\right)  \right]  \right\rangle
=\exp\left[-\frac{A \left( \br-\brp \right) ^2}{\hbar^2} \right],
\label{DiagPhaseS}
\ee
where
\be
A=\frac{\overline{\Sigma^2}}{v^2} \int_0^t d\bt \int_{-\infty}^{\infty} d\tau
e^{2\lambda (\bt-t)} C_T(\tau)
 \left[ \frac{1-d}{\tau}\frac{\partial C_S(v\tau)}{\partial \tau}
-\frac{\partial^2 C_S(v\tau)}{\partial \tau^2} \right].
\label{AGeneral}
\ee 
In the regime where $C_T$ decays slowly, 
\be
A_S=\overline{\Sigma^2}\frac{1-e^{-2\lambda t}}{2 \lambda v}
\int_{-\infty}^{\infty} dq \left[ \frac{1-d}{q}\frac{\partial C_S(q)}{\partial q}
-\frac{\partial^2 C_S(q)}{\partial q^2} \right],
\label{AS}
\ee
and on the other end, when $C_T$ dominates the decay of $C_\nabla$, 
\be
A_T=\overline{\Sigma^2} \tau_0 \frac{(1-e^{-2\lambda t})}{2 \lambda}
\label{AT}
\ee

Using this result, the expression for the diagonal part of the Loschmidt echo is
\ba
M^{\mathrm{d}}(t)&=&\left(  \frac{\sigma^{2}}{\pi\hbar^{2}}\right)  ^{d}%
\int\mathrm{d}\mathbf{r}\int\mathrm{d}\mathbf{r}^{\prime}\ \sum_{s}\ C_{s}%
^{2}\ \nonumber \\
&\times& \
\exp\left[  -\frac{2\sigma^{2}}{\hbar^{2}}\left(  {\overline{\mathbf{p}%
}}_{s}-\mathbf{p}_{0}\right)  ^{2}\right]  \ \exp{\left[  -\frac{A}{2\hbar
^{2}}\left(  \mathbf{r}-\mathbf{r}^{\prime}\right)  ^{2}\right]  }\ .
\label{SCDAlmost}
\ea

\noindent A Gaussian integration over $(\mathbf{r}-\mathbf{r}^{\prime})$
gives

\ba
M^{\mathrm{d}}(t)=\left(  \frac{\sigma^{2}}{\pi\hbar^{2}}\right)  ^{d}%
\ \int\mathrm{d}\mathbf{r}\ \sum_{s}\ C_{s}^{2}\ \left(  \frac{2\pi\hbar^{2}%
}{A}\right)  ^{d/2}\ \exp\left[  -\frac{2\sigma^{2}}{\hbar^{2}}\left(
{\overline{\mathbf{p}}}_{s}-\mathbf{p}_{0}\right)  ^{2}\right]  \ .
\label{SCDAlmost2}
\ea

\noindent The factor $C_{s}^{2}$ reduces to $C_{s}$ when we make the change of
variables from $\mathbf{r}$ to ${\overline{\mathbf{p}}}$. In the long-time
limit $C_{s}^{-1}\propto e^{\lambda t}$, since it is basically a second derivative of the action with respect to initial and final positions. For the same reason, for short times it should obey a ballistic behavior $C_{s}^{-1}=(t/m)^{d}$. Using a form that interpolates between these two limits we
finally obtain the main result of this section

\ba
M^{\mathrm{d}}(t)&=&\left(  \frac{\sigma^{2}}{\pi\hbar^{2}}\right)  ^{d}%
\int\mathrm{d}{\overline{\mathbf{p}}}\ \left(  \frac{2\pi\hbar^{2}}{A}\right)
^{d/2}\left(  \frac{m}{t}\right)  ^{d}\ \exp\left[  -\lambda t\right]
\ \exp\left[  -\frac{2\sigma^{2}}{\hbar^{2}}\left(  {\overline{\mathbf{p}}%
}-\mathbf{p}_{0}\right)  ^{2}\right]  \nonumber \\
&=&\overline{A}\exp\left[  -\lambda t\right]  \ , 
\label{SCLEFinal}%
\ea 

\noindent with $\overline{{A}}=[\sigma m/(A^{1/2}t)]^{d}$. Since the integral
over ${\overline{\mathbf{p}}}$ is concentrated around $\mathbf{p}_{0}$, the
exponent $\lambda$ is taken as the phase-space average value on the
corresponding energy shell. The coupling $\Sigma$ appears only in the
prefactor (through $\overline{{A}}$) and therefore its detailed description is
not crucial in discussing the time dependence of $M^{\mathrm{d}}$.

\section{Decay regimes of the Loschmidt echo}
\label{sec:DvsND}

In the previous sections we studied the time dependence of two different
types of terms arising in the semiclassical expression of the LE from the
separation of two sets of trajectories. The final expression for Eq.
(\ref{DiagNoDiag}) is then

\be
M(t)=\overline{A} \exp{(-\lambda t)} + B \exp{(-\Gamma t/\hbar)},
\label{DecompositionLE}
\ee
where $B$ is a constant and $\Gamma=\hbar / \tilde{\tau}$.
From this expression one concludes that $M(t)$ in the
semiclassical regime presents an exponential decay with a rate given by 
the minimum between $\lambda$ and $\Gamma / \hbar$. 

For strong perturbations, 
when the diagonal terms dominate, the decay rate is given by $\lambda$ and we
say that the LE is in the Lyapunov regime. On the other end, for smaller
perturbations when the non diagonal terms prevail, we showed that the decay rate is related to that given by a Fermi golden rule (FGR) approach to the problem. Thus, this regime is called the FGR regime. The crossover between regimes at $\Gamma / \hbar=\lambda$ is an important issue and will be discussed in the next chapter.

The Lyapunov regime is of particular interest not only because it presents a
perturbation independent decay rate, but more importantly because the decay rate
is given by a classical quantity. As noted in previous discussions, the quantum
mechanics of classically chaotic systems rarely presents dynamical evidence of
chaos, with a few notable exceptions \cite{Heller84}.
The LE represents, in this context, a good starting point to develop a quantum theory of chaos. For this purpose, it needs to be well defined and, furthermore, it needs to recover the proper classical behavior in the semiclassical limit.

The limits of small $t$ and weak $\Sigma$ yield an infinite $\overline{{A}}$, 
and thus creates a divergence in Eq.~(\ref{DecompositionLE}). However, the calculations 
are only valid in certain intervals of $t$ and strength of the perturbation. 
The times considered should verify $\Gamma t\geq \hbar$.  
Long times, resulting in the failure of the diagonal approximations 
[Eqs.~(\ref{MComplete}) and (\ref{MDiagonal})], or the assumption that 
the trajectories are unaffected by the perturbation, are excluded from this 
analysis. Similarly, the small values of $\Sigma$ are not properly treated in 
the semiclassical calculation of the diagonal term $M^{\mathrm{d}}(t)$, while 
for strong $\Sigma$ the perturbative treatment of the actions is expected to 
break down and the trajectories become affected by the quenched disorder. This 
last condition translates into a \textquotedblleft transport 
mean-free-path\textquotedblright\cite{Jalabert96,Jalabert00} $\tilde{\ell}_{\mathrm{tr}%
}=4(k\xi)^{2}\tilde{\ell}$ being much larger than the typical dimension $L$ of our 
system. In the limit $k\xi\gg1$ that we are working with, it is not difficult to satisfy the condition $\tilde{\ell}_{\mathrm{tr}}\gg L\gg\tilde{\ell}$. 
 
It is worth noting that the width $\sigma$ of the initial wave-packet 
is a prefactor of the diagonal contribution. The non-diagonal term, on the other hand, is 
independent on the initial wave-packet. Therefore, as stated in 
Ref.~\cite{Jacquod02}, and numerically verified in \cite{Wang02}, changing our initial state [Eq.~(\ref{InitialPacket})] into a coherent superposition of $N$ wave-packets 
would reduce $M^{\mathrm{d}}$ by a factor of $N$ without changing 
$M^{\mathrm{nd}}$. The localized character of the initial state is then a key 
ingredient in order to obtain the behavior observed here. In particular, only a
FGR regime is observed when the initial state is random \cite{Wang02} or an
eigenstate of the Hamiltonian \cite{WisniackiCohen}

Let us thoroughly list the decay regimes of the LE in order of increasing
perturbation, thus summarizing and placing into context the results of this chapter (the regimes are depicted qualitatively in Fig.~\ref{fig:Schematics}):

\begin{figure}[htb]
\begin{center}
\leavevmode
\epsfxsize 4in
\epsfbox{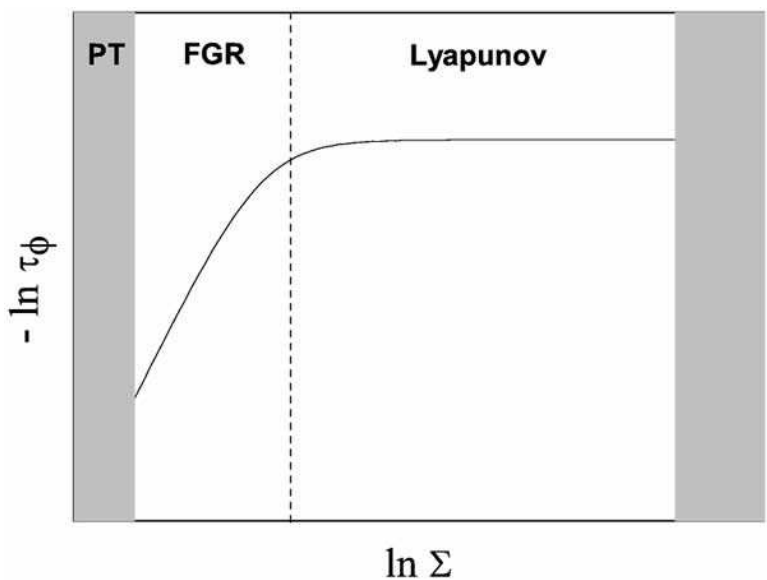}
\caption{Schematics of the different regimes of the LE viewed through the typical time of the exponential decay vs the strength of the perturbation (of course this plot highlights only the FGR and Lyapunov regimes). The gray area on the left is regime (1), where perturbation theory applies. The log-log scale shows that the FGR exponent is a power law of the perturbation [regime (2)]. After the Lyapunov regime, (3), the perturbation dominates the system and no general prediction is available.}
\label{fig:Schematics}
\end{center}
\end{figure}

\begin{enumerate}
\item For extremely small perturbations, where $\overline{\Sigma} < \Delta$ 
 ($\Delta$ is the mean level spacing), the LE can described by quantum perturbation theory  \cite{Peres84,PeresBook}.
The result is a Gaussian decay with a rate that depends quadratically on the perturbation
strength. This decay regime is also observable for very small times.
\item For $\Sigma > \Delta$, one enters the Fermi Golden rule regime. Actually,
as will see in the sequel, the decay observed in this regime is more
general than the cases where the FGR applies \cite{Wisniacki02,Casati04}. The general observation is an exponentially decaying LE with a rate given by the width of the local density of states of the perturbation (LDOS). In any case we denote this regime as a FGR regime, to adhere to common notation. It should be noted that the transition between the Gaussian
perturbative decay and this first exponential decay can be fully described by a
uniform semiclassical approach \cite{Cerruti03}.
\item When the underlying classical dynamics is chaotic, for stronger
perturbations (such that $\Gamma / \hbar > \lambda$ the Lyapunov exponent), the LE
decays exponentially but now with a rate $\lambda$ independent of the
perturbation strength and shape, determined only by the classical chaos \cite{Jalabert01}. The
observation of this regime usually requires that the initial state is localized.
The perturbation only enters as a prefactor, as well as a polynomial dependence
in time which deviates from the expected classical behavior. The smoothness of
transition from the FGR to this Lyapunov regime depends on the chaoticity of the
underlying classical system. For stronger chaos (larger $\lambda$), the
fluctuations in phase that cause the decay of the non diagonal terms are strong
and thus the diagonal term emerges dominant. In the opposite case, it has been
shown \cite{Casati04} that the decay rate can present strong oscillations around the Lyapunov exponent when the perturbation strength is near the critical one.
\item For extremely strong perturbations (when $\Sigma$ dominates the dynamics),
it has been noted that there is a saturation of the decay rate at the band
width of the unperturbed Hamiltonian \cite{Jacquod01}. This occurs when ${\cal H}_0$ ``cannot stand''
stronger perturbations which are much larger than the largest frequency in the
system, namely its bandwidth. In this regime there is evidence that the LE again
follows the autocorrelation function, the Fourier transform of the LDOS \cite{Barnett}. The shape of the decay then depends on the particular form of the LDOS.
\end{enumerate}

This thesis is focused on the FGR and the Lyapunov regimes for strongly classically chaotic systems.  Of course, this by no means is equivalent to saying that 
they are the most relevant regimes in all physical situations. In general,
this is a question whose answer lies in the eyes of the beholder. However, the fast
growing control over experimental systems in areas such as quantum dots, cold
atoms, or other insofar unthought of systems, let us imagine a near future where a simple knob will tune
the experiment to any of the above regimes.

The analytical results have been presented so far in a very general way, thus there is a need for examples to gain insight. For this, in the next section we will particularize the theory to different models. These specific results will also be useful to perform numerical tests of the theory, to be presented in the next chapter. 

\section{Semiclassical Loschmidt echo: examples}
\label{sec:SCExamples}

In this section we will see how the semiclassical theory for the LE
applies to particular examples. First we will consider a particular form of the
correlators for the perturbation which will allow us to obtain closed
expressions for the diagonal and non--diagonal terms of the previous section. 
Second, we will study the LE in a model purposely devised to break
many of the assumptions in the theory, such as
the continuous evolution of $\Hc_0$
and the presence of disorder in the perturbation. The model is a Lorentz gas
with a perturbation in the mass tensor of the particle, which will prove
numerically advantageous (compared to bound systems) in the next chapter.
Finally, we will study a toy model which, although it is not chaotic, will present instability. Its most important feature is that it is exactly soluble, and this always allows deep investigations of the inner aspects of a theory. 

\subsection{Gaussian decay of correlators}
\label{sec:GaussianCorrelators}

A general class of perturbations can be defined by the 
particular form of the correlators [Eq.~(\ref{Correlator})], 
\be
C_{S}(r)=\frac{2}{\sqrt{\pi}}\exp(-r^{2}/\xi^{2});\quad C_{T}(\tau)=\frac{2}{\sqrt{\pi}}\exp(-\tau^{2}/\tau_0^{2}).
\ee
Under the assumption that $t$ is large compared to $\tau_0$ and $\xi/v$, 
let us replace in Eq.~(\ref{VarianceS}) 
\be
\left\langle \Delta S_{s}(t)^{2}\right\rangle =
\frac{4 \overline{\Sigma^2}}{\sqrt{\pi}}
\int_{0}^{t}d\overline{t}%
\int_{-\infty}^{\infty}d\tau
\exp(-v^{2}\tau^{2}/\xi^{2})\exp
(-\tau^{2}/\tau_{0}^{2})
\ee
and obtain the decay rate for the FGR regime
\be
\frac{1}{\widetilde{\ell}} = \frac{4 \overline{\Sigma^2}}
{v\sqrt{\pi}\hbar^{2}\sqrt{(v/\xi)^{2}+1/\tau_{0}^{2}}}.
\ee

For the diagonal terms, let us note that 
\ba
C_\nabla (\left| \bq-\bqp \right|,\bt-\btp)&=&
\frac{\overline{\Sigma^2}}{\xi^2}
\left(d-\frac{|\bq-\bqp|^2}{\xi^2}\right)\frac{1}{\xi^2}C_S(|\bq-\bqp|)
C_T(\bt-\btp),
\label{GForceCorrelator}
\ea
Using the above expression we can also obtain the prefactor of the diagonal terms
\be
A=\frac{\sqrt{2\pi} \ \overline{\Sigma^2}(1-e^{-2\lambda t})
\left[(1+d)v^2+d \xi^2/\tau_0^2 \right]}
{2 \lambda \xi^4 \sqrt{(v/\xi)^{2}+1/\tau_{0}^{2}}}.
\label{GAS}
\ee

\subsubsection{Quenched disorder}
\label{sec:QuenchedDisorder}

A particular case of the correlators specified in this section is the 
{\em quenched disorder} model studied in the original paper by Jalabert and Pastawski \cite{Jalabert01}. Here the perturbation consists of $N_i$ impurities with a Gaussian 
potential characterized by the correlation length $\xi$,

\be 
\Sigma=\tilde{V}(\mathbf{r})=\sum_{\alpha=1}^{N_{i}} \ \frac{u_{\alpha}}%
{(2\pi\xi^{2})^{d/2}}\ \ \exp{\left[  -\frac{1}{2\xi^{2}}\left( 
\mathbf{r}\!-\!\mathbf{R}_{\alpha}\right)  ^{2} \right]  } \ . 
\label{QuenchedSigma}%
\ee 
  
\noindent The independent impurities are uniformly distributed (at positions 
$\mathbf{R}_{\alpha}$) with density $n_{i} =N_{i}/\Omega$, ($\Omega$ 
is the sample volume). The strengths $u_{\alpha}$ obey $\langle u_{\alpha 
}u_{\beta}\rangle=u^{2}\delta_{\alpha\beta}$. The correlation function 
$C_S$ of the above potential is given by%
 
\be 
C_{\tilde{V}}(|\mathbf{q}-\mathbf{q}^{\prime}|) = \frac{u^{2}n_{i}}{(4\pi 
\xi^{2})^{d/2}}\ \ \exp{\left[  -\frac{1}{4\xi^{2}}(\mathbf{q}-\mathbf{q}%
^{\prime})^{2}\right]  } \ , \label{QuenchedCorrelator}%
\ee 
and hence is a particular example of the general case of Gaussian correlators
presented above, with $C_T=1$. 
In particular, the mean free path of the perturbation writes
 
\be 
\frac{1}{\tilde{\ell}}=\frac{u^{2}n_{i}}{v_{0}^{2}\hbar^{2}(4\pi\xi^{2})
^{(d-1)/2}}
\ .
\label{QuenchedMFP}%
\ee 
 
The prefactor $A$ of the diagonal terms is 

\be 
A=\frac{(d-1)u^{2}n_{i}}{4\lambda v_{0}\xi^{2}(4\pi\xi^{2})^{(d-1)/2}} \ . 
\label{QuenchedA}%
\ee  

As mentioned earlier, we gave this particular example of quenched disorder
because it was the first analytical calculation that showed the existence of the
Lyapunov regime of the LE, and also because it will be treated numerically in
the following chapter. 
However, we have not yet seen the existence of a Lyapunov regime for a static and uniform perturbation (neither temporal nor spatial noise).
Moreover, no particular Hamiltonian $\Hc_0$ has been written. In the next
section we will produce such results for an experimentally relevant system under the
presence of a non-disordered perturbation.

\subsection{Loschmidt echo in a Lorentz gas}
\label{sec:LorentzGas} 
  
We will consider in this section the case where the system Hamiltonian 
$\mathcal{H}_{0}$ represents a two dimensional Lorentz gas, i.e. a particle 
that moves freely (with speed $v_0$) between elastic collisions (with specular 
reflections) with an irregular array of hard disk scatterers (impurities) of 
radius $R$. Such a billiard system is a paradigm of classical dynamics, and 
has been proven to exhibit mixing and ergodic behavior, while its dynamics for 
long distances is diffusive \cite{ArnoldBook,DorfmanBook,AleinerLarkin}. 
The existence of rigorous results for the Lorentz gas has made it a 
preferred playground to study the emergence of irreversible behavior out of 
the reversible laws of classical dynamics \cite{DorfmanBook}. Moreover, 
anti-dot lattices defined in a two dimensional electron gas 
\cite{Weiss91,Richter93,Portal01} constitute an experimentally 
realizable quantum system where classical features have been identified and 
measured. The terms anti-dot, impurity and disk will be used indistinctly. 
 
 The Lorentz gas has been thoroughly studied (for example, in Ref. \cite{DorfmanBook}), and to avoid straying away from the subject we shall not 
discuss here its classical dynamics in detail. A brief presentation can be found
in appendix \ref{appe:Lorentz}, where some of its quantum properties are also discussed. Here, we only need to recall the properties and assumptions that will be used in the analytical treatment of the LE. 

We require that each disk has an exclusion region $R_{\mathrm{e}}$ from 
its border, such that the distance between the centers of any pair of disks is 
larger than a value $2R_{\mathrm{e}}>2R.$ Such a requirement is important to 
avoid the trapping of the classical particle and the wave-function localization in the quantum case: both situations that would unnecessarily complicate the analysis. We will consider the anti-dots density to be roughly uniform. 
Within these restrictions, the exclusion distance $R_{\mathrm{e}}$ 
completely determines the dynamical properties of the Lorentz gas. Among them, 
we are interested in the Lyapunov exponent (measuring the rate of separation 
of two nearby trajectories) and the elastic mean free path $\ell$ (given by the 
typical distance between two collisions). Analytical and numerical methods to obtain the Lyapunov exponent are presented in Appendix \ref{appe:Lorentz}. For the distribution of lengths between successive collisions, a shifted Poisson distribution
 
\be 
P(s)=\left\{ 
\begin{array} 
[c]{cl}
{\displaystyle{
\frac
{\exp\left[  -\frac{s}{(\ell-2(R_{\mathrm{e}}-R))} \right]  } 
{
(\ell-2(R_{\mathrm{e}}-R))
\exp\left[  -\frac{2(R_{\mathrm{e}}-R)}
{(\ell -2(R_{\mathrm{e}}-R)) } \right]  
}
}} & \hspace{1cm}%
\mbox{if $s>2(R_{\rm e}-R)$}\ ,\\ 
0 & \hspace{1cm}\mbox{if $s<2(R_{\rm e}-R)$}\ , 
\end{array} 
\right.  \label{LDistribution}%
\ee 
 
\noindent is a reasonable guess, which yields $\left\langle 
s\right\rangle =\ell =v/\tau_{\mathrm{e}}$. This distribution is consistent with 
numerical simulations in the range of anti-dot concentration that 
we are interested in (see appendix \ref{appe:Lorentz}, Fig. \ref{fig:HistogramS}). Since velocity both\ $v_{0}$ and momentum 
$p_{0}$\ are conserved within this model (all collisions are elastic) we will omit their 
subindex. 
   
\subsubsection{The perturbation Hamiltonian} 
\label{sec:MassTensor}
 
In order to shed light on the dependence of the LE on the details of $\Sigma$, we contemplate a perturbation radically different to that considered in Sec.~\ref{sec:SCLE}: a distortion of the mass tensor, introduced in Ref.~\cite{Cucchietti02Lorentz} and briefly discussed in the sequel. 
 
The isotropic mass tensor of $\mathcal{H}_{0},$ of diagonal components 
$m_{0},$ can be distorted by introducing an anisotropy such that $m_{xx}%
=m_{0}(1+\alpha)$ and $m_{yy}=m_{0}/(1+\alpha).$ This perturbation is inspired 
by the effect of a slight rotation of the sample in the problem of dipolar 
spin dynamics \cite{Pastawski98}, which modifies the mass of the spin wave 
excitations. The kinetic part of the Hamiltonian is now affected by the 
perturbation, which can be written as
\be 
\Sigma(\alpha)=\alpha\ \frac{p_{y}^{2}}{2m_{0}}-\frac{\alpha} {1+\alpha} 
\ \frac{p_{x}^{2}}{2m_{0}}. \label{MTPerturbation}%
\ee 
  
\noindent In our analytical work we will stay within the leading order 
perturbation in $\alpha$. That is, 
\be 
\Sigma(\alpha)=\frac{\alpha}{2m_{0}} \ \left(  p_{y}^{2}-p_{x}^{2}\right)  . 
\label{PerturbationAprox}%
\ee 
 
Making the particle \textquotedblleft heavier\textquotedblright\ in the $x$ 
direction (i.e. we consider a positive $\alpha)$ modifies the equations of 
motion without changing the potential part of the Hamiltonian. It is important 
to notice that, unlike the case of quenched disorder, the perturbation 
(\ref{MTPerturbation}) is non-random, and will not be able to provide any 
averaging procedure by itself, but only through the underlying chaotic dynamics. 
 
For a hard wall model, such as the one we are considering, the perturbation 
(\ref{MTPerturbation}) is equivalent to having non-specular reflections. 
This allows to show (see appendix \ref{appe:Lorentz}) that the distortion of 
the mass tensor is equivalent to an area conserving deformation of 
the boundaries  $x\rightarrow x(1+\xi)$, $y\rightarrow 
y/(1+\xi),$ as used in other works on the LE \cite{Wisniacki02}, 
where $\xi=\sqrt{1+\alpha}-1$ is the stretching parameter. 
 
\subsubsection{Semiclassical Loschmidt echo} 
\label{sec:SCLG}
 
This section presents the calculations of the Loschmidt echo for the system previously described. $\mathcal{H}_{0}$ describes a Lorentz gas and $\Sigma$ is given by Eq.~(\ref{MTPerturbation}). Clearly, the approach is to adapt the semiclassical method  of Sec.~\ref{sec:SCLE} to this particular perturbation, as well as the modifications introduced by the discontinuity of the dynamics (elastic collisions) of the classical Hamiltonian. 

As before, we take as initial state a Gaussian wave-packet of width $\sigma$ [Eq.~(\ref{InitialPacket})].  
The semiclassical approach to the LE under a weak perturbation $\Sigma$ is given by Eq.~(\ref{MComplete}), with the extra phase%
 
\be 
\Delta S_{s}=\int_{0}^{t}\mathrm{d}\overline{t}\ \Sigma_{s}(\mathbf{q(}%
\overline{t}),\mathbf{\dot{q}(}\overline{t})). 
\ee 
 
\noindent The sign difference with Eq.~(\ref{DeltaS}) is 
because the perturbation is now in the kinetic part of the 
Hamiltonian. On the other hand, this sign turns out to be irrelevant because we will only consider the variance of $\Delta S$.
 
Using the perturbation of Eq.~(\ref{PerturbationAprox}), we only have to integrate 
a piecewise constant function (in between collisions with the scatterers), 
obtaining 
\be 
\Delta S_{s}=\frac{\alpha m_{0}}{2} \ \sum_{i=1}^{N_{s}}\tau_{i}\left( 
2v_{y_{i}}^{2}-v^{2}\right)  . \label{LGDeltaAction}%
\ee 
 
\noindent We have used $v_{x}^{2}+v_{y}^{2}=v^{2},$ and have 
defined $\tau_{i}$ as the free flight time ending with the $i$-th 
collision, $v_{y_{i}}$ is the $y$ component of the velocity in 
such an interval, and $N_{s}$ as the number of collisions that the 
trajectory $s$ suffers during the time $t$. 
 
As previously noted, the free flight 
times $\tau_{i}$ (or the inter-collision length $v\tau_{i})$ have 
a shifted Poisson distribution [Eq.~(\ref{LDistribution})].
This observation will turn out to 
be important in the analytical calculations that follow since the 
sum of Eq.~(\ref{LGDeltaAction}) for a long trajectory can be 
taken as composed of uncorrelated random variables following the 
above mentioned distribution. Unlike the case of 
Sec.~\ref{sec:SCLE}, the randomness is not associated with the 
perturbation (which is fixed), but with the diffusive dynamics generated by $\mathcal{H} 
_{0}$. 
 
\subsubsection{Non-diagonal contribution} 
\label{sec:NDLG}
 
As in the case of Sec.~\ref{sec:NDTerms}, the non-diagonal contribution is given 
by the second moment 
\be 
\left\langle \Delta S_{s}^{2}\right\rangle =\frac{\alpha^{2}m_{0}^{2}}{4} 
\ \left\langle \sum_{i,j=1}^{N_{s}}\tau_{i}\tau_{j}\left(  2v_{y_{i}}^{2}- 
v^{2}\right)  \left(  2v_{y_{j}}^{2}-v^{2}\right)  \right\rangle \ . 
\ee 
Separating in diagonal ($i=j$) and non-diagonal ($i\neq j$) contributions (in 
pieces of trajectory) we have 
\ba
\left\langle \Delta S_{s}^{2}\right\rangle &=&\frac{\alpha^{2}m_{0}^{2}N_{s}}
{4} \left[  \left\langle \tau_{i}^{2}\right\rangle \left(  4\left\langle 
v_{y_{i}}^{4}\right\rangle -4 v^{2}\left\langle v_{y_{i}}^{2}\right\rangle 
+v^{4}\right) \right. \nonumber \\
& & +\left. \left(  N_{s}-1 \right)  \left\langle \tau_{i} 
\right\rangle ^{2}\left(  4\left\langle v_{y_{i}}^{2}\right\rangle ^{2}
-4v^{2}\left\langle v_{y_{i}}^{2}\right\rangle +v^{4}\right)  \right]  . 
\label{DeltaSLG}
\ea 
We have assumed that different pieces of the trajectory ($i\neq j$) are 
uncorrelated, and that within a given piece $i,$ $\tau_{i}$ and $v_{y_{i}}$ 
are also uncorrelated. According to the distribution of time-of-flights 
(\ref{LDistribution}) we have%
 
\begin{subequations}
\label{AllTau}
\ba
\left\langle \tau\right\rangle &=&\tau_{\mathrm{e}} \ ,\label{Tau0} \\
\displaystyle \left\langle \tau^{2}\right\rangle  &=&2\tau_{\mathrm{e}}^{2} 
\ . \label{Tau1}%
\ea
\end{subequations}
 
Assuming that the velocity in the pieces of trajectories distribution is isotropic ($P(\theta)=1/2\pi$, where $\theta$ is the angle of the velocity with respect to a fixed axis) is in good agreement with numerical simulations, and results in%
 
\begin{subequations}
\label{AllVelocities}%
\ba
\left\langle v_{y}^{2}\right\rangle  &=&v^{2}\left\langle \sin^{2}%
\theta\right\rangle =\frac{v^{2}}{2} \ ,\label{VelY0} \\
\displaystyle \left\langle v_{y}^{4}\right\rangle  &  =&v^{4}\left\langle 
\sin^{4}\theta\right\rangle =\frac{3v^{4}}{8} \ . \label{VelY1}%
\ea
\end{subequations} 
 
Replacing in Eq.~(\ref{DeltaSLG}) we obtain that $4\left\langle v_{y_{i}}^{2}\right\rangle ^{2}%
-4v^{2}\left\langle v_{y_{i}}^{2}\right\rangle +v^{4}=0,$ implying a 
cancellation of the cross terms of $\left\langle \Delta S_{s}^{2}\right\rangle 
$, consistently with the lack of correlations between different pieces that we 
have assumed. We therefore get

\be 
\left\langle \Delta S_{s}^{2}\right\rangle =\frac{\alpha^{2}m_{0}^{2}N_{s}%
\tau_{\mathrm{e}}^{2}v^{4}}{4} \ . 
\ee 
For a given $t,$ $N_{s}$ is also a random variable, but for $t\gg 
\tau_{\mathrm{e}}$ we can approximate it by its mean value $t/\tau 
_{\mathrm{e}}$ and write 
\be 
\left\langle \Delta S_{s}^{2}\right\rangle =\frac{\alpha^{2}m_{0}^{2}v^{4}%
\tau_{\mathrm{e}} t}{4} \ . 
\ee 
We therefore have for the average echo amplitude
 
\ba
\left\langle m(t)\right\rangle &\simeq& \exp{\left[  -\frac{\alpha^{2}m_{0}%
^{2}v^{4}\tau_{\mathrm{e}}t}{8\hbar^{2}}\right]  }\left(  \frac{\sigma^{2}%
}{\pi\hbar^{2}}\right)  ^{d/2}\int d\mathbf{r}\sum_{s}C_{s}\exp{\left[ 
-\frac{\sigma^{2}}{\hbar^{2}}\left(  {\overline{\mathbf{p}}}_{s}%
-\mathbf{p}_{0}\right)  ^{2}\right]  } \nonumber \\
&=&\exp{\left[  -\frac{vt}{2\tilde{\ell}%
}\right]  }\ , \label{AmplitudeAverage}%
\ea
where we have again used $C_{s}$ as a Jacobian of the transformation from 
$\mathbf{r}$ to ${\overline{\mathbf{p}}}_{s}$ and we have defined an effective 
mean free path of the perturbation by 
\be 
\frac{1}{\tilde{\ell}}=\frac{m_{0}^{2}v^{2}\ell}{4\hbar^{2}}\ \alpha^{2}\ . 
\label{LGFGRexponent}%
\ee 
 
The effective mean free path $\tilde{\ell}=v~\tilde{\tau}$ should be 
distinguished from $\ell=v\tau_{\mathrm{e}}$ since the former is associated with 
the dynamics of $\Sigma$ and $\mathcal{H}_{0}$, while the latter is only fixed 
by $\mathcal{H}_{0}$. Obviously, these results are only applicable in the case 
of a weak perturbation verifying $\tilde{\ell}\gg\ell$. From 
Eq.~(\ref{AmplitudeAverage}) one re--obtains that the non-diagonal component of the LE  
\be 
M^{\mathrm{nd}}(t)=\left\vert \left\langle m(t)\right\rangle \right\vert 
^{2}=\exp\left[  -\frac{vt}{\tilde{\ell}}\right]  \ . 
\ee 
  
\subsubsection{Diagonal contribution} 
\label{sec:LGDTerms}
 
As in Sec.~\ref{sec:DTerms}, we have to discuss separately the contribution to 
the LE [Eq.~(\ref{MComplete})] originated by pairs of trajectories $s$ and 
$s^{\prime}$ that remain close to each other. In that case the terms $\Delta 
S_{s}$ and $\Delta S_{s^{\prime}}$ are not uncorrelated. The corresponding 
diagonal contribution to the LE is given by Eq.~(\ref{MDiagonal}), and therefore 
we have to calculate the extra actions for $s\simeq s^{\prime}$. Let us represent by $\theta$ ($\theta+\delta$) the angle of the trajectory $s$ ($s^{\prime}$) with a fixed direction (i.e. that of the $x$-axis). We can then write the perturbation 
[Eq.~(\ref{MTPerturbation})] for each trajectory as%
 
\begin{subequations}
\label{allSIG}%
\ba
\Sigma_{s}  &=& \frac{\alpha}{2m_{0}} \ p^{2} \ (2\sin^{2}\theta-1) 
\ ,\label{eq:SIG0} \\
\displaystyle \Sigma_{s^{\prime}}  &=& \frac{\alpha}{2m_{0}} \ p^{2} 
\ (2\sin^{2}\theta-2\delta\sin2\theta-1)+\mathcal{O}(\delta^{2}) \ . 
\label{eq:SIG1}%
\ea
\end{subequations} 
 
Assuming that the time-of-flight $\tau_{i}$ is the same for $s$ and 
$s^{\prime}$ (correct up to the same order of approximation in $\delta$) we have 

\be 
\Delta S_{s}-\Delta S_{s^{\prime}}=\frac{\alpha p^{2}}{m_{0}} \int_{0}^{t}d 
\overline{t} \ \delta(\overline{t}) \ \sin{\left[  2\theta(\overline 
{t})\right]  } \ . 
\ee

The angles $\delta$ alternate in sign, but the exponential divergence between 
nearby trajectories allows to approximate the angle difference after $n$ 
collisions as $\left|  \delta_{n}\right|  =\left|  \delta_{1}\right| 
e^{\lambda n\tau_{\mathrm{e}}}$. A detailed analysis of the classical dynamics \cite{DorfmanBook}
shows that the distance between the two trajectories grows with the number of 
collisions as $d_{1}=\left|  \delta_{1}\right|  v\tau_{1},$ $d_{2}%
=d_{1}+\left|  \delta_{2}\right|  v\tau_{2},$ and therefore%
 
\be 
d_{N_{s}} \simeq v\sum_{j=1}^{N_{s}}\left|  \delta_{j}\right|  \tau_{j} \simeq 
v\tau_{\mathrm{e}}\left|  \delta_{1}\right|  \ \sum_{j=1}^{N_{s}%
}e^{(j-1)\lambda\tau_{\mathrm{e}}} = \ell\ \left|  \delta_{1}\right| 
\ \frac{e^{N_{s}\lambda\tau_{\mathrm{e}}}-1} {e^{\lambda\tau_{\mathrm{e}}}-1} 
\ . 
\ee

\noindent By eliminating $\left|  \delta_{1}\right|  $ we can express an 
intermediate angle $\delta(\overline{t})$ as a function of the final 
separation $\left|  \mathbf{r}-\mathbf{r}^{\prime}\right|  =d_{N_{s}}$, 
\be 
\delta(\overline{t})\simeq\frac{\left|  \mathbf{r}-\mathbf{r}^{\prime}\right| 
}{\ell} \ \frac{e^{\lambda\tau_{\mathrm{e}}}-1}{e^{\lambda t}-1}%
e^{\lambda\overline{t}} \ , 
\ee 
where again we have used that $t=N_{s}\tau_{\mathrm{e}}$ is valid on average. 
Assuming that the action difference is a Gaussian random variable, in the 
evaluation of Eq.~(\ref{MDiagonal}) we only need its second moment%
 
\ba
\left\langle \left(  \Delta S_{s}-\Delta S_{s^{\prime}}\right)  ^{2} \right\rangle 
&\simeq& \alpha^{2}m_{0}^{2}v^{4}\ \frac{\left\vert \mathbf{r}%
-\mathbf{r}^{\prime}\right\vert ^{2}}{\ell^{2}}\ \left(  \frac{e^{\lambda 
\tau_{\mathrm{e}}}-1}{e^{\lambda t}-1}\right)  ^{2}\ 
\nonumber \\ &\times&
\left\langle \int_{0}%
^{t}\mathrm{d}\overline{t}\int_{0}^{t}d\overline{t}^{\prime}\ e^{\lambda 
\overline{t}+\lambda\overline{t}^{\prime}}\ \sin{\left[  2\theta(\overline 
{t})\right]  }\ \sin{\left[  2\theta(\overline{t}^{\prime})\right] 
}\right\rangle \ . 
\ea

As before, we assume that the different trajectory pieces are uncorrelated and the angles 
$\theta_{i}$ uniformly distributed.\ Therefore $\left\langle \sin\left[ 
2\theta_{i}\right]  \sin\left[  2\theta_{j}\right]  \right\rangle =\delta 
_{ij}/2$ and%
 
\ba
\left\langle \left(  \Delta S_{s}-\Delta S_{s^{\prime}}\right)  ^{2}%
\right\rangle  &  \simeq& \frac{\alpha^{2}}{2}\left(  \frac{m_{0}v^{2}}{\ell 
}\right)  ^{2}\left\vert \mathbf{r}-\mathbf{r}^{\prime}\right\vert ^{2}\left( 
\frac{e^{\lambda\tau_{\mathrm{e}}}-1}{e^{\lambda t}-1}\right)  ^{2}\sum 
_{i=1}^{N_{s}}\left\langle \int_{t_{i-1}}^{t_{i}}\mathrm{d}\overline 
{t}\ e^{\lambda\overline{t}}\right\rangle ^{2} \nonumber \\ 
& =& \frac{\alpha^{2}}{2}\left(  \frac{m_{0}v^{2}}{\lambda\ell}\right) 
^{2}\left\vert \mathbf{r}-\mathbf{r}^{\prime}\right\vert ^{2}\frac{\left( 
e^{\lambda\tau_{\mathrm{e}}}-1\right)  ^{4}}{\left(  e^{\lambda t}-1\right) 
^{2}}\frac{e^{2\lambda N_{s}\tau_{\mathrm{e}}}-1}{e^{2\lambda\tau_{\mathrm{e}%
}}-1}
\nonumber \\
&=&A\ \left\vert \mathbf{r}-\mathbf{r}^{\prime}\right\vert ^{2}\ , 
\label{DifActions}%
\ea

\noindent where we have taken the limit $\lambda t\gg1$, and defined%
 
\be 
A=\frac{\alpha^{2}}{2}\left(  \frac{m_{0}v^{2}}{\lambda\ell}\right)  ^{2}%
\frac{\left(  e^{\lambda\tau_{\mathrm{e}}}-1\right)  ^{3}}{e^{\lambda 
\tau_{\mathrm{e}}}+1} \ . 
\ee

The result of Eq.~(\ref{DifActions}) is analogous to Eq.~(\ref{AGeneral}) 
obtained in the case of a random perturbation. Obviously, the 
factor $A$ is different in both cases, but we use the same notation to stress 
the similar role as just a prefactor of $M^d$. Performing again a Gaussian 
integral of $M^{\mathrm{d}}$ over $\mathbf{r}-\mathbf{r}^{\prime}$ we obtain%
 
\be 
M^{\mathrm{d}}(t)=\left(  \frac{\sigma^{2}}{\pi\hbar^{2}}\right)  ^{d}%
\int\mathrm{d}\mathbf{r}\sum_{s}\ C_{s}^{2}\left(  \frac{2\pi\hbar^{2}}%
{A}\right)  ^{d/2}\exp{\left[  -\frac{2\sigma^{2}}{\hbar^{2}}\left( 
{\overline{\mathbf{p}}}_{s}-\mathbf{p}_{0}\right)  ^{2}\right]  }\ . 
\label{MDLG}
\ee

Under the same assumptions than in Sec.~\ref{sec:DTerms}, we obtain a 
result equivalent to that of Eq.~(\ref{SCLEFinal}),%
 
\be 
M^{\mathrm{d}}(t)\simeq\overline{A}e^{-\lambda t}, \label{MLGDFinal}%
\ee 
 
\noindent with, again, $\overline{A}=[\sigma m_{0}/(A^{1/2}t)]^{d}$.  
As we have shown, the form of the Loschmidt echo found in
Eq.~(\ref{DecompositionLE}) holds for the perturbation $\Sigma$ that 
we have discussed in this section [Eq.~(\ref{MTPerturbation})], 
as well as for the random one of Sec.~\ref{sec:SCLE}.
The only difference turned out to appear in the form of 
the \textquotedblleft elastic mean free path" $\tilde{\ell}$ and 
the prefactor $\overline{A}$, both of which are perturbation 
dependent. 
 
Recalling the discussion of Sec.~\ref{sec:DvsND} about the critical
perturbation between the FGR and the Lyapunov regimes, in the model discussed in this section, an explicit value of the perturbation parameter $\alpha$  is obtained by setting $v/\tilde{\ell}=\lambda$, and results
 
\be 
\alpha_{\mathrm{c}}=\frac{2\hbar}{m_{0}}\sqrt{\frac{\lambda}{v^3 \ell}}. 
\label{CriticalAlpha}%
\ee 
 
\noindent We will discuss in the next chapter the physical consequences of 
the above critical value and its dependence on various physical 
parameters. 

\subsection{An exact solution: the upside down harmonic oscillator}
\label{sec:UHO}

As a final example it will be very instructive to evaluate the LE in an exactly
solvable system, an upside down harmonic oscillator (UHO). Although the UHO is not
chaotic, it might be the simplest system with an unstable fixed point. In this
sense we are representing the chaotic behavior only by the stretching of the
probability density along the unstable manifold, while the folding
mechanism is discarded.

The Hamiltonian of the system is 
\be
\Hc_0=\frac{p^2}{2m}-\frac{m \omega_0^2 x^2}{2},
\label{HamiltonianUHO}
\ee
where the frequency $\omega_0$ plays the role of the Lyapunov exponent. The classical solution follows the path of the normal harmonic oscillator, giving
\be
x(t)=x_0 \cosh (\omega_0 t) + \frac{p_0}{m} \sinh (\omega_0 t),
\ee
where $x_0$ and $p_0$ are the initial position and momentum. 
In order to observe the instability as an exponential spreading of the wavepacket in the quantum version, one needs that initial state $\left| 0 \right>$ be a Gaussian wavepacket of width $\sigma$ located at $x=0$ and with mean momentum $p_0=0$. 

Let us consider a perturbation linear in the coordinate of the system,
\be
\Sigma(x,t) = \epsilon J(t) x.
\label{UHOPerturbation}
\ee
We write the amplitude of the LE in the interaction picture
\ba
m_J (t)&=&\left< 0 \right| U^\dagger_\Sigma (t) U_0(t) \left| 0 \right> \nonumber \\
&=& \left< 0 \right| \hat{T} \left( e^{-\ii \epsilon \int J(t')x(t')} \dd t' \right) \left| 0 \right>,
\label{UHO1}
\ea
with $\hat{T}$ the time ordering operator, and the subindex $J$ denotes the dependence on the perturbation. The system and the perturbation are quadratic in the coordinates of the system, and the initial state a Gaussian, therefore the solution of Eq. (\ref{UHO1}) can be obtained simply by completing squares in the argument of the exponentials and performing the Gaussian integral. The result is of course another exponential with a quadratic argument, a so called Gaussian functional of $J$ and $x$. It is not simple to do this in a straightforward fashion, however we can just write down the most general Gaussian functional in terms of unknown kernels $\nu$ and $\kappa$,
\be
m_J (t)=\exp{\left[ \ii \epsilon^2 \int \int \dd t_1 \dd t_2 \ J(t_1) \nu (t_1,t_2) J(t_2) + \ii \epsilon \int \dd t_1 \ J(t_1) \kappa (t_1)\right ]},
\label{UHO2}
\ee
and then find out what these kernels are by taking functional derivatives\footnote{In this case it suffices to use a working definition of the functional derivative, 
\be
\frac {\partial F \left[ f(x) \right]}{\partial f(x')} = \left. \frac{\partial F \left[ f(x)+\epsilon \delta(x-x') \right]}{\partial \epsilon} \right|_{\epsilon=0}.
\ee
}
of Eqs.~(\ref{UHO1}) and (\ref{UHO2}) with respect to $J$ and evaluating the results for $J=0$. For instance, from Eq. (\ref{UHO1})
\ba
\left. \frac{\partial m_J}{\partial J(t)} \right|_{J=0}&=& \left< 0 \right| (-\ii \epsilon x(t) \left. \hat{T} \left( e^{-\ii \epsilon \int J(t')x(t') \dd t'} \right) \right|_{J=0} \left| 0 \right> \nonumber \\
&=& \left< 0 \right| (-\ii \epsilon x(t) ) \left| 0 \right>  = 0,
\ea
where we have used that $\left< x(t) \right> =0$ for all $t$ because of the symmetry of the wavepacket around the origin. This result is to be compared to the derivative of Eq. (\ref{UHO2}),
\ba
\left. \frac{\partial m_J}{\partial J(t)} \right|_{J=0} &=&  \left[ \ii \epsilon^2 \int \dd t_1 \left( \nu (t_1,t)+\nu(t,t_1) \right) J(t_1) + \ii \epsilon \kappa (t) \right] \nonumber \\
& & \left. \exp{\left[ \ii \epsilon^2 \int \int \dd t_1 \dd t_2 \ J(t_1) \nu (t_1,t_2) J(t_2) + \ii \epsilon \int \dd t_1 \ J(t_1) \kappa (t_1) \right] } \right|_{J=0}  \nonumber \\
&=&  \ii \epsilon \kappa (t),
\ea
and therefore, $\kappa (t)=0$.
Similarly, taking the second derivative of $m(t)$ with respect to $J$ and using the classical solution for $x(t)$, one obtains 
\be
\nu(t_1,t_2)=\ii \ \left< 0 \right| (x(t_1)x(t_2)+x(t_2)x(t_1)  \left| 0 \right> = \frac{i \hbar}{2m\omega_0} \cosh{\omega_0(t_1+t_2)}.
\label{UHO3}
\ee
Using this in Eq.~(\ref{UHO2}) and the relation $\cosh (a+b)=\cosh a \cosh b + \sinh a \sinh b$, we can write the LE as
\be
M_J(t)=\exp{\left\{ -\frac{\hbar}{m \omega_0} \left[ \left( \epsilon \int \cosh (\omega_0 t) J(t) \right)^2 + \left( \epsilon \int \sinh (\omega_0 t) J(t) \right)^2 \right] \right\} }.
\ee
Completing squares and undoing the Gaussian integrals, we can write
\ba
M_J(t)&=&\frac {1}{2 \pi \sigma^2} \int \dd r_1 \int \dd r_2 \ e^{-\frac{r_1^2}{2 \sigma^2}} e^{-\frac{r_2^2}{2 \sigma^2}} e^{\ii r_1 \epsilon \int \dd t_1 \cosh (\omega_0 t_1) J(t_1)} e^{\ii r_2 \epsilon \int \dd t_2 \sinh (\omega_0 t_2) J(t_2)} \nonumber \\
&=& \frac {1}{2 \pi \sigma^2} \int \dd r_1 \int \dd r_2 \ e^{-\frac{r_1^2+r_2^2}{2 \sigma^2}} e^{\ii \epsilon \int \dd t' \left( r_1 \cosh (\omega_0 t') + r_2 \sinh (\omega_0 t') \right) J(t')},
\label{UHO4}
\ea
where $\sigma^2=\hbar/m\omega_0$ is the initial dispersion of a minimum uncertainty wave packet.

Let us now consider the average LE over realizations of the perturbation, that is we assume a distribution for $J$,
\be
P(J)=N \exp{\left(-\frac{1}{2}\int \int J(t_1) A^{-1}(t_1,t_2) J(t_2) \right) },
\label{UHO5}
\ee
where $A(t_1,t_2)=\left< J(t_1)J(t_2)\right>$ is the noise correlation function and $N$ is a normalization factor such that  $\int \mathcal{D}J P(J) = 1$, with $\mathcal{D}J$ represents that this is an integral over all functions $J(t)$. Denoting $x(t')=\left( r_1 \cosh (\omega_0 t') + r_2 \sinh (\omega_0 t') \right)$, a trivial Gaussian integration of Eq.~(\ref{UHO4}) gives the average LE
\be
\bar M=\frac {1}{2 \pi \sigma^2} \int \dd r_1 \int \dd r_2 \ e^{-\frac{r_1^2+r_2^2}{2 \sigma^2}} e^{-\frac{\epsilon^2}{2}\int \int x(t_1) A(t_1,t_2) x(t_2)}.
\label{UHO6}
\ee

To obtain definite results we have to give the correlation of the noise. A significant and simple case is the white noise correlation $A(t,t')=D \delta(t-t')$, where one obtains
\be
\bar M = \left(1+\frac{\epsilon^2 D \sigma^2}{2\omega_0}\sinh (2\omega_0 t)+\frac{\epsilon^4 D^2 \sigma^4}{4 \omega_0^2}\left(\sinh^2(\omega_0 t)
-\omega_0^2 t^2\right)\right)^{-1/2}.
\ee
This exact result shows that for long times ($\omega_0 t\gg \log(\epsilon^2 D \sigma^2/2\omega_0)$) the echo $\bar M$ decays as $\exp(-\omega_0 t)$, which in this example is the equivalent of the Lyapunov decay. 

For short times a decay with a rate determined by diffusion is observed,
\be
M(t) \simeq \left(1-\frac{\epsilon^2 D \sigma^2}{2}t \right)^{-1/2} \simeq e^{-\frac{\epsilon^2 D \sigma^2}{4} t}.
\ee
Although it looks like a FGR, this is only a transitory perturbation dependent
regime that always leads to a decay dominated by the Lyapunov ($\omega_0$) exponent. 

As we have seen, this apparently oversimplified example (by the absence of chaos) already captures the essence of the Lyapunov decay of the LE. Not only this provides insight into the LE problem, but also proves that further analytical progress can be made in more complex situations by using UHOs as building blocks for complicated environments \cite{Blume03}.

\section{Summary}

In this chapter we have applied the semiclassical approximation to calculate the behavior of the LE in classically chaotic systems. It was found that the semiclassical expression of $M(t)$ can be written as a sum of two terms that decay exponentially. One of these terms is given by a Fermi Golden Rule expression and therefore its decay rate $\Gamma$ depends quadratically on the perturbation strength. The other term has a decay rate not only independent of the perturbation but also, and more importantly, given by the Lyapunov exponent $\lambda$ of the classical system. Thus, $M(t)$ has a regime of parameters where it decays with the minimum between $\lambda$ and $\Gamma / \hbar$. 
The transition between both regimes occurs approximately when $\Gamma / \hbar=\lambda$, a condition we will explore in the next chapter. Some examples for particular systems were given, of relevance is the Lorentz gas for which the results were shown to be the same regardless the non--disordered perturbation. 

\section*{Original results}

\begin{itemize}
\item  Generalization of the original results of Jalabert and Pastawski \cite{Jalabert01} to any perturbation with spatial as well as temporal noise (with finite variance), Sec. \ref{sec:GeneralLE}. These results are being prepared for publication \cite{Cucchietti04}. The specific perturbation studied in \cite{Jalabert01} is given as a particularization in the examples, Sec.~\ref{sec:QuenchedDisorder}. 
\item Semiclassical analysis of the Lorentz gas with a mass tensor perturbation, Sec.~\ref{sec:LorentzGas}. This derivation is also an extension of \cite{Jalabert01} to a hard wall system (which implies non--continuous equations of motion) and a perturbation without disorder. These results were published in Ref.~(\cite{Cucchietti03B}).
\item The final result of $M(t)$ for the inverted harmonic oscillator was presented in \cite{Cucchietti03}, although the derivation of Sec.~\ref{sec:UHO} is presented for the first time here.
\end{itemize}

\chapter{Universality of the Lyapunov regime}
\label{chap:Universality}

\begin{quote}
{\em The most exciting phrase to hear in science , the one that heralds new discoveries, is not ``Eureka!'' (I found it!), but rather ``that's funny...''.}

Isaac Asimov.
\end{quote}

The exciting results shown in the previous chapter should quickly raise our attention and demand closer scrutiny. Many approximations are actually uncontrolled, assumptions are made that could probably be too strong, weak or plainly wrong, and plenty of questions arise regarding aspects difficult to contemplate using analytical tools. Among those questions, perhaps the most important are those that concern the range of validity of the Lyapunov regime. In order to clarify these issues the semiclassical theory presented before needs numerical support and exploration.

The chapter is divided in two parts: the first one will mainly provide numerical evidence of systems that present the transition from a FGR to the Lyapunov regime. For this, different combinations of system/perturbations give generality to the results of the previous chapter. The second part will focus on the universality of the Lyapunov regime, not only in its classical chaos interpretation but also on some extra quantum aspects.

\section{Correspondence between semiclassical and numerical calculations}
\label{sec:Numerical}

\subsection{The Lorentz gas}
\label{sec:LorentzNumeric}

Let us start by studying one of the systems considered at the end of the previous chapter, the Lorentz gas. The physical advantages of this system will be evident in Sec.~\ref{sec:Universality}. Although it is not the numerical example to be shown that has the most similarities with the theory, it is presented it at this point for historical reasons: it was the first model where numerical evidence of the Lyapunov regime was
observed. 

The classical dynamics of the system is described in detail in Sec.~\ref{sec:LorentzGas} and in appendix~\ref{appe:Lorentz} (see in particular the calculation of the classical Lyapunov exponent). In order to compute the quantum analog, the system is discretized using a small lattice unit $a$ (see App. \ref{appe:Suzuki}), and the whole system is embedded in a finite box of size $L$ with periodic boundary conditions. It is important to verify that $a$ is the smallest scale in the simulation, and that the results do not depend on it. Clearly small system sizes $L$ will not appropriately represent general results (because they cannot accommodate many impurities thus giving dynamics strongly dependent of the realizations). The smallest $L$ that allows the observation of an exponential decay of $M(t)$ over a large interval was found to be $L=200a$, which means the consideration of a Hilbert space of dimension $4\times10^4$ states. As standard diagonalization routines cannot manage such large matrices, the evolution was performed resorting to a Trotter-Suzuki algorithm (see appendix \ref{appe:Suzuki})
that does not provide energy or eigenvector information but computes the quantum
dynamics in the spatial base with high precision and efficiency.

The typical simulation used disks of radius $R=20a$, and with a de 
Broglie wavelength $\lambda_{dB}=2\pi/k_{dB}=16/3a$. Notice that $k_{dB}R\simeq
23\gg 1$, which assures we are sufficiently well in the semiclassical regime. Also, this value of $k_{dB}$ is at the limit where the dispersion relationship is still approximately quadratic (like the free particle's dispersion). Shorter wavelengths would strongly feel the discretization of the system.

The concentration of impurities $c$ is computed as the ratio of the area occupied by the  disks to the area of the box,  
\be 
c=N\pi R^{2}/L^{2}. 
\ee 

\begin{figure}[htb]
\begin{center}
\leavevmode
\epsfxsize 4in
\epsfbox{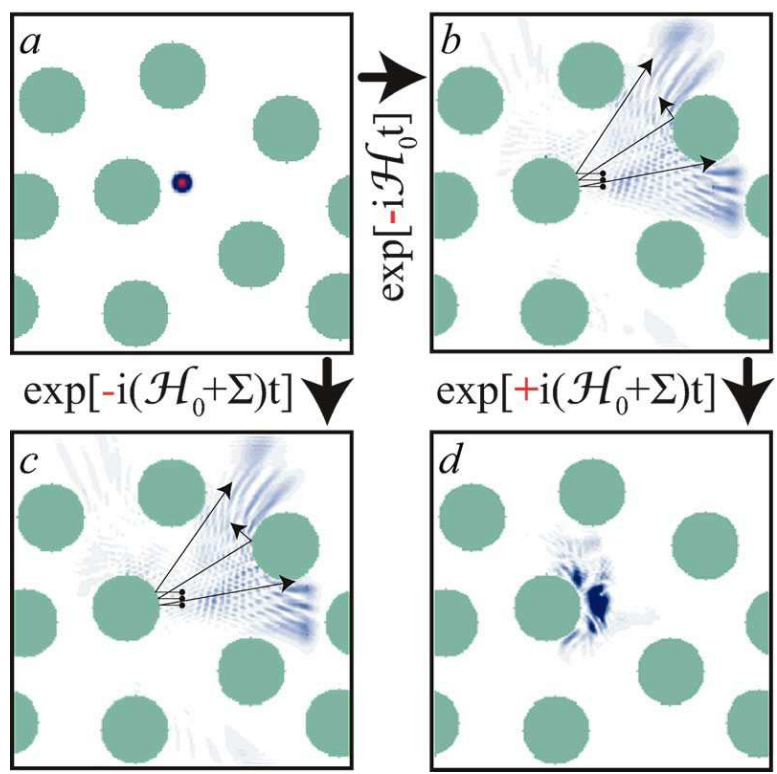}
\caption{Density of the wave function in the Lorentz gas. The boxes have sides $L=200a$ and the disks have $R=20a$. ($a$) Initial wavefunction, a Gaussian packet with momentum pointing to the left. ($b$) Evolution of state ($a$) with the unperturbed Hamiltonian of the system, $\Hc_0$, for a time $t=30$. The solid lines are classical trajectories. ($c$) Evolution of the same initial state but with the perturbed Hamiltonian, $\Hc_0+\Sigma$, where the perturbation is the mass tensor distortion. ($d$) State ($c$) evolved with $-\Hc_0+\Sigma$, which makes an imperfect time reversal. The Loschmidt echo is the overlap between states ($a$) and ($d$) or, equivalently, between ($b$) and ($c$). In this example $M=0.09$}
\label{fig:4PanelsLG}
\end{center}
\end{figure}

The perturbation in the mass tensor of Eq.(\ref{MTPerturbation}) is
easily implemented in the tight binding scheme (see App. \ref{appe:Suzuki}) by enlarging or reducing the hoping elements of the Hamiltonian in the respective directions. The perturbation strength is given by the (adimensional) parameter $\alpha$. An illustrative picture of the quantum dynamics of the Lorentz gas and the effect of the perturbation is shown in Fig.~\ref{fig:4PanelsLG}.

All results presented in this section for $M(t)$ are averaged over 100 realizations of the disorder potential. In following sections the effect of the averaging procedure on the results is discussed.

\begin{figure}
\begin{center}
\leavevmode
\epsfxsize 4in
\epsfbox{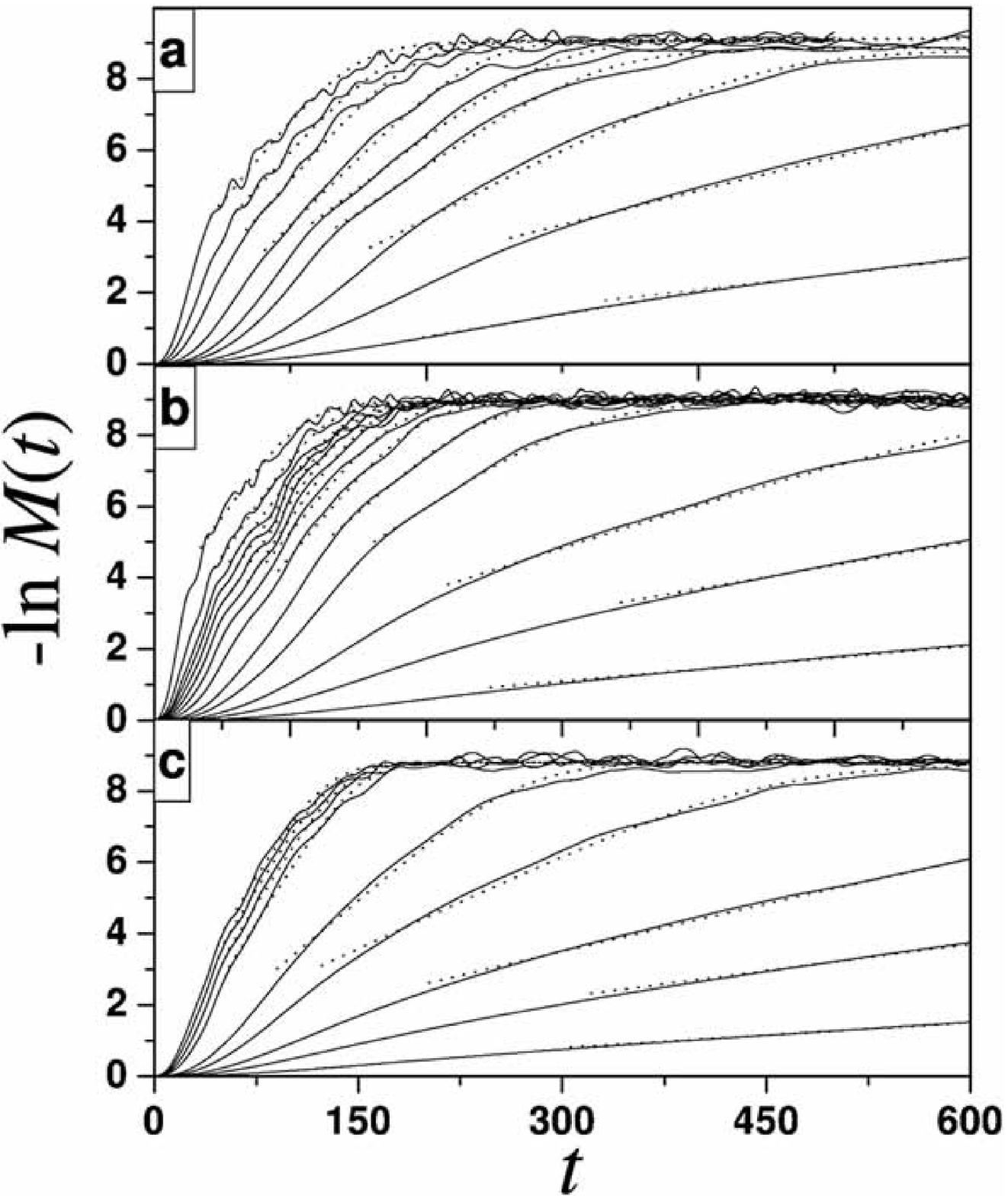}
\caption{Time decay of the Loschmidt echo $M(t)$ for different 
values of the perturbation strength $\alpha$ and concentration of 
impurities $c$ in the Lorentz gas. (top panel) $c=0.157$ and $\alpha=$0.004, 0.007, 0.01, 0.015, 0.02, 0.03, 0.05, 0.07, 0.1 (from top to bottom); (middle panel) 
$c=0.195$ and $\alpha=$0.004, 0.007, 0.01, 0.015, 0.02, 0.03, 
0.04, 0.05, 0.06, 0.07, 0.08, 0.1, 0.15; (lower panel) $c=0.289$ and 
$\alpha=$0.004, 0.007, 0.01, 0.015, 0.02, 0.03, 0.04, 0.05, 0.06, 
0.07. The time is measured in units of $\hbar/V$, where $V$ is the 
hopping term of the tight-binding model (see appendix 
\ref{appe:Suzuki}). The doted lines represent the best fits to the decay, as described in the text.} 
\label{fig:PanelsLG}
\end{center}
\end{figure}

$M(t)$ was calculated for different strengths of $\alpha$ and concentration of disks 
$c$. In Fig.~\ref{fig:PanelsLG} we can see the results for $c=0.157$, $0.195$ and $0.289$, and increasing values of $\alpha$. 

Since this is the first time we present numerical results of $M(t)$, let us discuss the various regimes present in the time evolution of the LE.  Firstly, for very short 
times, $M(t)$ exhibits a Gaussian decay, $M(t)=\exp\left[  -b\alpha^{2}%
t^{2}\right]  $, where $b$ is a parameter that depends on the initial state, 
the dynamics of $\mathcal{H}_{0}$ and the form of the perturbation $\Sigma$. 
This initial decay corresponds to the overlap of the perturbed and unperturbed 
wave-packets whose centers separate linearly with time by the sole effect of 
the perturbation. This regime ends approximately at the typical time of the 
first collision. 

Secondly, for intermediate times we find the region of interest for the 
semiclassical theory. In this time scale the LE decays exponentially with a 
characteristic time $\tau_{\phi}$. For small perturbations, $\tau_{\phi}$ 
depends on $\alpha$. We can see that for all concentrations there is a 
critical value $\alpha_c$ beyond which $\tau_{\phi}$ is independent 
of the perturbation. Clearly, the initial perturbation-dependent Gaussian 
decay prevents the curves to be superimposed. 
 
Finally, for very large times the LE saturates at a value 
$M_{\infty}$ that depends on the system size $L$, but could also depend
on the diffusion constant $D$. This regime will be discussed in detail in the next sections. 
 
\begin{figure}[htb]
\begin{center}
\leavevmode
\epsfxsize 3.5in
\epsfbox{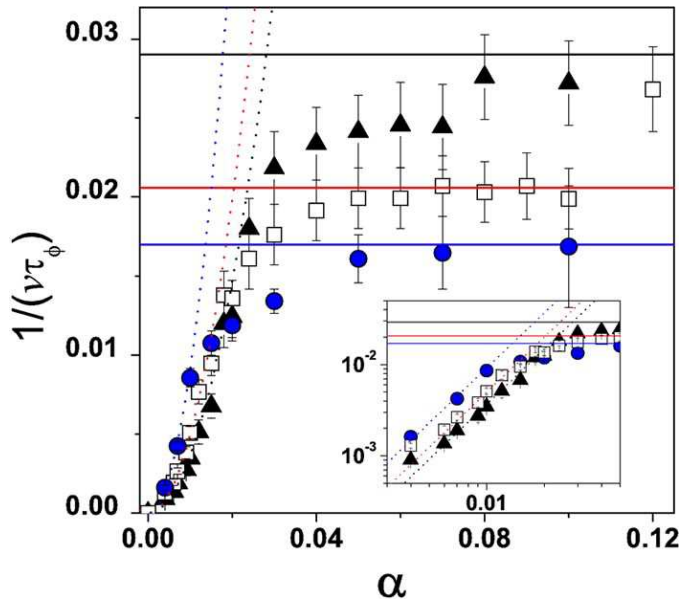}
\caption{Extracted values of the decay rate $1/\tau_{\phi}$ of the 
LE as a function of the perturbation strength $\alpha$ for the 
three concentrations of Fig.~\ref{fig:PanelsLG}. The rates  
(normalized to the group velocity of the initial wave-packet $v$) 
is given in units of $a^{-1}$; $c=0.157$ 
(circles), $0.195$ (squares) and $0.289$ (triangles). The solid 
lines are the corresponding classical Lyapunov exponents and the 
dashed lines are fits to the quadratic behavior predicted by 
Eq.~(\ref{LGFGRexponent}). The predicted coefficients for the 
three concentrations are $72 a^{-1}$, $55 a^{-1}$ and $33 a^{-1}$, 
while the obtained ones are $92 a^{-1}$, $50 a^{-1}$ and $37 
a^{-1}$ respectively. In the inset, a log-log scale of the same 
data to show the quadratic increase of 
$1/\tau_{\phi}$ for small perturbations.}
\label{fig:TauPhiLG}
\end{center}
\end{figure}

In order to compare the numerical results of $M(t)$ with the semiclassical 
predictions, let us extract $\tau_{\phi}$ by fitting $\ln M(t)$ to $\ln\left[ 
A\exp(-t/\tau_{\phi})+M_{\infty}\right]  .$ This logarithmic fit assures the correct weighting of the data for many orders of magnitude. The dashed lines in 
Fig.~\ref{fig:PanelsLG} correspond to the best fits obtained with this 
procedure. The extracted values of $\tau_{\phi}$ for the different 
concentrations are shown as a function of the perturbation strength in 
Fig.~\ref{fig:TauPhiLG}. In agreement with the analytical results of the 
previous chapter, we see that $1/\tau_{\phi}$ grows quadratically with the 
perturbation strength up to a critical value $\alpha_c$, beyond 
which a plateau appears at the corresponding Lyapunov exponent. The dashed 
lines are the best fit to a quadratic behavior. The values obtained in this 
way agree with those predicted by the semiclassical theory (Eq. 
[\ref{LGFGRexponent})] for the non-diagonal (FGR) term. The saturation values 
above $\alpha_c$ are well described by the corresponding Lyapunov 
exponents (solid lines), in agreement with the semiclassical prediction 
[Eq.~(\ref{MDLG})]. The very good quantitative agreement between the 
semiclassical and numerical calculations for the Lorentz gas (as well as in 
the case of other models \cite{Jacquod01,Wisniacki02,Cucchietti02Smooth}) strongly supports the generality of the saturation of $\tau_{\phi}$ at a critical value of the 
perturbation strength. 

The FGR exponent, which depends on $\mathcal{H}_{0}$ but not much on its 
chaoticity \cite{JacquodIntegrable}, is given by the typical squared matrix 
element of $\Sigma$, and the density of connected final states $1/\Delta$. That is why we observe that, for fixed perturbation strength $\alpha$, the factor $v/\tilde{\ell}$ 
depends on the concentration of impurities of $\mathcal{H}_{0}$ (see inset of 
Fig.~\ref{fig:TauPhiLG}, where a log-log scale has been chosen in order to 
magnify the small perturbation region). 
 
Notably, the dependence of $v/\tilde{\ell}$ with $\mathcal{H}_{0}$ leads to a 
counter-intuitive effect (clearly observed in the inset of 
Fig.~\ref{fig:TauPhiLG}), namely that the critical value needed for the 
saturation of $1/\tau_{\phi}$ is smaller for less chaotic systems (smaller 
$\lambda$). The reason for this is that in more dilute systems $\Sigma$ is 
constant over larger straight pieces of trajectories (in between collisions), 
leading to a larger perturbation of the quantum phase and resulting in a 
stronger effective perturbation. 

\subsection{Smooth Stadium billiard}
\label{SmoothNumerical}

The second model where we will investigate the dependence of the Loschmidt echo 
on the magnitude of an external perturbation is devised to correspond exactly with the system studied in the original work by Jalabert and Pastawski \cite{Jalabert01}. 

\begin{figure}[htb]
\begin{center}
\leavevmode
\epsfxsize 5in
\epsfbox{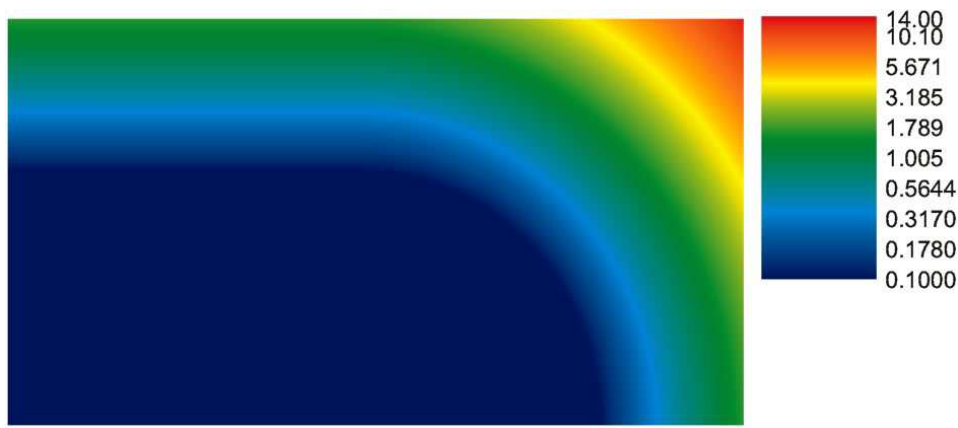}
\caption{Potential profile of the Smooth stadium billiard for $\nu=3$, in the scales mentioned in the text.}
\label{fig:SmoothBilliard}
\end{center}
\end{figure}

The unperturbed system is a smooth ``billiard'' stadium (also dubbed ``bathtube'') introduced in Ref. \cite{Vallejos99,Ortiz00}. This model consists of a 
two-dimensional Hamiltonian $\Hc_0={\bf p}^2/2m + U({\bf r})$ with the 
potential given by
\ba
U({\bf r}) = U_0\! \times \left\{ \begin{array}{cc} +\infty, & x < 0,
\\ (y/R)^{2\nu}, & 0 \le x < R, \\ \Big\{[(x-R)^2 + y^2]/R^2\Big\}^\nu, 
& x \ge R\,. \end{array} \right.
\ea 
In addition, $U({\bf r}) = +\infty$ whenever $y<0$ (See Fig.\ref{fig:SmoothBilliard}).
Actually, we should consider a quarter of a stadium in 
order to avoid features related to parity symmetries \cite{HaakeBook}.
The exponent $\nu$ sets the slope of the confining potential. For
$\nu=1$ the smooth stadium is separable and thus integrable. As the
value of $\nu$ is increased, the borders become steeper. In the limit
of $\nu \rightarrow \infty$, the stadium gains hard walls, becoming
the well-know Bunimovich billiard, one of the paradigms of classical
chaotic systems to be considered in the next section.
Thus, by varying $\nu$, we can tune the system dynamics from integrable 
to chaotic.

\begin{figure}[htb]
\begin{center}
\leavevmode
\epsfxsize 4in
\epsfbox{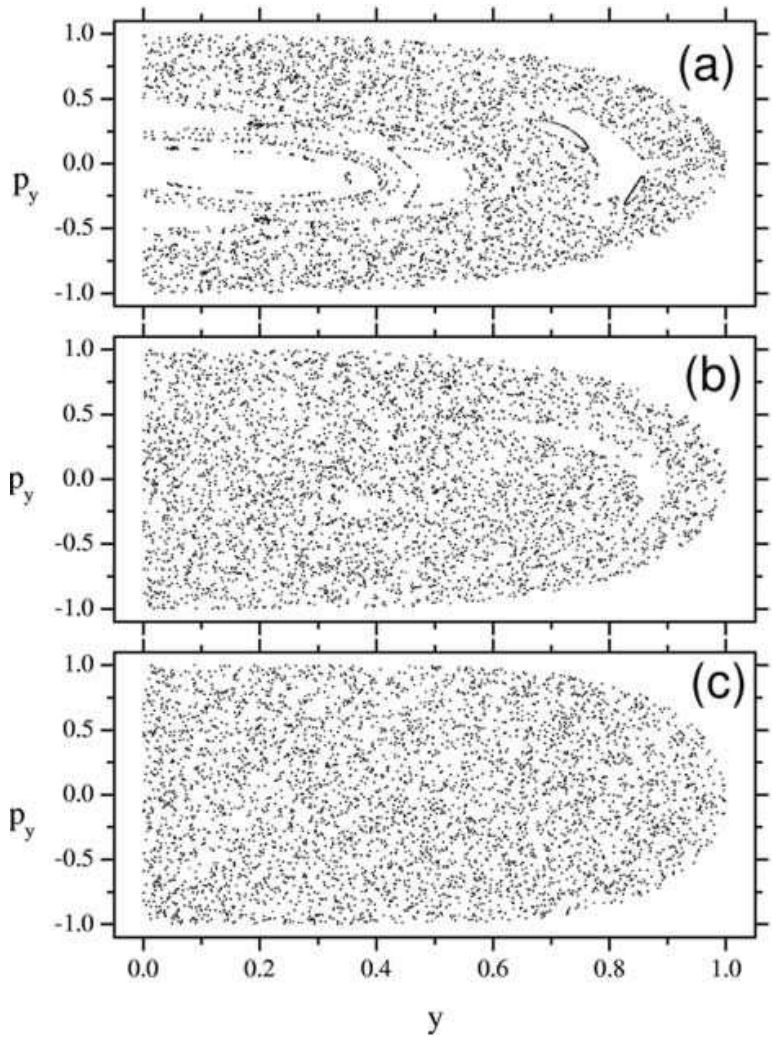}
\caption{Poincar\'e surface of section for the smooth stadium billiard
for $E=1$, $R=1$, and (a) $\nu=1.5$, (b) $\nu=2$, and (c) $\nu = 3$.}
\label{fig:PoincareSmooth}
\end{center}
\end{figure}

In order to make the presentation more concise, let us use units such
that $U_0 = 1$ and $m=1/2$. Thus, for $R=1$ the equipotential
$U(x,y)=1$ corresponds to the border of the stadium with unit radius
and unit length. For any value of the energy $E$ the equipotential
$U(x,y)=E$ gives the classical turning points, defining the allowed
area ${\cal A} \equiv {\cal A}(E)$. This area is an important
parameter of the classical and quantum dynamics of this system. Any exponent in the range $1<\nu\le 2$ already leads to a mixed phase space, i.e., a situation with both regular and chaotic motions present. In particular, for $\nu \ge 2$, $R=1$, and total
energy $E=1$ the classical dynamics is predominantly ergodic, although
small remnants of integrability still exist. These observations are
illustrated by the Poincar\'e surfaces of section displayed in
Fig.~\ref{fig:PoincareSmooth}.

\begin{figure}[tb]
\begin{center}
\leavevmode
\epsfxsize 3.5in
\epsfbox{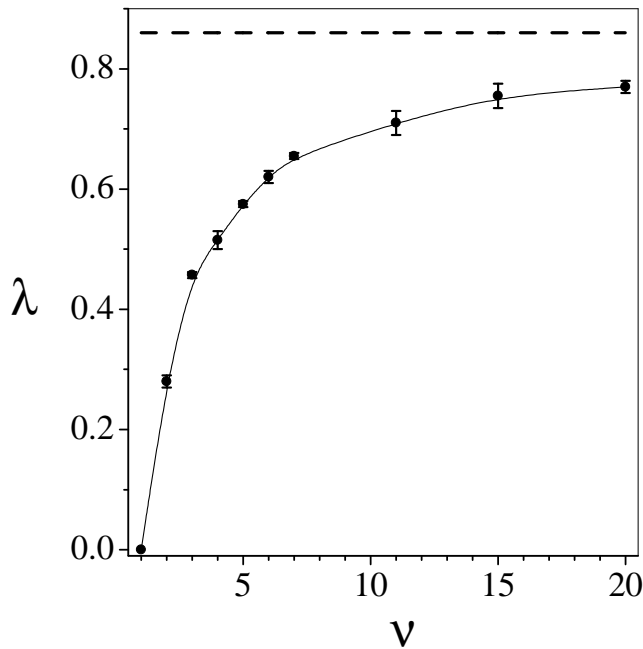}
\caption{The Lyapunov exponent of the smooth stadium for $E=1$ and
$R=1$ as a function of $\nu$. The circles are the results of the numerical
computations, while the continuous line serves as a guide to the
eye. The dashed line corresponds to the billiard limit,
$\lambda_{\mbox{\scriptsize Bunimovich}} = 0.86$.}
\label{fig:LyapunovSmooth}
\end{center}
\end{figure}

The global Lyapunov exponent $\lambda$ was computed using the algorithm by
Benettin {\it et al.} \cite{Benettin76}. The evolution of the
classical trajectories was carried out numerically\footnote{The computation of $\lambda$ for the Smooth billiard was carried out by R.O. Vallejos \cite{Cucchietti02Smooth}} using a symplectic algorithm \cite{Yoshida90}.
The Lyapunov exponent was computed for several values of $\nu$. At
$E=1$, $\lambda$ varies smoothly as a function of $\nu$, as shown in
Fig.~\ref{fig:LyapunovSmooth}. As expected, as $\nu$ becomes very
large $\lambda$ approaches the value of the Lyapunov exponent for 
the Bunimovich stadium billiard (see next section), namely $\lambda_{\mbox{\scriptsize 
hard}} = 0.86$.

The chosen perturbation for this system is a Gaussian static disordered potential (Eq. [\ref{QuenchedSigma})]. $M(t)$ was computed and afterwards
an ensemble average over different realizations of $\Sigma({\bf r})$ was taken. 
All impurities are uniformly distributed over an area ${\cal{A}}$ of the
two-dimensional plane where the stadium resides, with concentration
$n_i = {\cal N}_i/{\cal{A}}$. 

The quantum evolution was carried out through the
fourth-order Trotter-Suzuki algorithm \cite{Suzuki90,Suzuki93,deRaedt96} (appendix
\ref{appe:Suzuki}). As with the Lorentz gas, one must resort to a spatial discretization of the system. Within the energy range explored, it was found that a two-dimensional lattice of area $2.1 R\times 1.1 R$ provided very accurate results for $N=180$
sites per unit distance $R$ (with the intersite distance given by
$a=R/N$), corresponding to a total number of $378 \times 198$ lattice
sites.

The range of parameter values explored in the simulations is
limited by computational cost. Moreover, the choice of parameters
was guided by the constraints imposed by the semiclassical
calculations of Sec.~\ref{sec:SCLE}. First, in order
to include a large number of randomly located impurities, their
correlation width $\xi$ had to be taken much smaller than $R$. Second,
the semiclassical regime where Eq. (\ref{DecompositionLE}) applies
requires $\xi$ to be larger than the wave packet width $\sigma$,
which, in turn, has to be much larger than the particle wavelength
$\lambda_{dB}$. Other constraints arise from finite size effects. For
instance, the large-time saturation value of the Loschmidt echo, 
$M(t\rightarrow \infty )$, depends on the ratio $\sigma/N$. Thus, for a
fixed $N$, it is necessary to make $\sigma$ as small as possible in
order to guarantee a small value for $M(t \rightarrow \infty)$. In
addition, one can only accurately recover the dispersion relation of
the free particle, $E_{\bf k} = \hbar^{2}k^{2}/2m$, when $k a \ll 1$.
All these constraints are summarized by the inequalities
\begin{equation}
\label{Constraints}
a \ll \lambda_{dB} \ll \sigma < \xi \ll R.
\end{equation}

A reasonable compromise between a good accuracy and a feasible simulation time
was found for $\xi = 0.25R$, $\sigma = 0.18R$, $\lambda_{dB} = 0.07R$,
and $N = 180$. This choice, combined with the values of the classical
model parameters, $m = 1/2$ and $E = 1$, gave rise to units such that
$\hbar=0.011 R$. Thus, the inequalities of Eq. (\ref{Constraints})
were approximately observed in the simulations. For the quantum
evolution, a time step $\delta t = 2 m a^{2}/10 \hbar = 2.8 \times
10^{-4} E/\hbar$ proved to be sufficiently small.

It is important to make a few remarks about the averaging
procedure. In the simulations, besides averaging over impurity
configurations, it was also found important to average over initial
positions ${\bf r}_0$ and directions ${\bf p}_0$. The main reason is
that numerical simulations of billiards deal with relatively small,
confined systems and directionality has a strong influence in the
short-time dynamics. 

The initial conditions for the quantum evolution were chosen from a
subset that also minimized finite-size effects. That is, it is preferable to choose
initial conditions that allow for the observation of an exponential
decay before the saturation time. For that purpose, we took
$0.5R<x_0<R$, $0.2R<y_0<0.5R$, and initial momentum $\bf{p}_0$ such
that the first collision with the boundary occurred at $x>R$, avoiding
trajectories close to bouncing ball-like modes along $y$. (Such
trajectories were found to lead to strong non-exponential decays in
$M(t)$ for time intervals shorter than the saturation time.)

\begin{figure}[tb]
\begin{center}
\leavevmode
\epsfxsize 4in
\epsfbox{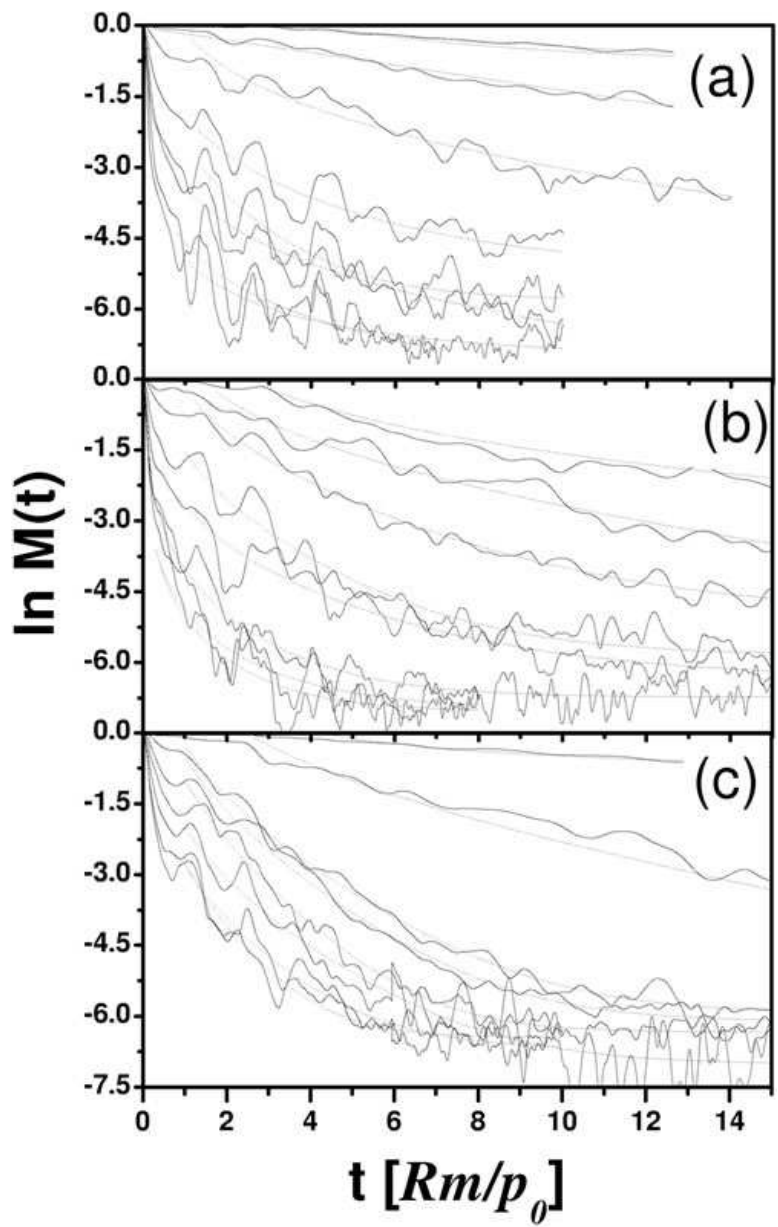}
\caption{$M(t)$ for the smooth stadium with $\nu=1.5$ (a), $2$ (b), and $3$ (c) for different values of the perturbation strength: $u = 0.002$, $0.005$, $0.01$,
$0.02$, $0.03$, $0.04$, $0.05$, and $0.06$.}
\label{fig:PanelsSmooth}
\end{center}
\end{figure}

In Fig.~\ref{fig:PanelsSmooth} we can see $M(t)$ for $\nu=1.5$, $2$, and $3$
for different values of the perturbation strength. In all graphs we
observe that the asymptotic decays are approximately exponential within a
certain ranges of $u$, as predicted in Sec.~\ref{sec:SCLE}. \
In order to obtain the characteristic
decay times, $\ln M(t)$ was fitted to the function $\ln [A \exp
(-t/\tau_\phi)/t + M_{\infty}]$. The fit was performed for times $t >
R/v$, where $v = \sqrt{2E/m} = 2$ is the wave packet velocity, to
exclude the initial, non-universal (and non-exponential) time
evolution. It is worth noting that the usual nonlinear fitting
procedures are rather insensitive to certain combinations of
parameters $\tau_\phi$ and $A$. Thus, while the parameter $M_{\infty}$
could be fixed by averaging the long-time tail of the data,
the uncertainty in $A$ and $\tau_\phi$ was avoided by fixing the value of the
fitted curve at the initial point to be exactly equal to the
respective data value. It was checked that such a procedure yields values for
$A$ proportional to $u^{-2}$, as expected.

The typical number of samples used in the averaging procedure (for
each trace of the $M(t)$ shown) was in the range 80-100. In fact, it is noticeable that the number of samples needed to obtain comparable statistical mean squares fluctuation for $M(t)$ scaled with the perturbation strength $u$. That is, the larger the perturbation, the
larger were the fluctuations in $M(t)$. This fact set another practical limit to the range of perturbation strengths $u$ one can investigate in numerical simulations.

\begin{figure}[htb]
\begin{center}
\leavevmode
\epsfxsize 3.5in
\epsfbox{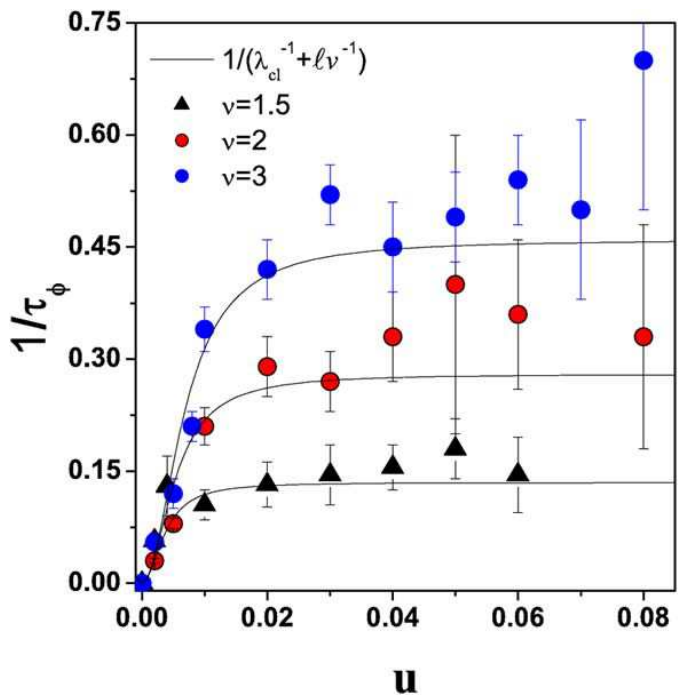}
\caption{The characteristic decay rates for the smooth stadium obtained from
Fig. \ref{fig:PanelsSmooth} as a function of perturbation strength. 
The solid curves correspond to the phenomenological expression,
Eq. (\ref{eq:phenom}).}
\label{fig:TauPhiSmooth}
\end{center}
\end{figure}

In Fig. \ref{fig:TauPhiSmooth} the inverse
characteristic decay times $1/\tau_\phi$ obtained in the fittings are plotted as a
function of the impurity strengths $u$ for the three values of
$\nu$ (from Fig.~\ref{fig:PanelsSmooth}). 

For comparison, notice in the plot the phenomenological curve that should interpolate between the minimum of the two exponents
\begin{equation}
\label{eq:phenom}
\tau_{\rm phenom}(u) = \frac{1}{\lambda} + \frac{\hbar}{\Gamma}
\end{equation}
where $\lambda$ is the classical Lyapunov ($u$ independent) and
$\Gamma=\hbar v_0/\tilde{\ell}$ is the characteristic decay rate obtained in
Sec.~\ref{sec:SCLE}, Eq.~(\ref{QuenchedMFP}). Such a curve matches the expected
asymptotic behaviors for $1/\tau_\phi$ at small and large values of
$u$.

The plateau around the classical Lyapunov exponent $\lambda$ clearly confirms the theoretical prediction of Sec.~\ref{sec:SCLE}. For weak perturbations, the data is also consistent with the quadratic behavior of $1/\tilde{\tau}$.

\subsection{The Bunimovich stadium billiard: when the FGR does not apply}
\label{sec:Bunimovich}

So far, we have seen numerical evidence of the Lyapunov regime in two systems. 
The Smooth billiard is a model exactly described by the original presentation of Jalabert and Pastawski \cite{Jalabert01} (see Sec.~\ref{sec:QuenchedDisorder}) with a quenched disorder perturbation. The Lorentz gas, on the other hand, did not have any disorder in the perturbation, but there was some in the dynamics of the unperturbed Hamiltonian. Therefore, the question remains if disorder plays a relevant role in the decay of the LE, whether in the perturbation or in the Hamiltonian. As anticipated in the previous chapter, disorder is a practical tool that allows analytical progress, and results do not depend strongly on it. This section shows a system that is completely free of disorder to provide numerical evidence on this aspect.

As an extra feature, it will be seen that the LDOS in this system is not a Lorentzian as obtained for random matrices (\ref{sec:RMtheory}). Therefore we will be able to observe the behavior of the LE for weak perturbations in this situation where the theory does not apply. 

\begin{figure}[htb]
\begin{center}
\leavevmode
\epsfxsize 3.5in
\epsfbox{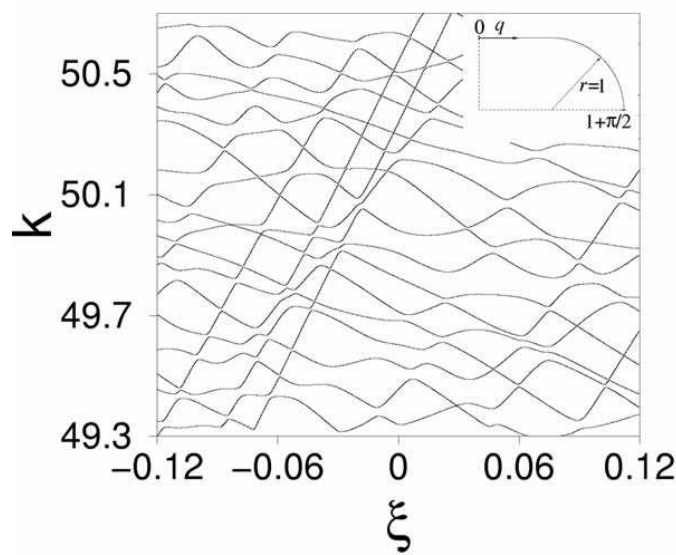}
\caption{Spectrum of the desymmetrized Bunimovich stadium billiard with mixed boundary conditions controled by the parameter $\protect\xi$ [Eq. (\ref{SigmaBoundary})]. The wave numbers $k_{\protect\mu}(\protect\xi)$ run between 49.3 and 50.7. Inset:
Schematic figure of the system. In solid line the boundary of the
stadium billiard where the mixed boundary conditions are applied [Eq. (\ref{MixedBC})].
The coordinate $q$ on the boundary is also shown. Dashed lines correspond to
the symmetries axis with Dirichlet boundary conditions. }
\label{fig:Bunimovich}
\end{center}
\end{figure}

We consider the desymmetrized Bunimovich stadium billiard\footnote{The numerical results of the LE in the Bunimovich stadium were obtained by Diego. A. Wisniacki \cite{Wisniacki02} using the method described in this section.}
\cite{Bunimovich74}, one of the paradigms of classical chaos theory.
It consists of a free particle inside a 2-dimensional planar region whose boundary ${\cal C}$ (shown in Fig.~\ref{fig:Bunimovich}) is a quarter of a circle of radius $r$ with a square box of side $r$ next to it.
If we take $r$ equal to unity then the enclosed area is $1+\pi /4$. This system not only has a great theoretical importance by being a very well known fully chaotic system, but also is of experimental relevance \cite{MarcusScience99,MarcusPRL99}. 

The classical dynamics is completely defined once the boundary is given. On the other
hand, to address the quantum mechanics, it is necessary to solve the
Helmholtz equation, $\nabla ^{2}\phi _{\mu }=k_{\mu}^{2}\phi _{\mu }$ with
appropriate boundary conditions. $k_{\mu }$ is the wave number and by
setting $\hbar =2m=1$, $k_{\mu }^{2}$ results the energy. The most commonly
used boundary conditions are the Dirichlet (hard walls) and the Neumann
(acoustics) conditions. However, we are interested in the possibility of
perturbing the quantum system without breaking the orthogonal symmetry and
leaving the classical motion undisturbed \cite{Szafer93}. This is possible
using more generalized boundary conditions:
\begin{equation}
\phi (q)\;+\;\xi \;g(q)\frac{\partial \phi }{\partial {\bf n}}(q)=0,
\label{MixedBC}
\end{equation}
where $q$ is a coordinate along the boundary of the billiard (see Fig.~\ref
{fig:Bunimovich}), and ${\bf n}$ is the unit vector normal to the boundary. 
$g(q)$ is a real function and $\xi $ the parameter controlling the strength 
of the perturbation. Dirichlet boundary conditions are recovered when $\xi =0$
while Neumann conditions are satisfied in the limit 
$\xi \rightarrow \infty $. The eigenfunctions and eigenenergies for the case $\xi =0$ are readily obtained by using the scaling method \cite{Vergini95}.

In order to compute the LE in this system, a relation between the
eigenvalues and eigenfunctions for different values of the parameter $\xi $
is needed. Based on a recently developed Hamiltonian expansion for deformed
billiards \cite{Wisniacki99}, it is easy to show that the eigenvalues and
eigenfunctions for different values of the parameter $\xi $ can be obtained
from the Hamiltonian ${\cal H}_{0}+\Sigma (\xi )$ which is expressed in the
basis of eigenstates at $\xi =0$ (hereafter referred to as  $\phi _{\mu }$),
\begin{equation}
\Sigma _{\mu \nu }^{{}}=\xi \times {\rm \Phi }_{\mu \nu }\;\oint_{{\cal C}%
}g(q)\;\frac{\partial \phi _{\mu }}{\partial {\bf n}}\frac{\partial \phi
_{\nu }}{\partial {\bf n}}{\rm d}q.  \label{SigmaBoundary}
\end{equation}
The function $g(q)$ measures the strength of the change in the boundary
condition along the contour. Within a perturbation theory it
would represent the direction and strength of a distortion of the stadium
\cite{Wisniacki99}, and it can be shown to be equivalent to the mass tensor 
perturbation introduced in Sec.~\ref{sec:MassTensor} (see App.~\ref{appe:Lorentz}. 
Here we shall use
\[
g(q)=\left\{
\begin{array}{ccc}
\alpha &  & 0\leq q\leq 1, \\
(1+\alpha )\sin (q-1)+\alpha &  & 1<q\leq 1+\pi /2
\end{array}
\right.
\]
with $\alpha =-1/(2+\pi /2)$ that could be assimilated to a dilation along
the horizontal axis and a contraction along the perpendicular one. \ Notice
that the integral above could be viewed as an inner product among the wave
functions $\frac{\partial \phi _{\mu }}{\partial {\bf n}}$ defined over 
${\cal C}$. This relation defines an effective Hilbert space in a window 
$\Delta k \approx$ Perimeter/Area \cite{Wisniacki99}. The cut-off function
${\rm \Phi }_{\mu \nu }=\exp \left[ -2\;(k_{\mu }^{2}-k_{\nu
}^{2})^{2}/(k_{0} \Delta k)^{2}\right] $ restricts the effect of the
perturbation to states in this energy shell of width $B\simeq k_{0} \Delta k$. 
It allows us to deal with a basis of finite dimension with wave numbers
around the mean value $k_{0}$ and restricting to a particular region $\Delta k$  of interest.

Figure~\ref{fig:Bunimovich} shows the dependence of the energy levels on the
perturbation. They exhibit many avoided crossings as $\xi $ is
varied. While the energy levels show the typical behavior of a general
system without constants of motion, we also recognize that some small
avoided crossings are situated along parallel tilted lines. These energies
correspond to the well known ``bouncing ball'' states which are highly
localized in momentum. The selected perturbation does not modify
substantially those states.

\begin{figure}[htb]
\begin{center}
\leavevmode
\epsfxsize 5in
\epsfbox{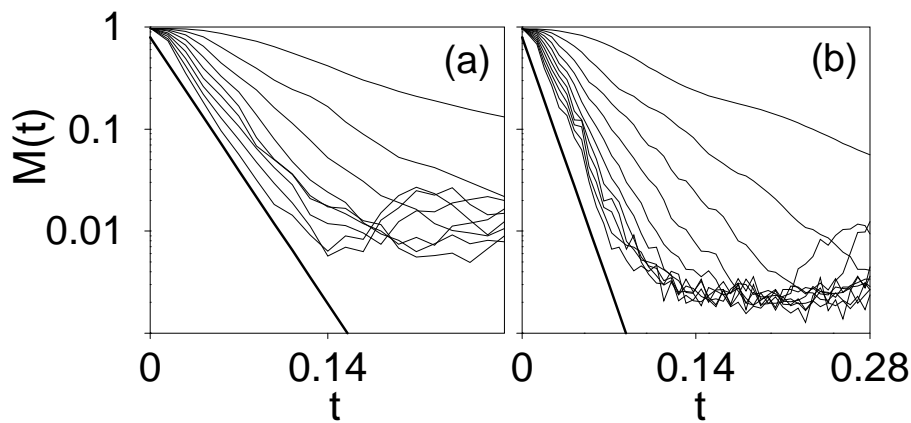}
\caption{$M(t)$ for the desymmetrized stadium billiard perturbed by a change
in the boundary conditions. The calculations are shown in two different
energy regions. (a) Corresponds to the region around $k_{0}=50$. The value
of $\protect\xi $ is, from the top curve to bottom: 0.019, 0.038, 0.057,
0.075, 0.094, 0.11, 0.13, 0.15 and 0.17. (b) Corresponds to the region
around $k_{0}=100$. The value of $\protect\xi $ is, from the top curve to
bottom: 0.0066, 0.0131, 0.020, 0.0262, 0.0327, 0.0393, 0.0458, 0.0524,
0.0589, 0.066, and 0.072. The thick lines corresponds to an exponential
decay with decay rate $\protect\tau_{\protect\phi}=1/\protect\lambda$.}
\label{fig:PanelsBunimovich}
\end{center}
\end{figure}

While a global exponential decay of $M(t)$ can be clearly identified in
almost any individual initial condition, the fluctuations for a system with 
$k_{0}$ not too large can introduce error in the estimation of the rate.
Hence, we have taken an average over $30$ initial states. Fig.\ref{fig:PanelsBunimovich} (a) and (b) show typical sets of curves of $M(t)$ for $k_{0}=50$ and $k_{0}=100$ respectively. It can be seen clearly that after the initial transient, $M(t)$ decays exponentially, $\sim \exp [-t/\tau _{\phi }]$. 
For $\xi >\xi_{c}\simeq 4.5/k$ the decay rate $\tau _{\phi }$ becomes independent of
the perturbation and $1/\tau _{\phi }\approx \lambda $ with $\lambda $ the
Lyapunov exponent of the classical system \cite{Dellago95} in accordance
with Sec.~\ref{sec:SCLE}. On the other hand, for large times $M(t)$
saturates to the finite value $M_{\infty }\approx 1/N$ with $N$ the effective
dimension of the Hilbert space \cite{Peres84}. 

\begin{figure}[htb]
\begin{center}
\leavevmode
\epsfxsize 3.5in
\epsfbox{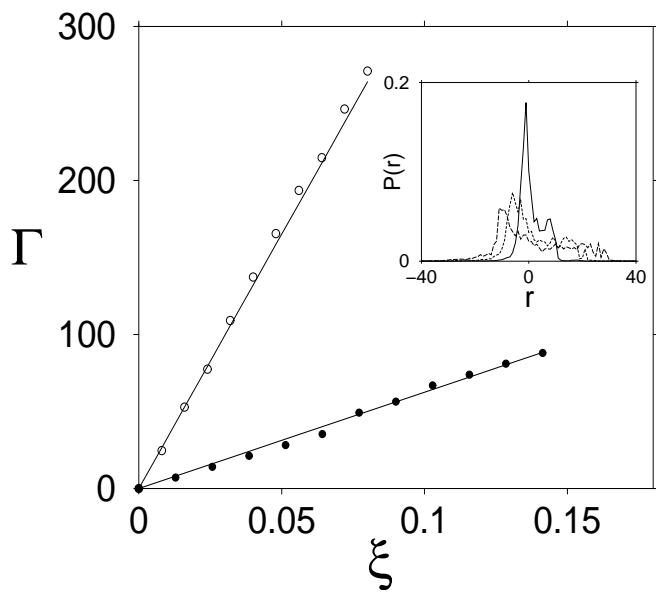}
\caption{Width $\Gamma$ of the local density of states as a function of the
perturbation strength $\protect\xi$ for $k_{0}=50$ (filled circles) and $%
k_{0}=100$ (circles). The solid lines are the best linear fit. Inset: Local
density of states $P(r)$ for different perturbations in $k_{0}=50$ (r is
measured in mean level spacing units).}
\label{fig:LDOSBuni}
\end{center}
\end{figure}

For smaller perturbations, the system is predicted to be in the FGR regime (Sec. \ref{sec:NDTerms}), with an exponential decay given by the Fourier transform of the local density of states (LDOS). However, it is not general for any system or perturbation that the LDOS is a Lorentzian. In particular, the LDOS for the Bunimovich stadium with the perturbation presented above is shown in the inset of Fig.~\ref{fig:LDOSBuni} for three different perturbation strengths, all showing they are clearly not Lorentzian. This is related to the fact that the function $g(q)$ that determines the perturbation does not connect all different regions of phase space; for instance, the bouncing ball states
are practically undisturbed by $\Sigma $ determining the non-generic nature
of the perturbation. In particular, one can evaluate the width $\Gamma$ of the LDOS as its second moment, showing the spreading of the unperturbed eigenstates when expressed in terms of the new ones. The results show a {\bf linear} dependence of $\Gamma $ on $\xi \,$ shown in Fig.~\ref{fig:LDOSBuni} [opposed to the quadratic behavior expected by the semiclassical theory, Eq.~(\ref{SpatialFGR}), and random matrix theory, Eq.~(\ref{GammaFGR})]. A best linear fit to the data results in $\Gamma \simeq 0.36\xi k^{2}$.
Taking into account that $\lambda \simeq 0.86k$, the critical
value $\xi _{c}$ for the crossover from the FGR regime to the Lyapunov
one is expected at $\xi _{c}=2.4/k$. However, from Fig. ~\ref{fig:PanelsBunimovich} we can see that the saturation occurs at $\xi _{c}\approx 4.5/k$. 
For the Bunimovich stadium then, the crossover between regimes occurs when the Lyapunov exponent is equal to the {\it half} width of the LDOS. 

These results contrast with the FGR dependence observed in the previous
examples. In general, when the FGR is not applicable, one cannot make the connection to the
Fourier transform of the LDOS as in Sec.~\ref{sec:RMtheory}. Therefore, this is direct evidence that both quantities may have quite different underlying physics \cite{WisniackiCohen}.
On the other hand, it was shown here that the LE decays exponentially with a rate 
given by the perturbation dependent width of the LDOS. This is a topic that deserves further investigation.

In summary, the Bunimovich stadium is a valuable example that sets off from the
analytical approach of previous sections in two important aspects: First, it is
a system where there is no disorder at all, neither in the Hamiltonian nor in
the perturbation. The observation of the Lyapunov regime in such a case positively answers
 the question that disorder is not an essential ingredient of the
theory. Second, in the Bunimovich stadium that the FGR is not applicable due to the non-general origin of the perturbation. In this case we observed that a FGR-like regime
exists in the sense that some general properties of the LDOS are related to the
decay of the LE, however the particular form of that relationship could in principle be
quite model-dependent.

\section{Universality}
\label{sec:Universality}

We turn to study a very important aspect of the Lyapunov regime observed in
classically chaotic systems, namely, its robustness and generality.
In particular, we will claim that these properties, along others, be regarded as the universal character of the Lyapunov regime. Universality, however, is not a well defined mathematical concept, and some discussion on its meaning is therefore needed.

In classical mechanics there exist many examples of physical systems that can be described by deterministic equations of motion, albeit they present clearly non-predictable behavior. The success of chaos theory is the finding that on many occasions these systems (which in general have very different particular characteristics) can be classified into large classes where some quantitative and qualitative predictions can be made. This feature is commonly referred to as the {\em universality} of chaotic behavior \cite{UniversalityChaotic}.

The same attribute in the Loschmidt echo would certainly heighten its rank as a powerful tool in quantum chaos. The results of the previous chapter using semiclassical techniques, along with numerical evidence, clearly show that the Lyapunov regime is present in classically chaotic systems and that it describes their behavior independently  of details of the Hamiltonian dynamics or the perturbation.
However, more requirements in addition to those described above are needed to make a fair claim of universality. Some of these conditions are set by the limitations of the theory, for instance the need for averaging or the presence of disorder in the perturbation or the Hamiltonian. The latter was dealt with in the previous example of the Bunimovich stadium, while the former will be shown to be irrelevant in the sequel. Other requirements are more explicitely related with the quantum nature of the LE. On one hand, we need to ascertain that its features are not trivially inherited from some classical counterpart by the quantum--classical correspondence before the Ehrenfest time, but that they are intrinsic to the LE itself. On the other hand, we also demand that the LE recover the appropriate classical form in the limit of high energies (or $\hbar$ going to zero), so that it smoothly describes chaotic systems for all energy regimes up to the classical level.

In this section we will deal with these topics by resorting to numerical results in the Lorentz gas, that as we will see for these purposes has many advantages over other models. The last two issues above mentioned, however, will leave some questions opened that will need the more profound analysis presented in the next chapter.

\subsection{Individual vs. ensemble-average behavior} 
\label{sec:Averages}
 
In order to make analytical progress in our semiclassical calculations, 
an ensemble average was introduced (over realizations of the quenched disordered perturbation or over initial conditions). This tool raises the question of whether the exponential decay of $M(t)$ is already present in individual realizations or, 
on the contrary, the averaging procedure is a crucial ingredient in obtaining 
a relaxation rate independent of the perturbation \cite{Silvestrov03}. 
 
As discussed in Sects.~\ref{sec:SCLE} and \ref{sec:SCLG}, for 
trajectories longer than the correlation length $\xi$ of the perturbation, the 
contributions to $\Delta S$ from segments separated by more than $\xi$ are 
uncorrelated. This leads us to consider that the decay observed for a single 
initial condition will be equivalent to that of the average. In this section 
we test this hypothesis numerically. 
 
\begin{figure}[htb]
\begin{center}
\leavevmode
\epsfxsize 3.5in
\epsfbox{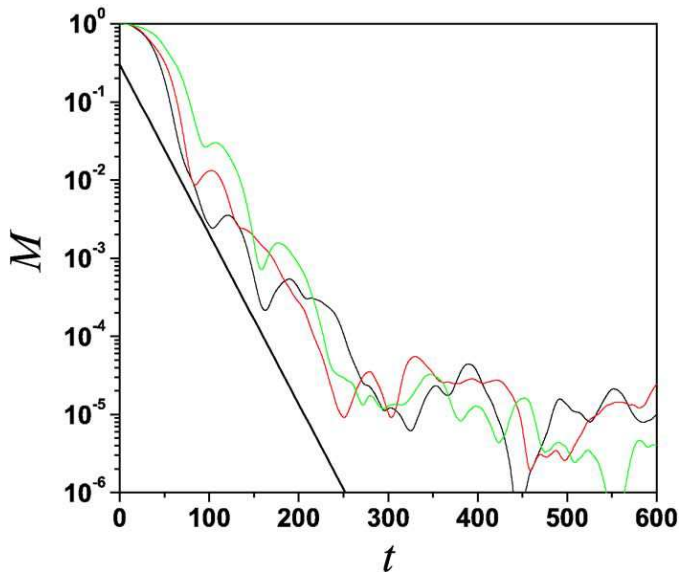}
\caption{$M(t)$ for three different single initial conditions of 
the wave-packet in the Lorentz gas. All the curves oscillate around the decay corresponding to the Lyapunov exponent, whose slope is shown in the straight 
line for comparison.}
\label{fig:IndividualCurves}
\end{center}
\end{figure}

For large enough systems presenting a large saturation time, we 
expect $M(t)$ to fluctuate around an exponential decay. This 
expectation is clearly supported by the numerical results shown in 
Fig.~\ref{fig:IndividualCurves}, where we can see $M(t)$ for 
three different initial conditions in a Lorentz gas with $L=800a$ and 
fixed $\alpha=0.024.$ An exponential decay with the semiclassical 
exponent is shown for comparison (thin solid line). 
 
In order to obtain the exponent of the decay with a good precision, we could 
calculate $M(t)$ for a single initial condition in a large enough system. 
Alternatively, Fig.~\ref{fig:IndividualCurves} shows that it is correct to obtain the exponent through an ensemble average to reduce the size of the fluctuations. As the former method is computationally much more expensive, one typically resorts to the latter. 
 
This situation is analogous to the classical case where one obtains the 
Lyapunov exponent from a single trajectory taking the limit of the initial 
distance going to zero and the time going to infinity, or else resorts to more 
practical methods \cite{Benettin76} that average distances over short evolutions. 
 
Notice that in the Lorentz gas the average over initial conditions and the 
average over realizations of the impurities positions are equivalent. Because of this, the term initial conditions is used to refer also to realizations of $\mathcal{H}_{0}$. All 
the cases shown in Sec.~\ref{sec:Numerical} and this one are averaged over realizations.
In particular for these calculations, the average is constrained to those 
systems where the classical trajectory of the wave-packet collides with at 
least one of the scatterers. This restriction helps avoiding those 
configurations where a ``corridor'' exists, in which case $M(t)$ presents a 
power-law decay possibly related to the behavior found in integrable systems 
\cite{JacquodIntegrable}. 
 
As a side note, let us remark on the averaging procedure used numerically. The averaging of quantities that fluctuate around an exponential decay is a delicate matter, since fluctuations can affect the result dramatically. In particular, for the LE it has been noted that averaging $M(t)$ over initial conditions can result in an exponential decay different than the one for a single initial condition \cite{Wang02,Silvestrov03}. 
Given the exponential dependence 
of $M(t)$ in $\lambda$, the phase space fluctuations of the Lyapunov exponent 
will induce a difference between the average $\ln M(t)$ and that of $M(t)$. 
The former procedure is more appropriate in order to have averages of the 
order of the typical values. On the other hand, if the fluctuations of the 
exponent are small, both procedures give similar results. This is the 
situation found in the Lorentz gas.  
$\left\langle M(t)\right\rangle $ and $\left\langle \ln M(t)\right\rangle $ were calculated and later the decay rates of the exponential regime extracted using the fit described in 
Sec.~\ref{sec:Numerical}. In the Lorentz gas at the range of parameters of interest both averaging procedures give values of $\tau_{\phi}$ that are indistinguishable from each 
other within the statistical error.  However, although the rates are similar, the actual values of $\left\langle M(t)\right\rangle $ and $\left\langle \ln M(t)\right\rangle $ are 
different, usually the later being larger. 

\subsection{Ehrenfest time and thermodynamic limit}  
\label{sec:Ehrenfest} 

Let us consider the extension of the time regime where the Lyapunov
decay is observed.  After the initial Gaussian decay, the following relevant time scales for the LE are the so called Ehrenfest time $t_E$, and the saturation or breakdown of exponential decay time $t_s$ (given by the size of the available Hilbert space).
 
The Ehrenfest time is such that up to it one expects 
the propagation of a quantum wave-packet to be described by the classical 
equations of motion. After $t_E$ the quantum-classical correspondence breaks down \cite{Berman78} and interference effects become relevant. In a classically chaotic system $t_E$ scales as $\ln[\hbar]$, for which it is also known as the {\it log} time. 

Strictly speaking, the semiclassical calculation presented in Sec.~\ref{sec:SCLE} is valid only until the Ehrenfest time\footnote{Heller et al. \cite{Heller91,Heller93} have shown examples where the semiclassical approximation remains very good for times polynomial in $\hbar$, much longer than Ehrenfest's time. A particular source of quantum effects that cannot be captured by the SC theory are diffraction due to discontinuities in the potential or its derivatives, and the density of these points reduce or increase the precision of the SC approximation dramatically.}. 
In addition to this, if the LE has a classical counterpart (sometimes defined as
the overlap between classical distributions in phase space \cite{Benenti02,Benenti03}), the very same definition of $t_E$ indicates that the quantum LE would follow the
classical behavior simply because of the quantum--classical correspondence.

It is then of importance to test numerically the behavior of the LE after $t_E$. 
However, in other systems where the Lyapunov regime of the LE has been observed, such as bounded systems like the Bunimovich \cite{Wisniacki02} or the smooth \cite{Cucchietti02Smooth} stadiums, chaotic maps \cite{Benenti02} or kicked systems \cite{Jacquod01}, $t_{E}$ coincides with the saturation time 
$t_{s}=1/\lambda$ $\ln[N]$. This is because in these systems the number of 
states $N$ plays the role of an effective Planck's constant $\hbar_{\mathrm{eff}}=1/N$. Therefore, when in these systems the LE is governed by a 
classical quantity, the whole range of interest occurs before the Ehrenfest 
time. It is then impossible from that evidence to conclude if the independence
of the decay rate on the perturbation strength  has some quantum origin at all, or if it is more general than the regime of validity of the semiclassical theory.
 
In the Lorentz gas, presented in Sec.~\ref{sec:SCLG}, we can differentiate between the time scales $t_{s}$ and $t_{E}$ by appropriately controlling the parameters.  
This is a property not shared by finite systems, but robust for extended ones like the Lorentz gas (this does not imply an unbounded exponential decay of the LE, as discussed below). The saturation time for an initial wave packet of width $\sigma$ in a Lorentz gas embedded in a box of size $L$ is given by 
\be 
t_{s}\simeq\frac{2}{\lambda}\ln\frac{L}{\sigma},
\ee 
 
\noindent while the Ehrenfest time, defined as the time it takes for a 
minimal wave-packet of wavelength $\lambda_{dB}$ to spread over a 
distance of the order of $R$ \cite{AleinerLarkin}, is given by 
\be 
t_{E}\simeq\frac{1}{\lambda}\ln\frac{2R}{\lambda_{dB}} \ . 
\ee 
  
The numerical calculations shown in Sec.~\ref{sec:LorentzNumeric} show that $M(t)$ decays exponentially after the Ehrenfest time without much further ado. 

\begin{figure}[htb]
\begin{center}
\leavevmode
\epsfxsize 4in
\epsfbox{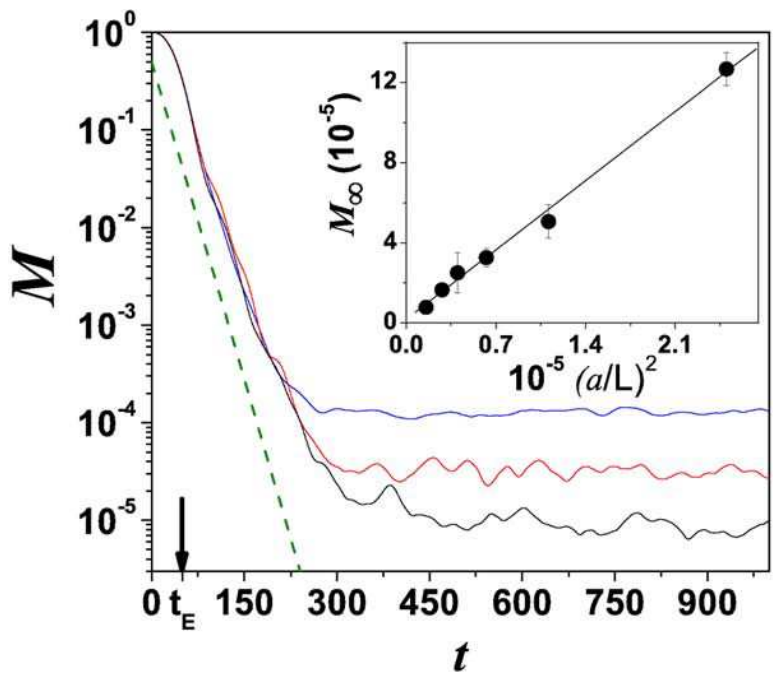}
\caption{Solid lines: Average $M(t)$ for different sizes $L=200a$ (blue), $400a$ (red) and $800a$ (black) for a fixed perturbation $\alpha=0.024$. The Ehrenfest time is shown with an arrow. In dashed line, shown for reference, exponential decay with the Lyapunov exponent of the classical LG. In the inset, $M_\infty$ is shown as a function of $(L/a)^{-2}$. The straight lines is the best fit $M_{\infty}=(0.6\pm0.1)\left(  \sigma/L\right)^{2}$.}
\label{fig:LGDifferentL}
\end{center}
\end{figure}

Furthermore, in Fig.~\ref{fig:LGDifferentL} we can see that, as expected, increasing the size of the system for fixed concentration simply increases the range of the exponential, while $t_E$ remains fixed. The dependence of the saturation value $M_{\infty}$ as a function 
of the inverse system size $1/L^{2}$ was previously studied by Peres
\cite{Peres84}. Supposing that for long times the 
chaotic nature of the system will equally mix the 
$\tilde{N}=\left( L/\sigma\right)  ^{2}$ levels appreciably 
represented in the initial state with random phases $\phi_{j},$ we 
write 
\ba
M_\infty&=&\lim_{t \rightarrow \infty} M(t) \nonumber \\
&=& \frac{1}{\tilde{N}^{2}} \left|  \sum_{j}\exp\left[  \mathrm{i} 
(\phi_{j}-\phi_{j}^{\prime})\right]  \right|  ^{2}=\frac{1}{\tilde{N}}  . 
\label{Minfinity}
\ea 
In the inset of Fig.~\ref{fig:LGDifferentL} the saturation value for long times Eq.~(\ref{Minfinity}) deduced by Peres \cite{Peres84} is plotted. A best linear fit to the data gives $M_{\infty}=(0.6\pm0.1)\left(  \sigma/L\right)^{2}$ which confirms the prediction.

According to these results, in the thermodynamic limit of $L\rightarrow \infty$ the Lyapunov regime persists for arbitrarily large times. However, this occurs only for times smaller than the critical time
where the saturation value coincides with the space explored by the
particle. In other words, for infinite unbounded systems there could be a
saturation of the LE at the ``available" (time dependent) Hilbert space, 
which in the case of the Lorentz gas would follow a diffusive law. 
Therefore, the exponential decay
of the quantum LE ends when $\exp{(-\lambda t)}=\sigma^2/r^2 (t)$, where
$r^2(t)=2dDt$ and $D$ is the diffusion coefficient (see App.~\ref{appe:Lorentz}). The maximum possible saturation time $t_s^*$ of the Lorentz gas, independent of the box size $L$, is the solution of
\be
t_s^*\simeq\frac{1}{\lambda}\ln\frac{\ell v t_s^*}{\sigma^2}.
\ee

For times shorter than $t_s^*$, the expanding range of the exponential
with $L$ for times larger than $t_{E},$ where the correspondence principle 
does not prevail, as exemplified in Fig.~\ref{fig:LGDifferentL}. 

The survival of a classical signature in the quantum dynamics 
after the Ehrenfest time is due to a more complex effect, namely 
the environment which, through the perturbation, randomizes the phase of the
wave function and washes out terms of quantum nature. We will 
discuss this process and its relation to decoherence in detail in the next
chapter.
  
\subsection{Universality of the Lyapunov regime in the semiclassical limit}  
\label{sec:univ} 
 
\begin{figure}[tb]
\begin{center}
\leavevmode
\epsfxsize 4in
\epsfbox{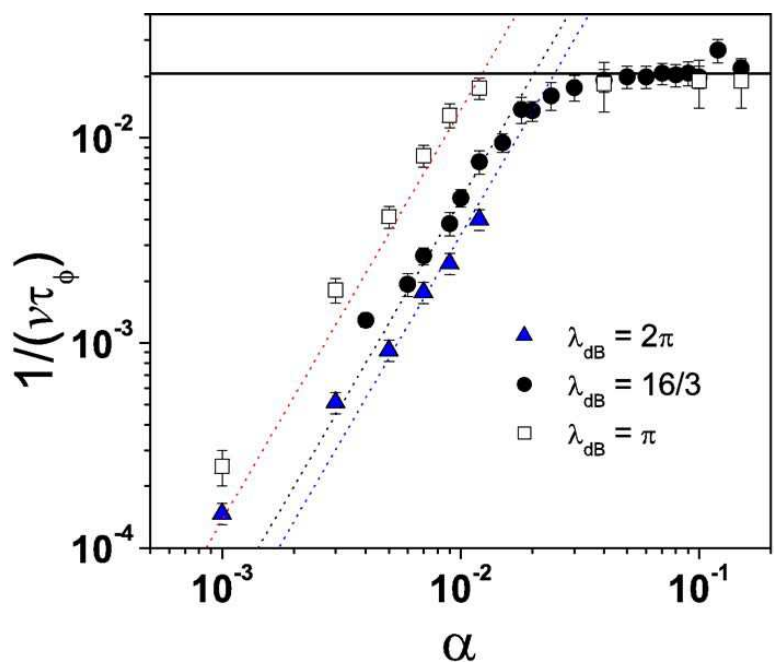}
\caption{Decay rates $1/\tau_{\phi}$ for different wavelengths 
$\lambda_{dB}$ of the initial wave-packet for a concentration 
$c=0.195$ with the same units as in Fig~\ref{fig:TauPhiLG}. 
Solid line: classical Lyapunov exponent. Dashed lines: the FGR 
quadratic behavior. Note that for decreasing $\lambda _{dB}$ the 
critical perturbation diminishes, implying a collapse of the Fermi 
Golden Rule regime.}%
\label{fig:AlphacvsK}%
\end{center}
\end{figure}

The semiclassical analysis of the Lorentz gas yielded a critical value of the 
perturbation to enter in the Lyapunov regime [Eq.~(\ref{CriticalAlpha})], that vanishes in the semiclassical limit, $\alpha_{\mathrm{c}}\rightarrow0$ for $\hbar$ (or $\lambda_{dB}$) $\rightarrow0$, implying the collapse of the Fermi Golden Rule 
regime. This behavior is reproduced by numerical calculations (Fig.~\ref{fig:AlphacvsK}). Here, $\lambda_{dB}$ is decreased while keeping fixed the size $\sigma$ of the initial wave packet. A point that should not be over--sighted is that the perturbation $\Sigma$ [Eq. (\ref{MTPerturbation})], for a given value of the parameter $\alpha$, scales with the energy in a way that the underlying classical trajectories are always affected in the same 
way by the perturbation. The extracted crossover values of $\alpha_{\mathrm{c}}$ are in quantitative agreement with Eq.~(\ref{CriticalAlpha}), decreasing with $\lambda_{dB}$ in the tested interval, where numerical computations take reasonable time to finish. 

Note that other choices of the perturbation $\Sigma$, such as the quenched disorder of Eq.~(\ref{QuenchedSigma}) \cite{Jalabert01,Cucchietti02Smooth}, can be shown to give critical values that decrease with decreasing $\hbar$ as in Eq.~(\ref{CriticalAlpha}), provided that the perturbation is scaled to the proper semiclassical limit. That is, for a fixed 
perturbation potential, we should take the limit of $\lambda_{dB}\rightarrow0$. As a result, if we keep $\hbar$ constant and decrease $\lambda_{dB}$ by increasing the particle energy, we should scale up the perturbation potential consistently (assuming that $\mathcal{H}_{0}$ generates the same dynamics at all energies). 
 
The strong conclusion to be extracted from this result is the main one of this chapter: in the semiclassical limit, any perturbation will be strong enough to put us in the Lyapunov regime, in consistency with the hypersensitivity expected for a classical system. In this limit the Lyapunov regime of the LE shares the universality of classical chaos. 
However, this is not such an unexpected result, as in this limit the Ehrenfest time diverges and the correspondence principle should prevail at all times. 
 
\begin{figure}[tb] 
\begin{center}
\leavevmode
\epsfxsize 4in
\epsfbox{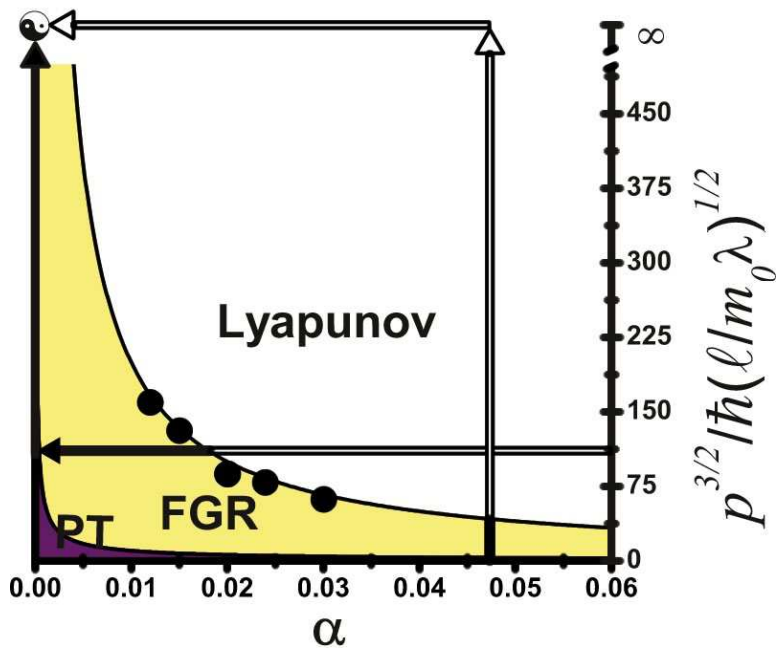}
\caption{Regime diagram for the Loschmidt echo as a function of the 
perturbation and the energy (or inverse $\hbar$). The yellow area is the FGR 
regime, while the clear one is the Lyapunov regime. The line that divides both 
regimes is Eq. (\ref{CriticalAlpha}). The dots are the numerical values 
obtained from Figs. \ref{fig:TauPhiLG} and \ref{fig:AlphacvsK}. The blue region is where perturbation theory (PT) applies, although for increasing energy this regime collapses faster than the FGR. The arrows 
schematize the possible ordering of the classical double limit of the 
perturbation and the wavelength going to zero. Notice how the lower one gives 
always zero while the upper (correct) one gives $\lambda$ since it remains 
always in the Lyapunov regime.}
\label{fig:PhaseDiagram}
\end{center}
\end{figure}

The sound evidence presented above allows other perspectives which will provide useful insight on the different regimes of the LE. In particular, we can devise a plot of a scaling parameter proportional to the particle's energy and the inverse of $\hbar$ as a function of the perturbation strenght. In such a plot, the critical perturbation Eq.~(\ref{CriticalAlpha}) is a curve that separates the FGR from the Lyapunov regime. In a sense, this perspective offers a ``phase'' or regime diagram for the LE. We can see this plot in Fig.~\ref{fig:PhaseDiagram}. The shaded region corresponds to the Fermi Golden Rule regime and the clear one to the Lyapunov regime, while the line that divides both phases is  Eq.~(\ref{CriticalAlpha}). Note that the dots are the numerical values of $\alpha_{\mathrm{c}}$, extracted from Figs.~\ref{fig:TauPhiLG} and \ref{fig:AlphacvsK}. There is of course another transition from FGR to perturbative regime (dotted area) appearing when $\Sigma\simeq\Delta$. This perturbative value also goes to zero in the semiclassical limit of $\lambda_{dB}\rightarrow0.$ Finally, the Lyapunov regime is bounded from above by an $\hbar$ independent critical value $\alpha_{\mathrm{p}}$ marking the classical breakdown discussed bellow (dashed area). 
 
A remarkable conceptual feature highlighted by Fig.~\ref{fig:PhaseDiagram}, is the importance of the order in which we take the limits of $\Sigma$ and $\lambda_{dB}$ going to zero. Two distinct results are obtained for the different order in which we can take this double limit. As depicted in the figure (with arrows representing 
the limits), 
\be
\lim_{\lambda_{dB}\rightarrow 0}\lim_{\Sigma\rightarrow0}1/\tau_{\phi}=0,
\label{WrongLimit}
\ee 
for the FGR exponent always goes to zero with the perturbation.
On the other hand, taking the inverse (more physical) ordering
\be
\lim_{\Sigma\rightarrow0}\lim_{\lambda_{dB}\rightarrow0}1/\tau_{\phi}=\lambda,
\label{CorrectLimit}
\ee 
the semiclassical result is obtained because the FGR regime has collapsed.
 
The semiclassical theory clearly fails when the perturbation is strong enough 
(or the times long enough) to appreciably modify the classical trajectories. 
This would give an upper limit (in perturbation strength) for the results of 
Sec.~\ref{sec:SCLE}. A more stringent limitation comes from the finite value of $\hbar$, 
due to the limitations of the diagonal approximations and linear expansions of 
the action that we have relied on. In other systems, like the quenched 
disorder in a smooth stadium \cite{Cucchietti02Smooth}, the upper critical 
value of the perturbation (for exiting the Lyapunov regime) can be related to 
the transport mean free path of the perturbation $\tilde{\ell}_{\mathrm{tr}}$, 
which is defined as the length scale over which the classical trajectories are 
affected by the disorder\cite{Jalabert96}. 
 
We can obtain in the Lorentz gas an estimate of $\tilde{\ell}_{\mathrm{tr}}$ by 
considering the effect of the perturbation on a single scattering event. The 
difference $\delta\theta$ between the perturbed and unperturbed exit angles 
after the collision can be obtained using Eqs. (\ref{ReflectionLaw}), which 
results in 
\be 
\delta\theta\sim4n_{x}n_{y}\left(  \frac{\mathbf{v}\cdot\mathbf{n}}{v}\right) 
^{2} \ \alpha\ , \label{DeltaAngle}%
\ee 
where $\mathbf{v}$ is the initial velocity of the particle and $\mathbf{n}$ is 
the normal to the surface. 
  
Assuming that the movement of the particle is not affected by chaos 
(non-dispersive collisions), one can do a random walk approach and 
estimate the mean square distance after a time $\tau_{tr}$ from the 
fluctuations of the angle in Eq.~(\ref{DeltaAngle}). Estimating
the transport mean free time as that at which the fluctuations are of the 
order of $R$, one obtains 
\be 
\tilde{\ell}_{\mathrm{tr}}\simeq \frac{4 R^2}{3 \alpha^2\ell}\ , 
\label{TransportL}%
\ee 
where a uniform probability for the angle of the velocity is assumed.
Eq.~(\ref{TransportL}) is used to get the upper bound 
perturbation $\alpha_{\mathrm{p}}$ for the end of the Lyapunov plateau, 
\be 
\alpha_{\mathrm{p}}=\sqrt{\frac{4\lambda R^2}{3 \ell v}}\ . \label{AlphaMax}%
\ee 
For the parameters used in the examples of this chapter, $\alpha_{\mathrm{p}}\simeq 0.23,$ $0.29$ and $0.43$ respectively for increasing magnitude of the three 
concentrations shown in Fig.~\ref{fig:TauPhiLG}. It is rather
difficult to reach numerically these perturbations in our system, 
since the initial Gaussian decay drives $M(t)$ very quickly 
towards its saturation value, preventing the observation of an 
exponential regime. Moreover, one should keep in mind that Eq.~(\ref{AlphaMax}) is just an upper bound. Despite these difficulties, we can see in 
Fig.~\ref{fig:TauPhiLG} that the Lyapunov regime plateau appears
to end for sufficiently strong perturbations. For the range explored, the limiting values are in qualitative agreement with the estimation from Eq.~(\ref{AlphaMax}). 

\section{Summary}

The LE was studied numerically for three different chaotic systems. The results strongly support the analytical predictions of Chap.~\ref{chap:Semiclassical}. Furthermore, some of the approximations of the theory were tested numerically showing that the result is more general than expected. 

The LE was computed in a Smooth billiard, the Bunimovich stadium and a Lorenz gas. The different nature of the perturbations in the examples implies that its details are generally irrelevant for the LE in chaotic systems. The Bunimovich stadium also shows that disorder in the perturbation or in the Hamiltonian is just an artifact needed by the theory, and that results do not depend on it. In the same line, it was also shown that while averages over the disorder (or over initial states) can be used to obtain better precision of the decay rates, individual curves decay oscillating slightly around the average exponent. 

The robustness of the Lyapunov regime against these effects, plus the persistence of the exponential decay after the Ehrenfest time and the recovery of the classical hypersensitivity in the limit of $\hbar \rightarrow 0$ are evidence of the universality of the Lyapunov regime, much like the universality ascribed to classical chaos theory.

The \textquotedblleft phase diagram\textquotedblright representation for the different regimes of the LE, Fig. \ref{fig:PhaseDiagram}, is a simple conceptual tool that sums up most of the results presented in this section in a straightforward yet profound way. 

\section*{Original results}
\begin{itemize}
\item First numerical evidence of the Lyapunov regime, presented in \cite{Cucchietti02Lorentz}, later expanded in \cite{Cucchietti04}.
\item Persistence of the effect for times longer than $t_E$ \cite{Cucchietti02Lorentz,Cucchietti04}.
\item Study of the LE in the Bunimovich stadium and observation of a FGR-like regime for non Lorentzian LDOS \cite{Wisniacki02}.
\item Existence of the Lyapunov regime for the Smooth billiard, with a mixed phase space \cite{Cucchietti02Smooth}.
\item Observation of exponential decay in individual curves and independence of averages \cite{Cucchietti02Lorentz,Cucchietti04}.
\item High energy ($k_{dB} \rightarrow \infty$ or $\hbar \rightarrow 0$) limit of the LE, showing the collapse of the FGR in the classical limit and the recovery of classical hypersensitivity to perturbations \cite{Cucchietti04}.
\end{itemize}

\chapter{The Loschmidt echo, decoherence and the quantum-classical transition}

\begin{quote}
{\em
`The name of the song is called ``Haddock's Eyes.'' '

`Oh, that's the name of the song, is it?' Alice said, trying to feel interested.

`No, you don't understand,' the knight said, looking a little vexed. `That's what the name is {\bf called}. The name really {\bf is} ``The Aged Aged Man.'' '

`Then I ought to have said ``That's what the {\bf song} is called" ?' Alice corrected herself.

`No, you oughtn't: That's quite another thing! The {\bf song} is called ``Ways and Means'': but that's only what it's {\bf called}, you know !''

`Well, what {\bf is} the song, then?' said Alice, who was by this time completely bewildered.

`I was coming to that,' the knight said. `The song really {\bf is} ``A--sitting on A gate'': and the tune's my own invention.'.
}

Lewis Carrol, {\em Through the Looking Glass}.
\end{quote}

A remarkable feature of the Loschmidt echo observed in the previous chapters is the fact that a quantity related to the classical dynamics (the Lyapunov exponent) emerges out of a quantum magnitude. What is quantum, what is classical and what is the interplay between them for a given system are questions at the core of the so called quantum--classical transition.
It is well known that in our everyday life quantum mechanics is more the exception than the rule: we hardly ever encounter phenomena such as superpositions or matter interference--related effects\footnote{This paradox deeply intrigued Schr\"{o}dinger and was summed up in his famous paradox of the cat in a box with an atom that triggers a killing mechanism when it decays. The observation is that after a while, the cat would entangle with the atom and would therefore exist in a superposition of dead and alive states.}. 
How is it then that, although the underlying laws are quantum, the resulting reality is classical? Is there a point where quantum effects are lost and all that is left is classicality, or we could never expect to be in a regime
where they are important?

This second point of view could be justified due to the smallness of
$\hbar$, since the quantum classical correspondence principle assures us that
quantum mechanics is irrelevant up to quite large times. However, as we saw in the
previous chapter, the breakdown time for the correspondence (the Ehrenfest
time $t_E$) is actually short, because it depends logarithmically on $\hbar$. 
A crude calculation of $t_E$ for Hyperion, a moon of Saturn, showed \cite{Zurek95B} that $t_E\simeq20$ years, that is, at this point in time Hyperion should be a gigantic quantum superposition of moons all over its orbit!

A possible circumnavigation of this problem is provided by decoherence theory \cite{ZurekPT,JoosBook,Zurek03}. It shows that even the faintest interaction of a quantum system with an environment causes, in the end, a randomization of its phase and therefore a suppression of all quantum effects related to it.

We are interested in studying the relationship between decoherence and the LE because the latter presents many features that have similar counterparts in decoherence theory. In the following sections we will first
see some basic concepts of decoherence to illustrate these similarities.
Thereafter, we will show how a formal relationship between both
fields can be developed both ways, using techniques of the LE to connect with decoherence and vice-versa. For the former path, we will develop a semiclassical theory of Wigner functions and reinterpret the results of Sec.~\ref{sec:GeneralLE} in terms of the
emergence of classical behavior. In the latter
approach, using a formalism typical of decoherence
studies, we will be able to obtain a master equation for the LE and show that its decay rate
is equal to the rate of suppresion of coherence.

\section{Decoherence and the transition from quantum to classical}
\label{sec:Decoherence}

Decoherence is an essential 
ingredient in the explanation of the quantum--classical transition \cite{PazZurekLesHouches,JoosBook,ZurekPT,Zurek03}.
One considers a system coupled to an external environment,
over which the observers have neither information nor control. This is introduced in
the theory as the postulate that one has access only to the {\it reduced}
density matrix of the system $\rho$, which is obtained by tracing out the 
environmental degrees of freedom from the total density matrix,
\be
\rho={\rm Tr}_{\cal E} \rho_T.
\ee

\begin{figure}[htbp]
\begin{center}
\leavevmode
\epsfxsize 5in
\epsfbox{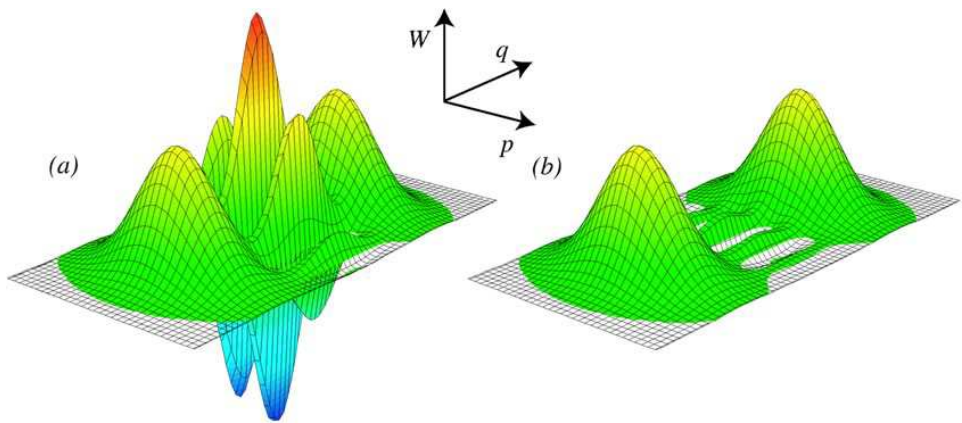}
\caption{$(a)$ Wigner function of the superposition of two Gaussian wave packets centered at $\pm q_0$ with zero momentum. Notice the oscillation pattern in between the two peaks that correspond to the classical distribution. $(b)$ The same Wigner function but in an open system, where decoherence has set in and suppressed the quantum oscillations.}
\label{fig:TwoWigners}
\end{center}
\end{figure}

Another object that has all the accessible information is the Weyl representation of the density matrix, the Wigner function
\be
W(\bq,\bp,t)=\frac{1}{2\pi\hbar}\int d\delta \bq e^{\ii \bp\dot\delta\bq}
\left<\bq-\frac{\delta\bq}{2}\right| \rho(t) \left| \bq+\frac{\delta\bq}{2}\right>.
\label{Wigner}
\ee
The real valued Wigner function provides a phase-space representation for quantum 
states, although it is not strictly a probability density since it can take negative
values. The regions where this happens are actually a signature of the presence of 
definite quantum phase correlations in the wave function, and, as we shall soon see, are suppresed by the environment induced decoherence. When this happens, the Wigner function becomes positive everywhere and becomes similar to
the classical probability distribution.
We can see some of this fascinating behavior already in a simple example, the
superposition of two Gaussian wave packets. The Wigner function for such a state is
\be
W(q,p)=\frac{1}{\sqrt{8}\pi\hbar}e^{-\frac{p^2\sigma^2}{2\hbar^2} }
\left[ e^{-\frac{(q-q_0)^2}{\sigma^2}} +  e^{-\frac{(q+q_0)^2}{\sigma^2} }+
2 e^{-\frac{q^2}{\sigma^2}} \cos \left( \frac{p q_0}{2\hbar} \right) \right]
\label{WignerSuperposition}
\ee
We can see in Fig. \ref{fig:TwoWigners}-$a$ that $W(q,p)$ is composed of two Gaussians centered in the classical positions $\pm q_0$ with zero momentum, and in between them a strong pattern of oscillations forms. These oscillations are the signature that our quantum state is a true superposition of states, and not a statistical mixture.

One of the simplest systems with analytical solution, and yet general
enough to have non--trivial behavior, is a quantum oscillator surrounded by a bath of independent oscillators (called the linear quantum Brownian motion model \cite{PazZurekLesHouches}).
Going through the detailed calculations 
is beyond the need of this section: it will suffice to present the results, the
hypothesis involved and some subsequent relevant additions to the theory.

The total Hamiltonian of the model is 
\ba
\Hc_{CL}&=&\Hc_{S}+\Hc_{Bath}+\Hc_{int} \nonumber \\
&=& \frac{p^2}{2m}+\frac{\Omega q^2}{2}+\sum_k
\frac{p_k^2}{2m}+\frac{\omega_k q_k^2}{2}+\sum_k V(q) q_k,
\label{CaldeiraModel}
\ea
where $q$ and $p$ are the conjugate coordinates of the system, and $q_k$ and
$p_k$ are the coordinates of the $k$-th oscillator in the bath. Assuming a linear coupling, $V(q)=q$, the reduced density matrix can be shown to obey a master equation \cite{HuPazZhang,PazZurekLesHouches}
\ba
\frac{\partial \rho}{\partial t}&=&-\frac{i}{\hbar}\left[\Hc_S+\frac{1}{2}M\tilde{\Omega}^2(t)q^2,\rho\right] \ -
\frac{i}{\hbar}\gamma(t)\left[q,{p,\rho}\right] \nonumber \\
& & 
-\ D(t) \left[q,\left[q,\rho\right]\right] \ - \frac{1}{\hbar}f(t) \left[q,[p,\rho] \right],
\label{MasterEquationGeneral}
\ea
where $\tilde{\Omega}(t)$ is a frequency renormalization, $\gamma(t)$ is a damping coefficient and $D(t)$ and $f(t)$ are diffusion coefficients.
In the limit of high temperatures and absence of dissipation \cite{PazZurekLesHouches}, $\gamma$ and $f$ can be neglected and $D$ considered constant, which gives
\be
\frac{\partial \rho}{\partial t}=-i\hbar\left[\Hc_S,\rho\right] \ -
\ D\left[V(q),\left[V(q),\rho\right]\right].
\label{MasterEquationDissipative}
\ee
Notice that this equation can be derived for a general coupling $V(q)$. The first term on the rhs of Eq.~(\ref{MasterEquationDissipative}) generates unitary 
evolution, the second one is responsible for decoherence: It induces a 
tendency towards diagonalization in position basis and, in the Wigner 
representation, it gives rise to a diffusion term \cite{PazZurekLesHouches}. For the simplest case of $V(q)=q$ the master equation (\ref{MasterEquationDissipative}) can be cast into an equation for the Wigner function,
\begin{equation}
\dot W(q,p)=\left\{\Hc_S,W\right\}_{MB}\ +\ D\ \partial^2_{pp}W(q,p),
\label{WignerMasterEquation}
\end{equation}
where $\left\{...\right\}_{MB}$ is the so--called Moyal bracket, responsible
for unitary evolution.

Eq. (\ref{WignerMasterEquation}) is useful to understand the effect of decoherence on the Wigner function. As we discussed above, interference effects become evident in the Wigner function as rapid oscillations between positive and negative values. If for a given region of phase space we can characterize this oscillations by a well defined wave--number $k_p$ along the momentum direction ($W(q,p,t)~\simeq~A(q,t)~\cos{(k_p p)}$), we can see that the decoherence term in (\ref{WignerMasterEquation}) gives an exponential decay with a rate $\Gamma_D=D k_p^2$. Thus, the oscillations are suppressed and the Wigner function becomes positive (see Fig. \ref{fig:TwoWigners}-$b$). In particular, $\hbar/\Gamma_D$ is then the decoherence time, the typical time it takes the Wigner function to lose all its quantum properties and become equivalent to the classical distribution.

Decoherence is a dynamical process through which the couplings to an environment causes the suppression of phase correlations in a quantum system. After a long time the system is left in a statistical mixture of those states that are more resilient to the decoherence process, the so called ``pointer states''. Generally, the pointer states can be described classically without difficulty, and therefore classical behavior emerges from the system in a natural way. As a last remark, the scale $k_p$ is usually very small for macroscopic systems [think of the inverse of $q_0$ in Eq.~(\ref{WignerSuperposition})], which prevents quantum effects from lasting for long.

The rate $\Gamma_D$ obtained above depends explicitly on $D$, the coupling with the environment. However, for a quantum system with a classically chaotic Hamiltonian the rate at which the environment degrades information about the initial state can be independent of the system--environment coupling strength \cite{Zurek94,Zurek95}. This rate (e.g., as measured by the von Neumann or the linear entropy production rate computed from the reduced density matrix of the system) is set by the classical Lyapunov exponents \cite{Patanayak,Monteoliva00,Monteoliva01}, provided that the coupling strength is within a certain (wide) range.

To see this, we can use the master equation (\ref{WignerMasterEquation}) to obtain 
the time derivative of the purity 
\be
\Pc={\rm Tr} \rho^{2}.
\label{Purity}
\ee
The purity is related to the linear entropy $H=-{\rm ln} \Pc$, analytically simpler to treat than von Neumann's entropy. Rewriting $\Pc$ in terms of the Wigner function and applying Eq.~(\ref{WignerMasterEquation}) \cite{Zurek94},
\ba
\dot{\Pc}&=&\frac{d}{dt} \int dqdp W^{2}(q,p) \nonumber \\
&=& 2 \int dqdp W(q,p) \left[ \left\{\Hc_{S},W\right\}_{MB}\ +\ D\ \partial^2_{pp}W(q,p) \right] \nonumber \\
&=&  2 D \int dqdp W(q,p)  \partial^2_{pp}W(q,p),
\label{PurityDotDerivation}
\ea
where in the last line we used that the unitary part of the evolution integrates out to zero.
After integrating by parts, equation (\ref{PurityDotDerivation}) can be rewritten as 
\be
\frac{\dot{\Pc}}{ \Pc}=-\frac{2D}{\sigma^{2}},
\label{PurityDot}
\ee
where $\sigma$ characterizes the dominant wavelength in the spectrum of the Wigner function,  
\be
\sigma^{-2}=\frac{\int (\partial_p W)^2}{\int  W^2}.
\label{SigmaPurity}
\ee 
Notice that if $\sigma^2$ results proportional to $D$, the rate of change of the purity becomes independent of the diffusion constant.
This happens indeed as the typical width $\sigma$ depends on the competition between two effects \cite{Zurek94}. The first one is the tendency of chaotic evolution to generate (exponentially fast, at a rate set by the Lyapunov exponent $\lambda$) small scale structure in the Wigner function. 
For a system with only one Lyapunov exponent, the classical distribution basically expands in one direction (the unstable one) and contracts in another (the stable), such that the total area remains constant. The expansion/contraction is proportional to $e^{\lambda t}$ and $e^{-\lambda t}$ respectively.
The former would therefore be the typical width of the Wigner function if chaotic motion were the only process acting on it.
However, there is still the effect of the diffusion term of Eq.~(\ref{WignerMasterEquation}), which tends to smear small scales
exponentially fast at a rate determined by the product $D k_p^2$. These two effects reach a balance when $\bar\sigma^{2}=2D/\lambda$ \cite{Zurek94}. Hence, in this regime the purity $\Pc$ decreases exponentially at a rate fixed by $\lambda$. For this behavior to take place $D$ should be above a threshold \cite{Zurek94}, otherwise the critical width is not established (the implicit assumption is that the time scale for diffusion to wash out a $k_p$--oscillation is shorter than the time scale for the oscillations to be regenerated by the dynamics). Therefore, already in a simple scenario the purity has a regime of strong enough perturbations where it decreases exponentially with a Lyapunov rate. 

All these considerations have shown features of decoherence (as quantified by purity) that bear a striking resemblance to the results  obtained for $M(t)$ in previous chapters. Now, let us return to the LE to elucidate the physical origin of these similarities.

\section{Loschmidt echo through semiclassical approximation of the Wigner function} 
\label{sec:wigner} 

In this section we will employ semiclassical techniques to study the evolution of the Loschmidt echo expressed as the integral or overlap between two Wigner functions evolved with slightly different Hamiltonians,
\ba
M(t)=\int \dd q \dd p \ W_0(q,p) W_\Sigma(q,p).
\label{LEWigner}
\ea

Such a framework will be particularly useful to identify links between the LE with decoherence theory. Moreover, the phase space representation of $M(t)$ will allow me to ascribe the different regimes of the LE to the behavior of the Wigner functions in different regions of phase space.
 
\subsection{Classical evolution of the Wigner function} 
 
Let us consider first the semiclassical approximation for a single Wigner function and its time evolution. Using the wave-function propagators of Eqs.~(\ref{QuantumPropagator}), we can express the time-dependence of the Wigner function as
 
\ba 
W(\mathbf{r},\mathbf{p};t)  
&=&\frac{1}{(2\pi\hbar)^{d}}\int\mathrm{d} 
\delta\mathbf{r}\int\mathrm{d}\overline{\mathbf{r}}\int\mathrm{d} 
\delta{\overline{\mathbf{r}}}\int\mathrm{d}{\overline{\mathbf{p}} 
}\ W(\overline{\mathbf{r}},{\overline{\mathbf{p}}};0)\ \exp\left[ 
\frac{\mathrm{i}}{\hbar}(\mathbf{p}\cdot\delta\mathbf{r}-{\overline 
{\mathbf{p}}}\cdot\delta{\overline{\mathbf{r}}})\right] \nonumber\\ 
&  &  \times\ K\left( 
\mathbf{r}-\frac{\delta\mathbf{r}} 
{2},\overline{\mathbf{r}}-\frac{\delta{\overline{\mathbf{r}}}}{2};t\right) 
\ K^{\ast}\left( 
\mathbf{r}+\frac{\delta\mathbf{r}}{2},\overline{\mathbf{r} 
}+\frac{\delta{\overline{\mathbf{r}}}}{2};t\right)  \ . 
\label{ExpansionWignerCompleta} 
\ea 

\begin{figure}[htbp]
\begin{center}
\leavevmode
\epsfxsize 3in
\epsfbox{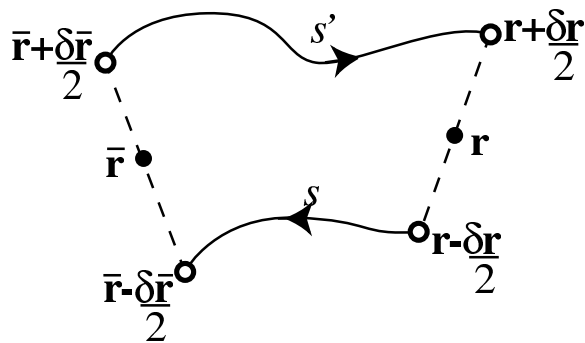} 
\caption{Two semiclassical trajectories needed to expand the evolution of one Wigner function.}
\label{fig:1SCWigner}
\end{center}
\end{figure}
 
\noindent where 
$W(\overline{\mathbf{r}},{\overline{\mathbf{p}}};0)$ is the 
initial Wigner function. The semiclassical expansion of the 
propagators [Eq.~(\ref{SemiclassicalPropagator})] leads to the 
propagation of the Wigner function by ``chords" 
\cite{Ozorio98,Toscano02}, where pairs of trajectories $(s,s^{\prime})$ 
traveling from $(\overline{\mathbf{r}}-\delta{\overline 
{\mathbf{r}}}/2,\overline{\mathbf{r}}+\delta{\overline{\mathbf{r}}}/2)$ 
to $(\mathbf{r}-\delta\mathbf{r}/2,\mathbf{r}+\delta\mathbf{r}/2)$ 
have to be considered (see Fig. \ref{fig:1SCWigner}). In the leading order in $\hbar$ we can 
approximate the above propagators by sums over trajectories going 
from $\overline{\mathbf{r}}$ to $\mathbf{r}$, and the 
semiclassical evolution of the Wigner function is given by 
\ba
W(\br,\bp;t)&=&(2\pi\hbar)^{d} \int \dd \brb
\int \dd \bpb W(\brb,\bpb;0) 
\sum_{s,s^{\prime}}\ \delta \left(  \bpb -\frac{ \bpb_s+\bpb_{s^\prime} } {2} \right)  \nonumber \\
&  & \times \delta \left(\bp-\frac{\bp_s+\bp_{s^\prime} }{2} \right)  
K_s \left( \br,\brb;t \right)   K_{s^\prime}^{\ast}\left( \br,\brb;t \right).
\label{ExpansionWignerSC}
\ea 
 
\noindent where ${\overline{\mathbf{p}}}_{s}$ ($\mathbf{p}_{s}$) 
and ${\overline{\mathbf{p}}}_{s^{\prime}}$ 
($\mathbf{p}_{s^{\prime}}$) are the initial (final) momenta of the 
trajectories $s$ and $s^{\prime}$, respectively. The dominant 
contribution arises from the diagonal term $s=s^{\prime}$,
 
\be 
W_{\mathrm{c}}(\mathbf{r},\mathbf{p},t)=\int\mathrm{d}\overline{\mathbf{r} 
}\sum_{s(\overline{\mathbf{r}},\mathbf{r},t)}\ C_{s}\ \delta\left( 
\mathbf{p}-\mathbf{p}_{s}\right)  \ 
W(\overline{\mathbf{r}},{\overline {\mathbf{p}}}_{s};0)\ . 
\label{SCWigner} 
\ee 
 
Using the fact that $C_{s}$ is the Jacobian of the transformation 
from $\overline{\mathbf{r}}$ to $\mathbf{p}_{s}$, we have 
 
\be 
W_{\mathrm{c}}(\mathbf{r},\mathbf{p};t)=\int\mathrm{d}\mathbf{p}_{s} 
\ \delta\left(  \mathbf{p}-\mathbf{p}_{s}\right)  \ 
W(\overline{\mathbf{r} },{\overline{\mathbf{p}}}_{s};0)\ , 
\label{SCWigner2} 
\ee 
  
\noindent where the trajectories considered now are those that 
arrive to $\mathbf{r}$ with momentum $\mathbf{p}$. We note 
$(\overline{\mathbf{r} },{\overline{\mathbf{p}}})$ the pre-image 
of $(\mathbf{r},\mathbf{p})$ by the classical equations of motion acting on 
a time $t$. That is, $(\mathbf{r},\mathbf{p}) = X_{t} 
(\overline{\mathbf{r}},{\overline{\mathbf{p}}})$. The momentum 
integral 
is trivial, and we obtain the obvious result%
 
\be 
W_{\mathrm{c}}(\mathbf{r},\mathbf{p};t)= W(\overline{\mathbf{r}} 
,{\overline{\mathbf{p}}};0) \ , \label{SCWigner3} 
\ee 
  
\noindent with $(\overline{\mathbf{r}},{\overline{\mathbf{p}}}) = 
X_{t} ^{-1}(\mathbf{r},\mathbf{p})$. Since $X_{t}$ conserves the 
volume in phase-space, we have shown that at the classical level the Wigner function 
evolves by simply following the classical flow. Although this seems like a dead-end result, it actually is what one expects for the semiclassical approximation for only one Wigner function. In order to introduce quantum phase effects, one needs to consider higher order expansions of Eq.~(\ref{SemiclassicalPropagator}) like in \cite{Ozorio98} or, alternatively, consider two Wigner functions and the relative phase between the trajectories of their semiclassical expansion. The Loschmidt echo is then appropriate for this analysis, which we will undertake in the following section.
 
\subsection{Semiclassical approximation of the LE 2: the Wigner function} 
\label{sec:SCWignerLE}
 
As indicated in Eq.~(\ref{LEWigner}), the Loschmidt echo is given by the 
phase-space trace of two Wigner functions associated with slightly different 
Hamiltonians ($\Hc_{0}$ and $\Hc_{0}+\Sigma$). In order to 
facilitate the discussion, let us introduce the density (or partial trace) 
$f_{\Sigma}$ writing the LE as%
 
\be 
M(t)=\int\mathrm{d}\mathbf{r}\ f_{\Sigma}(\mathbf{r},t)\ , 
\label{DefinitionF}%
\ee 
 
\noindent with%
 
\ba 
f_{\Sigma}(\mathbf{r},t)  &=& \frac{1}{(2\pi\hbar)^{d}}\ \int\mathrm{d}%
\mathbf{p}\int\mathrm{d}\delta\mathbf{r}\int\mathrm{d}\overline{\mathbf{r}%
}\int\mathrm{d}\delta{\overline{\mathbf{r}}}\int\mathrm{d}{\overline 
{\mathbf{p}}}\int\mathrm{d}\delta\mathbf{r}^{\prime}\int\mathrm{d}%
{\overline{\mathbf{r}}}^{\prime}\int\mathrm{d}\delta{\overline{\mathbf{r}}%
}^{\prime}\int\mathrm{d}{\overline{\mathbf{p}}}^{\prime}\ \nonumber\\
& \times& 
 \exp{\left[  \frac{\mathrm{i}}{\hbar}\left( 
\mathbf{p}\cdot\delta\mathbf{r}-{\overline{\mathbf{p}}}\cdot\delta 
{\overline{\mathbf{r}}}\right)  \right]  }
 \exp{\left[  -\frac{\mathrm{i}%
}{\hbar}\left(  \mathbf{p}\cdot\delta\mathbf{r}^{\prime}-{\overline 
{\mathbf{p}}}^{\prime}\cdot\delta{\overline{\mathbf{r}}}^{\prime}\right) 
\right]  }
 \nonumber\\ & \times&
W(\overline{\mathbf{r}},{\overline{\mathbf{p}}};0)\ W^{\ast}({\overline 
{\mathbf{r}}}^{\prime},{\overline{\mathbf{p}}}^{\prime};0) 
K\left(  \mathbf{r}-\frac{\delta\mathbf{r}}{2},\overline 
{\mathbf{r}}-\frac{\delta{\overline{\mathbf{r}}}}{2};t\right)  K^{\ast}\left(  \mathbf{r}+\frac{\delta\mathbf{r}%
}{2},\overline{\mathbf{r}}+\frac{\delta{\overline{\mathbf{r}}}}{2};t\right) 
\nonumber\\ 
&  \times & 
 K^{\ast}\left(  \mathbf{r}-\frac{\delta\mathbf{r}^{\prime}}{2}%
,{\overline{\mathbf{r}}}^{\prime}-\frac{\delta{\overline{\mathbf{r}}}^{\prime 
}}{2};t\right)  \ K\left(  \mathbf{r}+\frac{\delta\mathbf{r}^{\prime}}%
{2},{\overline{\mathbf{r}}}^{\prime}+\frac{\delta{\overline{\mathbf{r}}%
}^{\prime}}{2};t\right)  \ . \label{F}%
\ea 
  
\begin{figure}[tb]
\begin{center}
\leavevmode
\epsfxsize 3in
\epsfbox{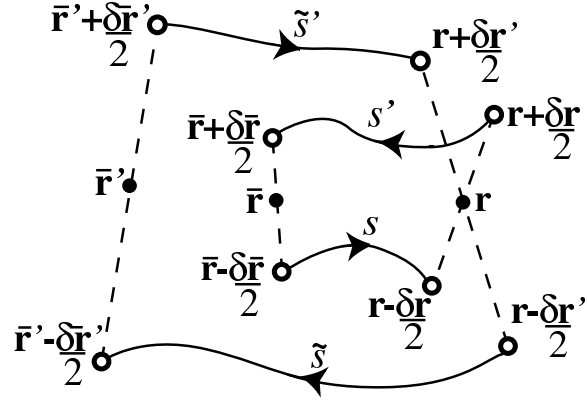} 
\caption{ Four classical trajectories used to compute semiclassically the 
Loschmidt echo through the evolution of two Wigner functions associated with 
different Hamiltonians.}%
\label{2Wigners}%
\end{center}
\end{figure}

The semiclassical evolution of $f_{\Sigma}$ is given by sets of four trajectories, as 
illustrated schematically in Fig.~\ref{2Wigners}. 
 
As previously done in other semiclassical calculations in this work, 
let us take Gaussian wave-packet (of width $\sigma$) as initial state. Its associated Wigner function reads%
 
\be 
W(\overline{\mathbf{r}},{\overline{\mathbf{p}}};0) = \frac{1}{(\pi\hbar)^{d}} 
\ \exp{\left[  -\frac{(\overline{\mathbf{r}}-\mathbf{r}_{0})^{2}}{\sigma^{2}} 
- \frac{({\overline{\mathbf{p}}}-\mathbf{p}_{0})^{2}\sigma^{2}}{\hbar^{2}} 
\right]  } \ . \label{InitialWigner}%
\ee 
  
Note that the integral on $\bp $ gives $\delta(\delta \br-\delta \brp)$, rendering the integral on $\delta \brp$ trivial. Assuming that $\Sigma$ constitutes a small perturbation, after performing these integrations we can obtain
 
\ba 
f_{\Sigma}(\mathbf{r},t)  &=&\frac{\sigma^{2}}{(2\pi^{3}\hbar^{4})^{d/2}}%
\int\mathrm{d}\delta\mathbf{r}\int\mathrm{d}\overline{\mathbf{r}}%
\int\mathrm{d}\delta{\overline{\mathbf{r}}}\int\mathrm{d}{\overline 
{\mathbf{p}}}\int\mathrm{d}\delta{\overline{\mathbf{r}}}^{\prime}%
\int\mathrm{d}{\overline{\mathbf{p}}}^{\prime}\ 
\nonumber\\ &\times & 
\exp{\left[  \frac{\mathrm{i}%
}{\hbar}\left(  {\overline{\mathbf{p}}}^{\prime}\cdot\delta{\overline 
{\mathbf{r}}}^{\prime}-{\overline{\mathbf{p}}}\cdot\delta{\overline 
{\mathbf{r}}}\right)  \right]  }\ 
\exp{\left[  -\frac{2}{\sigma^{2}}%
(\overline{\mathbf{r}}-\mathbf{r}_{0})^{2}\right]  } 
\nonumber\\ &\times & 
\exp{\left[  -\frac{\sigma^{2}}{\hbar^{2}}\left( 
({\overline{\mathbf{p}}}-\mathbf{p}_{0})^{2}+({\overline{\mathbf{p}}}^{\prime 
}-\mathbf{p}_{0})^{2}\right)  \right]  }\ 
\sum_{s,s^{\prime}}\ \sum 
_{{\tilde{s}},{\tilde{s}}^{\prime}}\ 
\exp{\left[  -\frac{\mathcal{P}^{2}%
\sigma^{2}}{8\hbar^{2}}\right]  }\ 
\nonumber\\ &\times & 
K_{s}\left(  \mathbf{r}-\frac 
{\delta\mathbf{r}}{2},\overline{\mathbf{r}}-\frac{\delta{\overline{\mathbf{r}%
}}}{2};t\right)  
 K_{{\tilde{s}}}^{\ast}\left(  \mathbf{r}%
+\frac{\delta\mathbf{r}}{2},\overline{\mathbf{r}}+\frac{\delta{\overline 
{\mathbf{r}}}}{2};t\right) 
\nonumber\\ &\times & 
K_{s^{\prime}}^{\ast}\left(  \mathbf{r}-\frac 
{\delta\mathbf{r}}{2},\overline{\mathbf{r}}-\frac{\delta{\overline{\mathbf{r}%
}}^{\prime}}{2};t\right)   
K_{{\tilde{s}}^{\prime}}\left(  \mathbf{r}%
+\frac{\delta\mathbf{r}}{2},\overline{\mathbf{r}}+\frac{\delta{\overline 
{\mathbf{r}}}^{\prime}}{2};t\right)  \ . \label{FSC1}%
\ea 
  
\begin{figure}[tb] 
\begin{center}
\leavevmode
\epsfxsize 3in
\epsfbox{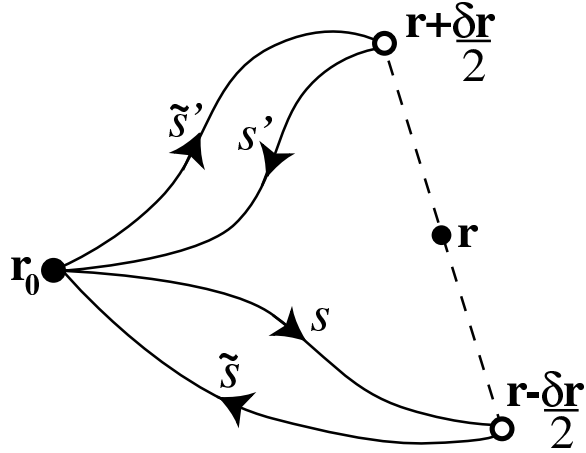} 
\caption{ For fairly localized initial wave-packet, the four classical 
trajectories contributing to the LE can be reduced to those starting at its 
center $\mathbf{r}_{0}$. }%
\label{2WwSIP}%
\end{center}
\end{figure}

\noindent Where we have defined%
 
\be 
\mathcal{P} = {\overline{\mathbf{p}}}_{s}+{\overline{\mathbf{p}}}_{s^{\prime}%
}-{\overline{\mathbf{p}}}_{{\tilde s}}-{\overline{\mathbf{p}}}_{{\tilde 
s}^{\prime}} \ ,
\ee 
and, after changing variables $\xi=(\brb+\brbp)/2$, $\delta \xi=\brb-\brbp$, the Gaussian integral on $\delta \xi$ was performed.
 Now the trajectories $s$ and $s^{\prime}$ (${\tilde s}$ and ${\tilde 
s}^{\prime}$) arrive to the same final point $\overline{\mathbf{r}}%
-\delta{\overline{\mathbf{r}}}/2$ ($\mathbf{r}+\delta\mathbf{r}/2$). Since the 
initial wave-packet is concentrated around $\mathbf{r}_{0}$, we can further 
simplify and work with trajectories $s$ and $s^{\prime}$ (${\tilde s}$ and 
${\tilde s}^{\prime}$) that have the same extreme points. Therefore, we have%
 
\ba 
f_{\Sigma}(\mathbf{r},t)  &=& \frac{\sigma^{2}}{(2\pi^{3}\hbar^{4})^{d/2}%
}\ \int\mathrm{d}\delta\mathbf{r}\int\mathrm{d}\overline{\mathbf{r}}%
\int\mathrm{d}\delta{\overline{\mathbf{r}}}\ 
\exp{\left[  -\frac{2}{\sigma 
^{2}}(\overline{\mathbf{r}}-\mathbf{r}_{0})^{2}-\frac{\delta{\overline 
{\mathbf{r}}}^{2}}{2\sigma^{2}} \right]  }\ 
\nonumber\\ &  &  \times\
\sum_{s,s^{\prime}%
}\ \sum_{{\tilde{s}},{\tilde{s}}^{\prime}}\ \exp{\left[  -\frac{\mathcal{P}%
^{2}\sigma^{2}}{8\hbar^{2}}-\frac{2\sigma^{2}}{\hbar^{2}}\left( 
\frac{\mathcal{R}}{4}-\mathbf{p}_{0}\right)  ^{2} \right]  }
\nonumber\\ &  &  \times\
K_{s}\left(  \mathbf{r}-\frac{\delta\mathbf{r}}{2}%
,\overline{\mathbf{r}}-\frac{\delta{\overline{\mathbf{r}}}}{2};t\right) 
K_{s^{\prime}}^{\ast}\left(  \mathbf{r}-\frac{\delta\mathbf{r}}{2}%
,\delta{\overline{\mathbf{r}}}-\frac{\delta{\overline{\mathbf{r}}}}%
{2};t\right)  
\nonumber\\ &  &  \times\
K_{{\tilde{s}}}^{\ast}\left(  \mathbf{r}+\frac{\delta 
\mathbf{r}}{2},\overline{\mathbf{r}}+\frac{\delta{\overline{\mathbf{r}}}}%
{2};t\right)  \ K_{{\tilde{s}}^{\prime}}\left(  \mathbf{r}+\frac 
{\delta\mathbf{r}}{2},\delta{\overline{\mathbf{r}}}+\frac{\delta 
{\overline{\mathbf{r}}}}{2};t\right)  \ , \label{FSC2}%
\ea 
  
\noindent with%
 
\be 
\mathcal{R} = {\overline{\mathbf{p}}}_{s}+{\overline{\mathbf{p}}}_{s^{\prime}%
}+{\overline{\mathbf{p}}}_{{\tilde s}}+{\overline{\mathbf{p}}}_{{\tilde 
s}^{\prime}} \ . 
\ee 
The integrals on $\bpb$ and $\bpbp$ are trivial, while the integral on $\delta \brbp$ involves a change of variables as above to the mean and the difference with $\delta \brb$. By the same considerations as before, we can reduce all four trajectories to 
start at the center $\mathbf{r}_{0}$ of the initial wave-packet 
(Fig.~\ref{2WwSIP}) 
 
\ba 
f_{\Sigma}(\mathbf{r},t)  & =&(4\pi\sigma^{2})^{d}\ \int\mathrm{d}%
\delta\mathbf{r}\ \sum_{s,s^{\prime}}\ \sum_{{\tilde{s}},{\tilde{s}}^{\prime}%
}\ \exp{\left[  -\frac{(\mathcal{P}^{2}+\mathcal{S}^{2}+\mathcal{T}^{2}%
)\sigma^{2}}{8\hbar^{2}}\right]  }
\nonumber \\ 
&  &  \times\ 
\exp{\left[  -\frac{2\sigma^{2}}{\hbar^{2}%
}\left(  \frac{\mathcal{R}}{4}-\mathbf{p}_{0}\right)  ^{2}\right] 
}K_{s}\left(  \mathbf{r}-\frac{\delta\mathbf{r}}%
{2},\mathbf{r}_{0};t\right)  \ K_{s^{\prime}}^{\ast}\left(  \mathbf{r}%
-\frac{\delta\mathbf{r}}{2},\mathbf{r}_{0};t\right)  
\nonumber\\ 
&  &  \times\  K_{{\tilde{s}}}^{\ast 
}\left(  \mathbf{r}+\frac{\delta\mathbf{r}}{2},\mathbf{r}_{0};t\right) 
\ K_{{\tilde{s}}^{\prime}}\left(  \mathbf{r}+\frac{\delta\mathbf{r}}%
{2},\mathbf{r}_{0};t\right)  \ , \label{FSC3}%
\ea 
 
\noindent with%
 
\ba 
\mathcal{S}  &  = &{\overline{\mathbf{p}}}_{s}-{\overline{\mathbf{p}}%
}_{s^{\prime}}+{\overline{\mathbf{p}}}_{{\tilde s}}-{\overline{\mathbf{p}}%
}_{{\tilde s}^{\prime}} \ ,\label{DEFST0}\\ 
 \mathcal{T}  &  = &{\overline{\mathbf{p}}}_{s}+{\overline 
{\mathbf{p}}}_{s^{\prime}}-{\overline{\mathbf{p}}}_{{\tilde s}}-{\overline 
{\mathbf{p}}}_{{\tilde s}^{\prime}} \ . \label{DEFST1}%
\ea 
  
Given that%
 
\ba
\mathcal{P}^{2}+\mathcal{S}^{2}+\mathcal{T}^{2} &=&\left(  {\overline 
{\mathbf{p}}}_{s}-{\overline{\mathbf{p}}}_{s^{\prime}}\right)  ^{2}+ \left( 
{\overline{\mathbf{p}}}_{s}-{\overline{\mathbf{p}}}_{{\tilde s}}\right)  ^{2}+ 
\left(  {\overline{\mathbf{p}}}_{s}-{\overline{\mathbf{p}}}_{{\tilde 
s}^{\prime}}\right)  ^{2}+ 
\nonumber\\ 
& & \times\ 
\left(  {\overline{\mathbf{p}}}_{s^{\prime}%
}-{\overline{\mathbf{p}}}_{{\tilde s}}\right)  ^{2}+ \left(  {\overline 
{\mathbf{p}}}_{s^{\prime}}-{\overline{\mathbf{p}}}_{{\tilde s}^{\prime}%
}\right)  ^{2}+ \left(  {\overline{\mathbf{p}}}_{{\tilde s}}-{\overline 
{\mathbf{p}}}_{{\tilde s}^{\prime}}\right)  ^{2} \ , \label{SumPST}%
\ea
  
\noindent and since the pairs of trajectories $(s,s^{\prime})$ and $({\tilde 
s},{\tilde s}^{\prime})$ have the same extreme points, the dominant 
contribution to $f_{\Sigma}$ will come from the terms with $s=s^{\prime}$ and 
${\tilde s}={\tilde s}^{\prime}$. Such an identification minimizes the 
oscillatory phases of the propagators, and corresponds to the first diagonal 
approximation of the calculation of Sec.~\ref{sec:SCLE} and 
Ref.~\cite{Jalabert01}. Within such an approximation we have%
 
\ba 
f_{\Sigma}(\mathbf{r},t)  &  =&\left(  \frac{\sigma^{2}}{\pi\hbar^{2}}\right) 
^{d}\ \int\mathrm{d}\delta\mathbf{r}\ \sum_{s,{\tilde{s}}}\ C_{s}%
\ C_{{\tilde{s}}}\ 
\exp{\left[  -\frac{\left(  {\overline{\mathbf{p}}}%
_{s}-{\overline{\mathbf{p}}}_{{\tilde{s}}}\right)  ^{2}\sigma^{2}}{2\hbar^{2}%
}\right]  } \nonumber\\ 
& \times & 
\exp{\left[ -\frac{2\sigma^{2}}{\hbar^{2}}\left(  \frac{{\overline{\mathbf{p}}}%
_{s}+{\overline{\mathbf{p}}}_{{\tilde{s}}}}{2}-\mathbf{p}_{0}\right) 
^{2} \right] }  \nonumber\\ 
& \times & 
 \exp{\left[  \frac{\mathrm{i}}{\hbar}\left(  \Delta 
S_{s}\left(  \mathbf{r}-\frac{\delta\mathbf{r}}{2},\mathbf{r}_{0},t\right) 
-\Delta S_{{\tilde{s}}}\left(  \mathbf{r}+\frac{\delta\mathbf{r}}%
{2},\mathbf{r}_{0},t\right)  \right)  \right]  }. \label{FSC4}%
\ea 
  
\noindent As in Eq.~(\ref{EchoAmplitudeAprox}), $\Delta S_{s,{\tilde s}}$ is 
the extra contribution to the classical action that the trajectory $s$ 
(${\tilde s}$) acquires by effect of the perturbation $\Sigma$. 
 
\begin{figure}[tb] 
\begin{center}
\leavevmode
\epsfxsize 3in
\epsfbox{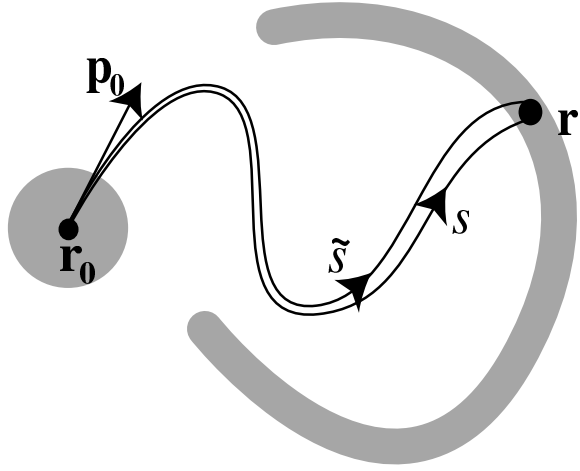} 
\caption{ Classical trajectories in the manifold that evolves classically from 
$\mathbf{r}_{0}$ to $\mathbf{r}$, representing the diagonal component of 
$f_{\Sigma}$. The action differences $\Delta S$ of trajectories $s$ and 
${\tilde s}$ are correlated. The shaded regions depict the initial and final 
classical densities. }%
\label{DiagonalTraj}%
\end{center}
\end{figure}

We have two different cases, depending on whether or not there are 
trajectories leaving from $\mathbf{r}_{0}$ with momentum close to 
$\mathbf{p}_{0}$ that arrive to the neighborhood of $\mathbf{r}$ after a time 
$t$. In the first case $\mathbf{r}$ is in the manifold that evolves 
classically from the initial wave-packet (Fig.~\ref{DiagonalTraj}). Such a 
contribution is dominated by the terms where the trajectory ${\tilde s}$ 
remains close to its partner $s$, and calling $f_{\Sigma}^{\mathrm{d}}$ this 
diagonal component, we get 
 
\ba
f_{\Sigma}^{\mathrm{d}}(\mathbf{r},t)&=&\left(  \frac{\sigma^{2}}{\pi\hbar^{2}%
}\right)  ^{d}\ \int\mathrm{d}\delta\mathbf{r}\ \sum_{s,{\tilde{s}}}%
\ C_{s}^{2}\ \exp\left[  -\frac{2\sigma^{2}}{\hbar^{2}}\left(  {\overline 
{\mathbf{p}}}_{s}-\mathbf{p}_{0}\right)  ^{2}\right]  \ 
\nonumber\\ &\times\ &
\exp\left[ 
\frac{\mathrm{i}}{\hbar}\left(  \Delta S_{s}\left(  \mathbf{r}-\frac 
{\delta\mathbf{r}}{2},\mathbf{r}_{0},t\right)  -\Delta S_{{\tilde{s}}}\left( 
\mathbf{r}+\frac{\delta\mathbf{r}}{2},\mathbf{r}_{0},t\right)  \right) 
\right] . \label{FSC5}%
\ea
 
Assuming, as in Sec.~\ref{sec:SCLE}, that $\mathcal{H}_{0}$ stands for a chaotic system and that the perturbation $\Sigma$ represents a spatial disorder (the more general case of a time dependent $\Sigma$ should follow easily), upon average we obtain spacial
  
\be 
\left\langle \exp\left[  \frac{\mathrm{i}}{\hbar}\left(  \Delta S_{s}\left( 
\mathbf{r}-\frac{\delta\mathbf{r}}{2},\mathbf{r}_{0},t\right)  -\Delta 
S_{{\tilde{s}}}\left(  \mathbf{r}+\frac{\delta\mathbf{r}}{2},\mathbf{r}%
_{0},t\right)  \right)  \right]  \right\rangle =\exp\left[  -\frac{1}%
{2\hbar^{2}}\ A\ \delta\mathbf{r}^{2}\right]  \ , \label{ImpurityAverage}%
\ee 
 
\noindent where $A$ is given by Eq.~(\ref{AS}). We therefore have%
 
\be 
f_{\Sigma}^{\mathrm{d}}(\mathbf{r},t) = \left(  \frac{2\sigma^{4}}{\pi 
\hbar^{2} A}\right)  ^{d/2} \ \sum_{s(\mathbf{r}_{0},\mathbf{r},t)} 
\ C_{s}^{2} \ \exp{\left[  -\frac{2 \sigma^{2}}{\hbar^{2}} \left( 
{\overline{\mathbf{p}}}_{s} - \mathbf{p}_{0}\right)  ^{2} \right]  } \ , 
\label{FSC6}%
\ee  
 
and the corresponding contribution to the Loschmidt echo is%
 
\be 
M^{\mathrm{d}}(t)=\int d\mathbf{r}\ f_{\Sigma}^{\mathrm{d}}(\mathbf{r}%
,t)=\left(  \frac{2\sigma^{4}}{\pi\hbar^{2}A}\right)  ^{d/2}\ \int 
\mathrm{d}{\overline{\mathbf{p}}}\ C\exp{\left[  -\frac{2\sigma^{2}}{\hbar 
^{2}}\left(  {\overline{\mathbf{p}}}-\mathbf{p}_{0}\right)  ^{2}\right]  }\ . 
\label{DiagonalContribution}%
\ee 
 
As in Eq.~(\ref{MSigma0}) we have used $C$ as the Jacobian of the transformation from $\mathbf{r}$ to ${\overline{\mathbf{p}}}$. 
Now the dominant trajectories are those starting from $\mathbf{r}_{0}$ and 
momentum $\mathbf{p}_{0}$. We are then back to the case of the diagonal contribution of Sec.~\ref{sec:DTerms}. 
 
\be 
M^{\mathrm{d}}(t) \simeq\overline{A} \ e^{-\lambda t}, \label{MDiagonalW}%
\ee 
 
\noindent where $C=(m/t)^{d} e^{-\lambda t}$ is assumed, and $\overline 
{A}=(m\sigma/A^{1/2}t)^{d}$. The decay rate of the diagonal contribution is 
set by the Lyapunov exponent $\lambda$, and therefore independent on the 
perturbation $\Sigma$. 
 
\begin{figure}[tb] 
\begin{center}
\leavevmode
\epsfxsize 3.5in
\epsfbox{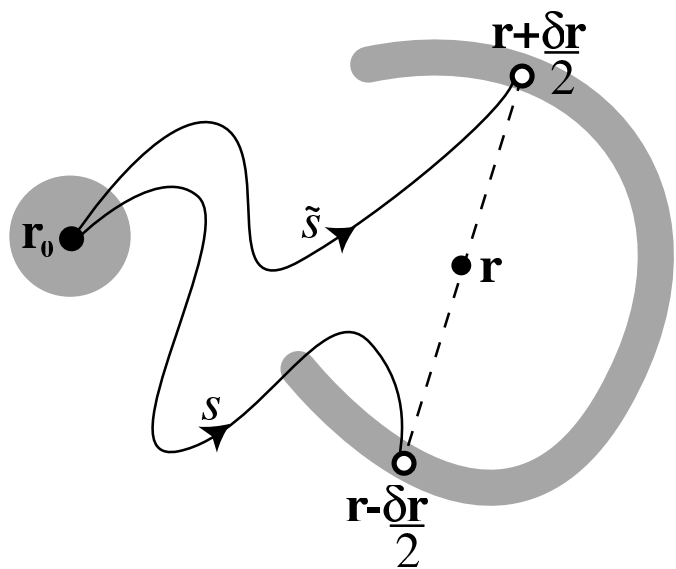} 
\caption{ Non-diagonal classical contribution to the LE given by trajectories 
departing from $\mathbf{r}_{0}$ and arriving to points equidistant from the 
point $\mathbf{r}$ where the Wigner function is evaluated. The action 
differences $\Delta S$ associated with both trajectories are uncorrelated. }%
\label{NDiagTraj}%
\end{center}
\end{figure}

The second possibility we have to consider is the case where there is not 
any trajectory leaving from $\mathbf{r}_{0}$ with momentum close to 
$\mathbf{p}_{0}$ that arrives to the neighborhood of $\mathbf{r}$ after a time 
$t$. It is a property of the Wigner function that in the region of phase space 
classically inaccessible by $X_{t}$ the points $\mathbf{r}$ half-way between 
branches of the classically evolved distribution will yield the largest values 
of $f_{\Sigma}$ (see Fig.~\ref{NDiagTraj}) and the discussion on Sec.~\ref{sec:Decoherence}). The trajectories $s$ and ${\tilde{s}}$ 
now visit different regions of the configuration space, and the impurity 
average can therefore be calculated independently for each of them. As in 
Eq.~(\ref{SpatialDeltaS}), we have 
 
\be 
\left\langle \exp\left[  \frac{\mathrm{i}}{\hbar}\Delta S_{s}\right] 
\right\rangle =\exp\left[  -\frac{1}{2\hbar^{2}}\left\langle \Delta S_{s}%
^{2}\right\rangle \right]  =\exp\left[  -\frac{v_{0}t}{\widetilde{2\ell}%
}\right]  _{.} \label{DeltaSW}%
\ee 
  
\noindent Such an average only depends on the length $L=v_{0}t$ of the 
trajectories. Thus, after average the non-diagonal term writes%
 
\ba
f_{\Sigma}^{\mathrm{nd}}(\mathbf{r},t)&=&\left(  \frac{\sigma^{2}}{\pi\hbar^{2}%
}\right)  ^{d}\exp\left[  -\frac{v_{0}t}{\widetilde{\ell}}\right] 
\ \int\mathrm{d}\delta\mathbf{r}\ \sum_{s,{\tilde{s}}}\ C_{s}C_{{\tilde{s}}%
}
\nonumber \\ & &
\exp{\left[  -\frac{\sigma^{2}}{\hbar^{2}}\left(  \left(  {\overline 
{\mathbf{p}}}_{s}-\mathbf{p}_{0}\right)  ^{2}+\left(  {\overline{\mathbf{p}}%
}_{{\tilde{s}}}-\mathbf{p}_{0}\right)  ^{2}\right)  \right]  }\ . 
\label{FSCND}%
\ea

\noindent The trajectory $s$ (${\tilde s}$) goes between the points 
$\mathbf{r}_{0}$ and $\mathbf{r} \mp\delta\mathbf{r}/2$. That is why the 
largest values of $f_{\Sigma}^{\mathrm{nd}}(\mathbf{r},t)$ are attained when 
$\mathbf{r}$ is in the middle of two branches of the classically evolved 
distribution. Other points $\mathbf{r}$ result in much smaller values of 
$f_{\Sigma}^{\mathrm{nd}}(\mathbf{r},t)$, since the classical trajectories 
that go between $\mathbf{r}_{0}$ and $\mathbf{r} \mp\delta\mathbf{r}/2$ 
require initial momenta ${\overline{\mathbf{p}}}_{s}$ (${\overline{\mathbf{p}
}}_{{\tilde s}}$) very different from $\mathbf{p}_{0}$. Thus, exponentially 
suppressed contributions result. 
 
The non-diagonal contribution to the Loschmidt echo can now be written as 
 
\ba 
M^{\mathrm{nd}}(t)&=&\int\mathrm{d}\mathbf{r}\ f_{\Sigma}^{\mathrm{nd} 
}(\mathbf{r},t)=\left(  \frac{\sigma^{2}}{\pi\hbar^{2}}\right)  ^{d}
\exp\left[  -\frac{v_{0}t}{\widetilde{\ell}}\right]  
\nonumber \\ 
& & \left\vert \int\mathrm{d}\mathbf{r}\ \sum_{s}\ C_{s}\ \exp{\left[  -\frac{\sigma^{2}
}{\hbar^{2}}\left(  \left(  {\overline{\mathbf{p}}}_{s}-\mathbf{p}_{0}\right) 
^{2}\right)  \right]  }\right\vert ^{2}=\exp\left[  -\frac{v_{0}t}
{\widetilde{\ell}}\right] .  \label{MNDW}
\ea 
 
\noindent Where again we 
have made the change of variables from $\mathbf{r}$ to ${\overline{\mathbf{p}}}$, 
and accordingly, we have obtained the non-diagonal contribution to the LE [Eq.~(\ref{MNonDiagonal})]. As discussed in Sec.~\ref{sec:DvsND}, such a contribution is a Fermi Golden Rule like \cite{Jacquod01}. In the limit of $\hbar\rightarrow0$ the 
diagonal term, Eq.~(\ref{MDiagonalW}), obtained from the final points who 
follow the classical flow, dominates the LE, consistently with the findings of 
Sec.~\ref{sec:Universality}. 
 
\subsection{Emergence of classicality in the Loschmidt echo} 
\label{sec:ClassicalityLE}

The cumbersome equations of the previous section might hinder the conclusions that can be extracted from the results, especially those regarding the connection between the LE and decoherence, and the phase space interpretation of the Lyapunov and FGR regimes. Therefore it is important to devote this special section to develop these conclusions, which will be confirmed later using a more illustrative approach.

However, before undertaking this analysis, we need to remind the historical purpose of the introduction of a unitary perturbation and the origin of the LE.
As discussed in Sec.~\ref{sec:Decoherence}, the traditional approach to decoherence is to introduce an environment ${\cal E}$ coupled to a system ${\cal S}$, perform the quantum evolution of the composed system ${\cal SE}$, and at the end obtain the reduced density matrix of ${\cal S}$ tracing out all degrees of freedom of ${\cal E}$. Alternatively, the Loschmidt echo approach is to consider the environment as all degrees of freedom over which we have little power to perform the time reversal operation. As an analytically treatable approximation, it is assumed that the effect of this uncontrolled degrees of freedom can be represented by a unitary perturbation to the
original Hamiltonian of the system. To connect the two approaches, one needs a correspondence between environments and Hamiltonian perturbations.
Nevertheless, this could be extremely difficult or even impossible to prove in a general case. 

Despite the lack of such a connection, research on the LE went on by interest on the object itself. However, 
varied results strongly hinted at the relationship between the LE and decoherence. For instance, the role of  $-\log M(t)$ as a measure of entropy was shown using geometric arguments by Usaj \cite{UsajThesis}. Furthermore, Saraceno and coworkers \cite{Saraceno03} recently considered two open systems whose self Hamiltonians are slightly different, and they were able to show that the rate of decay of the LE is the average of the decay rates of the purities of both systems. On a different approach, Zurek et. al. proposed \cite{Zurek02} to couple a simple spin $1/2$ with a chaotic environment, where the coupling depends on the two levels of the system. For the environment, therefore, there are two very similar evolutions, and after some simple algebra one can show that the purity of the reduced density matrix of the spin decays as the Loschmidt echo of the environment.
It is with the results showed in Sec. \ref{sec:SCWignerLE} and the ones of Sec.~\ref{sec:DecoAndLE} that a formal link between decoherence and the LE consolidated. As we will see in the sequel, this was done by showing a particular case of the aforementioned correspondence between environment and perturbation. 

Let us discuss the results we have obtained so far.
From the semiclassical evolution of the Wigner 
function we were able to identify the non-diagonal component $M^{\mathrm{nd}}$ 
as the contribution to the LE given by the values of the Wigner function 
between the branches of the classically evolved initial distribution 
(Fig.~\ref{NDiagTraj}). In this region both of the Wigner 
functions contributing to Eq. (\ref{LEWigner}) are highly oscillating. As their structures are very small, with high probability we can consider them to be quite different from each other. The overlap of this region, which is perfect for zero coupling (ensuring the unitarity requirement) is rapidly suppressed with increasing perturbation strength. Therefore, the non-diagonal part of the LE, associated with the Fermi Golden Rule regime, arises from the region of phase space where contributions to the the Wigner function have a quantum origin (interference patterns). In particular we have seen that the regime where these quantum effects dominate (FGR) collapses as $\hbar\rightarrow0$.
 
Beyond a critical perturbation, such that the quantum contribution to $M$ is suppressed, the diagonal component $M^d$ takes over as the 
dominant contribution to the LE, and is given by the values of the Wigner 
function on the regions of phase space that result from the classical 
evolution of the initial distribution. This is the Lyapunov regime, where the 
decay rate of $M(t)$ is given by $\lambda$. In particular, it is possible to observe this ``classical'' behavior only when the coupling to the environment (perturbation) is strong enough to suppress the quantum contribution. 

The previous calculations have allowed us to identify the two salient regimes of the LE with two different contributions to Wigner functions. The Lyapunov regime is given by the classical region of the Wigner function, while the FGR obtains from the oscillating interference patterns in between the classical regions. 

Notice that, despite its classical association, the diagonal terms of the LE are still 
of quantum origin, as we are comparing the increase of the actions of nearby 
trajectories by the effect of a small perturbation, assuming that the 
classical dynamics is unchanged. The behavior in the Lyapunov regime does not 
simply follow from the classical fidelity, where the change in the classical 
trajectories is taken into account, and the finite resolution with which we 
follow them plays a major role. The upper value of the perturbation strength 
for observing the Lyapunov regime is a classical one, i.e. $\hbar$ independent 
[$\ell_{\mathrm{tr}}\simeq L$ in Sec.~\ref{sec:SCLE} and 
Eq.~(\ref{AlphaMax})]. 
 For stronger perturbations (see discussions in Sects.~\ref{sec:SCLE} and 
\ref{sec:Universality}) the classical trajectories are affected and the decay rate of the 
LE is again perturbation dependent. 

\section{Decoherence and the Loschmidt echo}
\label{sec:DecoAndLE}

In this section we will abandon the semiclassical approximation used many times previously to treat the LE. Instead, by finding a master equation for $M(t)$ like the one of Sec. \ref{sec:Decoherence}, the whole toolbox of decoherence theory will become available to study the problem. In the process the relationship between decoherence and LE will be formally demonstrated. Furthermore, the new perspective will allow simpler interpretations of previous sections' results.

The key result of this section is that for a classically chaotic system, the rate of decoherence is equal to the rate of decay of the average LE. In particular, that above a critical perturbation both quantities decay with a rate given by the Lyapunov exponent. Finally, the results of the previous section regarding the origin in phase space of the different regimes of the LE will be further analyzed and interpreted using know results from decoherence theory.

The results to be developed are restricted to the average over perturbations of $M(t)$. This, as we saw in Sec.~\ref{sec:Averages}, could be taken as a limitation give by the need to make analytical progress, with typical cases lying close to the average. However, to avoid confusions, we will distinguish the average echo for an ensemble of perturbations $\Sigma(x,t)$ with probability density $P(\Sigma)$ with the symbol
\begin{equation}
\bar M(t)=\int \Dc\Sigma\ P(\Sigma)\ |\langle\Psi_0|U_\Sigma^\dagger (t) U_0(t)|\Psi_0\rangle|^2,
\label{Mbar1}
\end{equation}
with $U_{\Sigma}(t)$ and $U_{0}(t)$ the evolution operators for a time $t$ of the perturbed and unperturbed Hamiltonians respectively.
Notice that this can be rewritten by moving inside the integral only the quantities that depend on $\Sigma$,
\ba
\bar M(t)&=&\int \Dc\Sigma\ P(\Sigma)\ \langle\Psi_0|U_0^\dagger (t) U_\Sigma(t)|\Psi_0\rangle\langle\Psi_0|U_\Sigma^\dagger (t) U_0(t)|\Psi_0\rangle
\nonumber \\
&=&\langle\Psi_0|U_0^\dagger (t) \left[ \int \Dc\Sigma\ P(\Sigma)\  U_\Sigma(t)|\Psi_0\rangle\langle\Psi_0|U_\Sigma^\dagger (t) \right] U_0(t)|\Psi_0\rangle.
\label{AverageInside}
\ea
Thus, $\bar M(t)$ is simply the overlap between the average state $\rho(t)$ and the unperturbed density matrix $\rho_0(t)$ evolved from the initial state with $U_0$:
\be
\bar M(t)={\rm Tr} \left( \rho(t) \rho_{0}(t) \right),
\label{MRhos}
\ee
with
\ba
\rho_0(t)=U_0(t)|\Psi_0\rangle\langle\Psi_0|U_0^\dagger (t), \nonumber \\
\rho(t)= \int \Dc\Sigma\ P(\Sigma)\  U_\Sigma(t)|\Psi_0\rangle\langle\Psi_0|U_\Sigma^\dagger (t).
\label{Rhos}
\ea

Equation (\ref{MRhos}) already can be used to establish an inequality between $\bar M(t)$ and the purity $\Pc(t)$ [Eq.~(\ref{Purity})], typically used to characterize decoherence. Using Schwartz inequality, $(\Tr AB)^2 \le \Tr A^2 \Tr B^2$, and assuming that the initial state is pure ($\Tr \rho_0^2=1$), we see that 
\be
\bar M^2(t)\le \Pc(t).
\label{Inequality}
\ee
Another argument to obtain similar equations is given in Refs. \cite{UsajThesis,PastawskiBook}. Inequality (\ref{Inequality}) was also noticed and used in \cite{Prosen03b} when studying the LE and the purity in composite systems, and in \cite{Saraceno03} for the problem of two similar open systems. Equation (\ref{Inequality}) implies that when the purity $\Pc (t)$ decays exponentially with a rate $\gamma_D$, then $\bar M(t)$ should also decay exponentially (or faster) with a rate at least $\gamma_D/2$. However, as we will see later, we can go much further in the relationship between both decay rates.

The key point to study now is the behavior of the average state $\rho (t)$. In particular, in order to place the evolution of $\bar M(t)$ in the context of decoherence, we need to find a master equation with non--unitary terms for $\rho (t)$ similar to Eq.~(\ref{MasterEquationDissipative}). For this, we will appeal to the technique of the influence functional developed by Feynman and Vernon \cite{Feynman}, which has been successfully used to find master equations of open systems \cite{PazZurekLesHouches}.
Let us expand the expression for $\rho$ using the full quantum propagators [Eq.~(\ref{QuantumPropagator})]
\ba
\rho(x,x^\prime,t)&=&\int \Dc\Sigma P(\Sigma) \rho_\Sigma(x,x^\prime,t) \nonumber \\
&=& \int \Dc\Sigma P(\Sigma) \int \dd x_0 K_\Sigma(x,x_0,t) \int \dd x^\prime_0 K^{*}_\Sigma(x^\prime,x_0^\prime,t) \rho(x_0,x_0^\prime,0).
\label{RhoExpansion}
\ea
Now, instead of performing a semiclassical approximation [Eq.~(\ref{SemiclassicalPropagator})], we use the path integral representation of the evolution operator \cite{Feynman48},
\be
K(x,x_0,t)=\int \Dc q e^{\ii S[q]/\hbar},
\label{PathIntegral}
\ee
where the integral runs over {\it all possible} paths (not just the classical) that satisfy the  boundary conditions $q(0)=x_0$ and $q(t)=x$, and $S[q]$ is the action along the path. The average density matrix can then be written as
\ba
\rho(x,x^\prime,t)&=&\int \dd x_0 \int \dd x_0^\prime \ {\bar K}(x_0,x_0^\prime,x,x^\prime,t) \rho(x_0,x_0^\prime,0),
\label{RhoExpansion2}
\ea
where the average propagator is given by
\ba
{\bar K}(x_0,x_0^\prime,x,x^\prime,t)&=&
\int \Dc\Sigma \ P(\Sigma) \int \Dc q \int \Dc q^\prime e^{\ii (S_\Sigma[q]-S_\Sigma[q^\prime])/\hbar} \nonumber \\
&=& \int \Dc q \int \Dc q^\prime e^{\ii (S_0[q]-S_0[q^\prime])/\hbar} F[q,q^\prime].
\label{AveragePropagator}
\ea
In the last equation we have used that the action $S_\Sigma[q]=S_0[q]+\int \Sigma[q(t^\prime),t^\prime]\dd t^\prime$, and $F[q,q^\prime]$ defined as the Feynman-Vernon influence functional \cite{Feynman},
\be
F[q,q^\prime]=\int \Dc\Sigma \ P(\Sigma) \exp\left[\ii \int_0^t (\Sigma[q(t^\prime),t^\prime]-\Sigma[q^\prime(t^\prime),t^\prime]) \dd t^\prime \right].
\label{InfluenceFunctional}
\ee
To make analytical progress, it is convenient to consider a simple form of the perturbation. Results do not depend strongly on it, provided we exclude situations where the perturbation changes substantially the nature of the Hamiltonian. Let us assume that the time and spatial dependence of the perturbation are uncorrelated, $\Sigma(x,t)=V(x)J(t)$, where $V(x)$ is a function of the coordinates of our system
and $J(t)$ is an external source. For this case, averaging over  $\Sigma$ consists of averaging over functions $J(t)$. We will further assume that the probability density $P(J)$ is a Gaussian whose width defines the temporal correlation function for the sources:
\begin{equation}
P(J)=N\exp\left[-\frac{1}{2}\int\int \dd t \dd t' \ J(t)\nu^{-1}(t,t')\ J(t')\right], 
\label{PJ}
\end{equation}
with $\nu(t,t')=\int DJ\ P(J) J(t)J(t')$ the noise correlation function and $N$ a normalization factor. 
Replacing in Eq.~(\ref{InfluenceFunctional}) gives a straightforward Gaussian integration, and one obtains
\begin{equation}
F[q,q^\prime]=\exp\left[-\frac{1}{2}\int\int \dd t \dd t' \ V_{-}(t) \nu(t,t')\ V_{-}(t')\right],
\label{InfluenceFunctional2}
\end{equation}
where $V_{-}(t)=V(q(t))-V(q^\prime(t))$. 

A simple but physically relevant case is when the noise is white, i.e. $\nu(t,t')=2D\delta(t-t')$. In this case, we can compute the time derivative of the averaged density  matrix (\ref{RhoExpansion2}),
\ba
\frac{\partial \rho}{\partial t}&=&\int \dd x_0 \int \dd x_0^\prime \frac{\partial {\bar K}}{\partial t}(x_0,x_0^\prime,x,x^\prime,t) \rho(x_0,x_0^\prime,0) \nonumber \\
&=& \int \dd x_0 \int \dd x_0^\prime \rho(x_0,x_0^\prime,0) \int \Dc q \int \Dc q^\prime 
\left[ \frac{i}{\hbar} ({\cal L}_0 (x,t)-{\cal L}_0 (x^\prime,t)) - D V_{-}^2 (t) \right] \nonumber \\ & & \times \ 
e^{\ii (S_0[q]-S_0[q^\prime])/\hbar} F[q,q^\prime], 
\label{Derivative1}
\ea
where ${\cal L}_0 (x,t)$ is the Lagrangian of the unperturbed system which gives the unitary evolution [like the first term in the rhs of Eq.~(\ref{MasterEquationDissipative})]. 
Using that 
\be
\left( V(x)-V(x^\prime)\right)^2 \left< x \right| \rho \left| x^\prime \right>=
\left< x \right| \left[V(x),\left[V(x),\rho\right] \right] \rho \left| x^\prime \right>,
\ee
it is trivial to obtain from (\ref{Derivative1}) the master equation for $\rho$,
\begin{equation} 
\dot{\rho}=\frac{1}{i\hbar}\left[\Hc_0,\rho \right] 
 -\ D\left[V(x),\left[V(x), \rho\right]\right].
\label{MasterEquationLE} 
\end{equation} 

We see that Eq.~(\ref{MasterEquationLE}) is just like the master equation that arose from considering a quantum system interacting with a quantum environment formed by a 
set of harmonic oscillators (Eq.~\ref{MasterEquationDissipative}) \cite{CaldeiraLegget,HuPazZhang,PazZurekLesHouches}. 
In such a case the modulus of the influence functional generated
by the environment is identical to (\ref{InfluenceFunctional2}) provided one chooses 
the spectral density and the initial state of the environment in such a 
way that its noise--kernel is equal to the kernel $\nu(t,t')$ in (\ref{InfluenceFunctional2}). 
However, as can be seen from the derivation of Eq.~(\ref{MasterEquationGeneral}) \cite{PazZurekLesHouches}, in general the influence functional is a complex number whose phase is responsible for dissipation. In the physically relevant limit (usually associated with high temperatures)  for decoherence studies aimed at understanding the quantum--classical correspondence, relaxation effects can be ignored \cite{Zurek03}. 

We have therefore demonstrated that averaging of the evolution over an ensemble of perturbations yields an effect analogous to the tracing out of the unobserved degrees of 
freedom of an environment.  
They are not completely equivalent, though. While the equivalence can be established 
for the average over an ensemble of noise realizations, it does not 
exist for individual members of the ensemble, which
follow unitary evolution with a given noise.
By contrast, a decohering system will lose purity after becoming 
entangled with the environment, even when the state of the environment
is known beforehand (see Ref.~\cite{Zurek03} for a detailed discussion).

We have seen that, in this limit, the evolution of the average state $\rho$ is identical to that of a quantum system interacting with an environment. Thus, the effect of a particular environment on its evolution can be represented by a perturbation in the Loschmidt echo picture. Hence, the evolution of the echo $\bar M(t)$ is directly placed in the context of open quantum systems and decoherence. 

The master equation (\ref{MasterEquationLE}) for the average state can be used to obtain the time derivative of $\bar M(t)$, along the same lines followed for $\Pc$ [Eqs.~(\ref{PurityDotDerivation}) through (\ref{SigmaPurity})],
\begin{eqnarray} 
\dot{\bar M}&=&\partial_t{\rm Tr}(\rho \ \rho_0)=
D\int dxdp\ W_0(x,p)\partial^2_{pp} W(x,p).
\label{Mdot}
\end{eqnarray} 
The same argument as before can also be used to analyze the decay of the Loschmidt echo. In fact, equations (\ref{PurityDotDerivation}) and (\ref{Mdot}) just differ by a factor of $2$ and by the presence of $W_0$ instead of $W$ inside the integral. As above, we can transform the evolution equation of the echo into 
\be
\frac{\dot{\bar M}}{\bar M}=-\frac{D}{\sigma_{M}^{2}},
\label{Mdot2}
\ee
with the typical width $\sigma_{M}$ defined similarly as the width $\sigma$ for the purity [Eq.~(\ref{SigmaPurity})],
\be
\sigma_{M}^{-2}=\frac{\int W_0\partial^2_{pp}  W}{\int W_0  W}. 
\label{SigmaM}
\ee
Notice that when decoherence is effective and the dominant structure in $W$ 
approaches the critical value, the smallest scales of the pure Wigner 
function $W_0$ continue contracting and developing smaller and 
smaller scales  (sub--Planck scales are reached quickly in chaotic
quantum systems \cite{ZurekNature}). 
To estimate the behavior of $\sigma_{M}$ in this situation let us approximate locally $W \sim \exp(-p^2/2\sigma^2)$  and $W_0 \sim \exp(-p^2/2 \sigma_0^2(t))$. In the large time limit $t \gg 1/\lambda$, the chaotic nature of the system contract the pure Wigner function and $\sigma_0 \sim \exp(-\lambda t)$. Under these assumptions, Eq.~(\ref{SigmaM}) gives $\sigma_{M}^{2}=2\sigma^{2}$. This factor $2$ compensates the one missing in Eq.~(\ref{Mdot2}) when compared to Eq.~(\ref{PurityDot}), and therefore we see that the decay rates of the LE and the purity are the same. 

This equivalence in the the decay rates of both quantities is the main result obtained in this section, given by the derivation of the master equation for the average state $\rho$. It is also the first formal proof of such a relationship between LE and decoherence. Apart from demonstrating this connection, we can now use the findings of this section to present a more illustrative picture of the origin in phase space of the different regimes of the LE discussed in Sec.~\ref{sec:ClassicalityLE}. 

\begin{figure}[htbp]
\begin{center}
\leavevmode
\epsfxsize 3in
\epsfbox{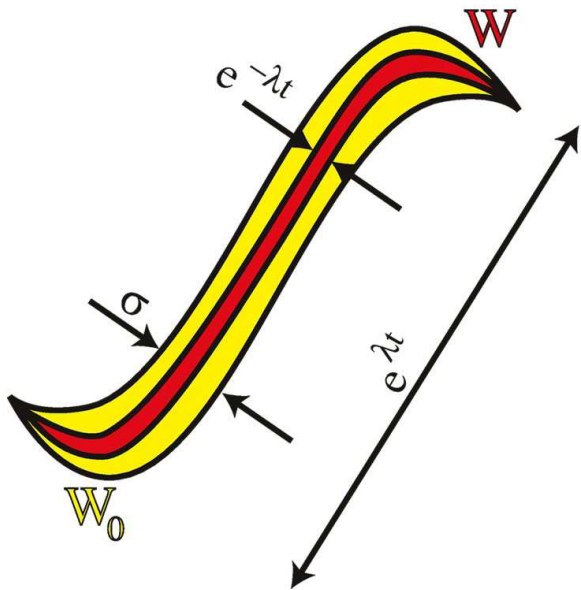} 
\caption{Schematics for the estimation of the integral over $A_C$ in Eq. (\ref{IntAC}), where the whole region in phase space is unfolded. The classical part of the Wigner function stretches along the stable direction with a total length proportional to $e^{\lambda t}$. The unperturbed Wigner function $W_0$ contracts in the stable direction so as to keep the area constant. The average (or decohered) Wigner function $W$, however, achieves a final width $\sigma_M$ given by the competition between the decoherence process and the chaotic compression.}
\label{fig:SchematicsIntegral}
\end{center}
\end{figure}

To compute the overlap $\bar M=\int \dd x \dd p \ W_0 W$ we can split the phase space integral into two regions: the region $A_C$ close to the classical unstable manifold of the initial state, where $W_0$ is positive, and the region $A_O$ over which $W_0$ oscillates:
\begin{equation} 
\bar{M}(t)=\int_{A_{O}} \dd x \dd p \ W_0 W 
+ \int_{A_{C}} \dd x \dd p \ W_0  W. 
\label{Integral}
\end{equation} 
In the oscillatory region we can estimate the value of the integral 
assuming that there is a dominant wave vector $k_p$. In such a case, 
from Eq. (\ref{MasterEquationLE}) [or its Weyl representation Eq. (\ref{WignerMasterEquation})] we assume 
$\bar W \simeq W_0 e^{-Dk_p^2 t}$. Using this,
\be
\int_{A_{O}} \dd x \dd p W_0 W \simeq e^{-Dk_p^2 t} \int_{A_{O}} \dd x \dd p W_0^2.
\ee
If more than one scale is present 
the result would be a sum of terms like this one. For the 
second integral, we can also use a 
crude estimate supposing that $W_0$ and 
$\bar W$ are constant over their respective effective support. In particular,
$W_0 \sim 1/A_C$ since its integral over $A_O$ cancels out. As $\bar W$ 
approaches the critical width $\bar\sigma$ along the stable manifold, 
the area of its effective support grows exponentially. Therefore, one 
gets that the second integral is (see Fig.~\ref{fig:SchematicsIntegral})
\be
\int_{A_C}W_0\bar W\sim A_{C}  W_0 \bar W \sim \bar W 
\sim \frac{e^{-\lambda t}}{\sigma_M}.
\label{IntAC}
\ee 
Thus, combining the two results we find that the 
expected behavior of the Loschmidt echo is 
\begin{equation} 
\bar{M}(t) = a\ \exp(-\lambda t)+b\ \exp(-Dk_p^2 t)
\label{Msemiclassic} 
\end{equation} 
for appropriate prefactors $a$ and $b$. We have thus re--obtained the result from Sec.~\ref{sec:SCLE} \cite{Jalabert01,Cucchietti03B}, included the perturbation dependence of factor $a$. 

The virtue of this analysis, entirely based on properties of the evolution of $\bar W$ derived in the context of decoherence studies, is the simplicity with which it enables us to demonstrate the behavior implied by the calculations of Sec. \ref{sec:SCWignerLE}: the FGR contribution arises from the decay of the interference fringes while the Lyapunov contribution is associated with the behavior of $\bar W$ near the classical unstable
manifold. 

Notice that the same arguments could have been applied to the purity, therefore its behavior in phase space should be the same as for the LE.
Also, it is important to remark that the treatment shown here is valid in a semiclassical regime where the evolution of the Wigner function is dominated by the classical Hamiltonian flow and the corresponding interference fringes generated when its phase space support folds. 

\section{Summary}

In this chapter we briefly introduced the theory of decoherence in open systems, often used to explain the transition from quantum to classical behavior. The treatment of the Loschmidt echo using Wigner functions enabled us to obtain a new interpretation of previously known results. In particular, the suppression of the FGR term can now be stated as the cancellation of the quantum contributions to the LE, and the emergence of the classical behavior in the shape of the Lyapunov regime. 

Afterwards we developed a master equation for the density matrix averaged over realizations of the perturbation. We saw that this master equation corresponds term by term to that obtained by coupling the system to an environment composed of quantum harmonic oscillators. Therefore, we demonstrated that the effect of an environment in the dynamics of the reduced density matrix can be assimilated by a perturbation in the self--Hamiltonian of the system.

Using this master equation we showed that the Loschmidt echo decays exponentially with the same rate as the purity, a quantity typically used to measure decoherence. 

On one hand, this connection brings the interesting possibility of measuring directly the rate of decoherence in a system by measuring the Loschmidt echo. This possibility is of special importance to the field of Quantum Information \cite{NielsenBook}, since it could provide a general scheme to obtain the decoherence rate in any of the possible implementations of a quantum computer.
Furthermore, we could implement an experimental determination of pointer states by measuring the rate of generated entropy and using the predictability sieve. 
 
On the other hand, the field of the LE can benefit greatly from the vast analytical tools existent in the more developed field of decoherence. We saw an example of this usefulness with a simpler demonstration of the results mentioned above on the origin in phase space of the two contributions to the LE. In particular this derivation of the result does not resort to classical trajectories, which makes the assignment of the regions in phase space much more direct and clear.

\section*{Original results}

\begin{itemize}
\item Semiclassical treatment of the Wigner function representation of the LE. With this derivation we were able to show an interpretation in phase space of the different types of pairs of trajectories contributing to the LE, which in turn assigns to the FGR and Lyapunov regimes to the quantum and classical contributions to the Wigner function, respectively. These results were published in \cite{Cucchietti03B}.
\item Derivation of a master equation for the average LE. Using this master equation, it was shown that the LE and the purity decay exponentially with the same decay rate. This was published in \cite{Cucchietti03}.
\item Using the master equation for the LE, a much simpler demonstration of the origin in phase space of the FGR and Lyapunov regimes was obtained \cite{Cucchietti03}. This derivation in particular can be applied directly to the purity with the same results.
\end{itemize}

\chapter{Conclusions}
\label{chap:Conclusions}

\begin{quote}
{\em Very interesting theory -- It makes no sense at all.}

Groucho Marx
\end{quote}

I started working on the Loschmidt echo just after Jalabert and Pastawski demonstrated the existence of the Lyapunov regime (later published in \cite{Jalabert01}), which might be considered the single biggest breakthrough in the subject. This by no means implies that later technical and conceptual achievements lack relevance, since they added substantially to the the deep understanding of the problem available today. I would like to stress in particular the role of the contributions presented in this thesis.

The first important step was the numerical verification of the predictions of \cite{Jalabert01}, shown in the first part of Chap. \ref{chap:Universality}. At the moment the first semiclassical calculation was done, serious doubts existed on how robust would the Lyapunov regime be for an actual model Hamiltonian. When the numerical evidence finally appeared, it provided not only support for the theory but also great insight on its range of validity.  In particular, the simulations shed light on the approximations regarding the semiclassical regime and the persistence after the Ehrenfest time.

These issues motivated further investigation, which resulted in the work shown in the second half of Chap. \ref{chap:Universality} regarding the universality of the Lyapunov regime. Clearly these are strong new results which demonstrated the validity of the theory for situations not available theoretically, and clarified the recovery of classical chaos in the limit of high energies. 

At the same time, I also developed the semiclassical theory for the Lorentz gas (Sec.~\ref{sec:SCLG}), showing analytically that the results were robust to non disordered perturbations. Even more, the derivation of the general case of Sec. \ref{sec:SCLE} is a useful generalization of \cite{Jalabert01} to any perturbation with noise in space or in time, setting off from the particular case shown in Sec.~\ref{sec:QuenchedDisorder}. 

From the results of Sec.~\ref{sec:Universality} it became clearer that the persistence of the Lyapunov regime after the Ehrenfest time needed thorough explanation, probably linked to the emergence of classicality in the LE. The semiclassical analysis of the Wigner function of Sec.~\ref{sec:wigner} and later the finding of a master equation for $M(t)$ (Sec.~\ref{sec:DecoAndLE}) are, to my opinion, the biggest conceptual leap in the subject after \cite{Jalabert01}. They provide not only  explanation and closure of various phenomena, but also a unification of fields. This is clearly to the benefit of the LE which benefits from the analytical resources developed in the more mature field of decoherence. Reciprocally, the experimental feasibility of the LE is a great attraction to other fields such as Quantum Information \cite{NielsenBook}, since it would allow to measure directly the rate of decoherence in possible implementations of a quantum computer.

The problem of the LE gained considerable attention right after the publication of \cite{Jalabert01}, and many aspects not considered here were the focus of many works in the literature. Some of these hitherto unmentioned papers are of great importance since they complement the results presented here in this thesis, therefore before discussing possible future directions of the investigation let me give a brief account.

Perhaps the most notorious absence in this work is the behavior of the LE in classically integrable systems. All results pertain chaotic systems where universal behavior is found, for example exponential divergence of trajectories or Lyapunov exponents. Integrable systems, on the other hand, have nothing that can be regarded as generalities. Actually, with some effort one could even design a system to have almost any behavior one might desire. Far less general is the effect perturbations have on integrable systems. Because of this, the investigation of the LE in integrable systems has encountered many controversies or even opposed results. For instance, power law decays \cite{JacquodIntegrable}, and faster than exponential decays \cite{Prosen03a,Prosen04} have been both predicted and observed. Another possibility is systems with a mixed phase space. Although the Smooth billiard could enter into this category, the initial states were chosen in the chaotic region. An stretch exponential decay has been observed \cite{Emerson02} for initial states in the border between the stable and unstable regions of phase space. Further research in these areas is highly desirable, mainly if one wishes to use the LE as a fair signature of quantum chaos.

I also focused on the FGR and the Lyapunov regimes, and little was said about the short time or weak perturbation regime. This is by all means an arbitrary decision, since for instance for quantum computing applications the most important range of decay is the first one or two percent. Wisniacki has shown \cite{Wisniacki03} that also in this regime a connection to the LDOS can be demonstrated. On another line, Cerruti and Tomsovic obtained \cite{Cerruti03} a uniform semiclassical approach that not only treats the weak perturbative regime, but also successfully describes the transition to the FGR regime. The perturbative regime is also of great importance to quantum information. In particular, it should be noticed that in that field a quantity very similar to the LE exists, the so called fidelity \cite{NielsenBook}. It is also the overlap between two wave functions evolved with similar Hamiltonians, and it is used to calculate how good is a computation with the incorrect quantum algorithm. Despite this similarity, I distinguished fidelity from the LE because the latter implies a perturbed time reversal, while  the former can actually be the overlap between any two states.

Another interesting issue not explored in this thesis is the behavior of the classical equivalent of the LE. Benenti and Casati have found \cite{Benenti02,Benenti03} a definition that is a logical extension of the quantum version, namely they use as classical echo the overlap between two classical distributions. Interestingly, they were able to show that after the Lyapunov decay, the classical LE follows a decay given by the Ruelle resonances. Similar behavior was noticed for the quantum version in \cite{Saraceno03}.

What are the remaining problems in the LE that deserve further investigation? Certainly some have already been mentioned, like the full characterization of integrable systems, or the classical version of the LE. Other case of interest is that of {\em disordered} integrable systems, for instance a Lorentz gas where the disks are replaced by squares (a wind-tree model). This would be required to understand the role of disorder separately from the effects of chaotic motion.

Apart from these obvious extensions, important fundamental issues have not been clarified yet. For instance, it is clear that the single particle theory developed semiclassically is not enough to describe the experiments in NMR. In particular, the experiments show a {\em Gaussian} decay of $M(t)$, opposed to the exponential obtained in Sec.~\ref{sec:SCLE}. It is not yet clear yet whether it is a general many body effect or a particular behavior of the spin system. In any case, research on both possibilities is highly desirable. Actually, that a Gaussian decay for the purity and the LE is possible for a single spin coupled to a bath of non interacting spins was recently shown in \cite{Cucchietti04b}. However, the time scale of the Gaussian is given by the coupling to the environment, understandable since the model studied there is too simple to take into account the intricacies of the experimental situation.

A very interesting line being developed \cite{Cory} is the use of the LE as a characterization tool. Using the results shown in this work, or others particularly developed for specific systems, the LE scheme can be implemented as a subroutine of a quantum algorithm. This could provide information on the system, the environment or the coupling between them, depending on what one wants to find out.

\appendix

\setcounter{chapter}{0}
\roman{chapter}

\chapter{Quantum dynamics of discrete systems}
\label{appe:Suzuki}

This appendix details the approximations and techniques used in Chap. \ref{chap:Universality} to simulate the dynamics in the Lorentz gas and the Smooth stadium billiard. 
For simplicity, all the examples are given for a one dimensional system, the generalization to higher dimensions is direct.

Let us start by considering a discrete and finite system of size $L$, divided in $N$ pieces such that $a=L/N$ is small compared to all other lengths in the problem. The wave function will be considered to exist only at discrete positions in space, $x=na$ with $n$ an integer from $0$ to $N-1$. To focus in only one example, we will consider periodic boundary conditions such that $\psi(Na)=\psi(0)$, although open boundaries are also simple.

Schr\"{o}dinger's equation for the discrete wave function $\psi$ can be obtained by rewriting the kinetic energy term of the Hamiltonian $\Hc$ using finite differences,
\ba
\ii \hbar \frac{\partial \psi}{\partial t}  &=& \Hc \psi (x,t) \nonumber \\ 
&=& -\frac{\hbar^2}{2m}\frac{\partial^2 \psi}{\partial x^2} + {\cal V}(x) \psi(x,t) 
\nonumber \\
& = & -\frac{\hbar^2}{2m}\frac{\psi[(n+1)a]+\psi[(n-1)a]-2\psi(na)}{a^2} + {\cal V}(na) \psi(na,t) 
\nonumber \\
& = & -\frac{\hbar^2}{2 m a^2} \left\{ \psi[(n+1)a]+\psi[(n-1)a] \right\} +  \left( {\cal V}(na)-\frac{\hbar^2}{m a^2} \right)  \psi(na,t).
\label{DiscreteS}
\ea
This equation can be readily written in matrix form,
\be
\ii \hbar \frac{\partial \psi}{\partial t} = \mathbf{H} \psi,
\ee
where $\psi$ now represents a vector with components $\psi_n=\psi(na)$, and the Hamiltonian matrix is given by the terms of Eq. (\ref{DiscreteS}), forming a tri-diagonal matrix
\be
\mathbf{H} =
\left(
\begin{array}{ccccccc}
E_0 & -V  &  0  &  0  &  0  & \ldots & -V \\
 -V  & E_1& -V &  0  &  0  & \ldots & 0 \\
   0 & -V & E_2 & -V &  0  & \ldots & 0 \\
   0 &  0 & -V  & E_3 & -V & \ldots & 0 \\
\vdots & \vdots & \vdots & \vdots & \vdots & \ddots & \vdots \\
-V &  0  &  0  & 0  & 0 & \ldots & E_{N-1}
\end{array} \right),
\ee
where $E_n={\cal V}(na)$ is the potential profile as a function of the coordinate, called the {\em on-site energies}, and the kinetic term transforms into {\em hopping elements} $V=\hbar^2/2ma^2$ outside the diagonal. Note that the constant term $-\hbar^2/ma^2$ in the potential energy has been dropped because it is just a redefinition of the zero of energy. 

The discretization scale $a$ has to be larger than two scales: First, the smallest wavelength used in the problem, and second,  the smallest scale of the potential energy features. The first condition is related to the appearance of diffraction effects, and can be estimated by analyzing the free particle problem, that is when $E_n = 0 \ \forall n$. In this case the eigenenergies of the Hamiltonian can be analytically obtained, giving a dispersion relation
\be
E(k)=2V[1- \cos (ka)], \ \ k=\frac{2 \pi m}{N}, \ \ m=0...N-1,
\ee
where $k$ is the momentum of the eigenstate $\left|k\right>$. When the wavelength $\lambda=2\pi/k$ is much smaller than $a$, one recovers the dispersion relation of the free particle,
\be
E(k) \simeq V k^2 a^2 = \frac{\hbar^2 k^2}{2 m}.
\ee

Typically, what units are used depend on how the problem is posed. In particular for the Lorentz gas, the convention $a=1$, $V=1$ and $\hbar=1$ was used. This determines how all other quantities are measured, for instance, time is given in units of $\hbar/V$. For the Smooth stadium, on the other hand, the size of the system $R$ is set to unity, as well as the energy scale given by $U_0=1$ so that unit energy is obtained where the boundary of the Bunimovich stadium is located. Furthermore, unit energy was assigned to the initial kinetic energy of the particle. This choice has the effect of setting an  {\em unorthodox}  value  $\hbar=1/k$, with $ka=0.5$ and $a=R/N=0.0055$. In both cases, $m$ was set to $1/2$.

To compute the quantum dynamics of Eq. (\ref{DiscreteS}) one could in principle resort to a diagonalization of $\mathbf{H}$ (if it is time-independent). This path, however, is not practical since it becomes easily intractable even for moderately large computers, restricting the problem to small values of $N$. Other methods to solve differential equations like Runge-Kutta, Crank-Nicholson, etc., are usually not unitary and therefore they have to be complemented with periodic renormalizations of the wave function. In general, these methods are unstable and require very small time steps to produce consistent results.

The method used in this thesis is a higher order version of the Trotter product decomposition \cite{Trotter59} of the evolution operator. It has the important features of being unitary by construction, and of being more efficient than direct diagonalization in resources and time. In contrast, it does not provide any spectral information like energies or eigenstates.

Let us denote $\mathbf{U}(\tau)=\exp (-\ii \mathbf{H} \tau)$ the exact evolution operator of the Hamiltonian for a given time $\tau$. The main approximation comes from the observation that if $\mathbf{H}=\mathbf{H}_1+\mathbf{H}_2$, then
\be
\| \mathbf{U}(\tau) - \mathbf{U}_1(\tau) \| \leq \frac{\tau^2}{2} \| \left[ \mathbf{H}_1, \mathbf{H}_2 \right] \|,
\ee
with $\mathbf{U}_1(\tau) = \exp (-\ii \mathbf{H}_1 \tau)\exp (-\ii \mathbf{H}_2 \tau)$ is the Trotter operator \cite{Trotter59}.
For very small $\tau$ this approximation can be quite good, where of course the choice of the decomposition of $\mathbf{H}$ plays an important role. A particularly clever (for numerical purposes) option is one where the $\mathbf{H}_n$'s are analytically diagonalizable, as we will see in the sequel. 

\begin{figure}[htbp]
\begin{center}
\leavevmode
\epsfxsize 3.5in
\epsfbox{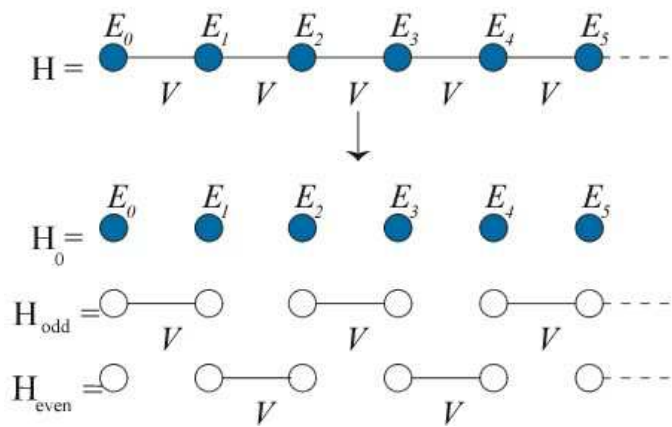}
\caption{Schematics of the decomposition of the Hamiltonian of a one dimensional system in three analytically solvable Hamiltonians. The colored (white) dots represent diagonal elements of the matrix different from (equal to) zero, while the links are the off-diagonal or {\em hopping} elements.}
\label{fig:Trotter}
\end{center}
\end{figure}

In the example of the one dimensional system above, a natural decomposition is 
\be
\mathbf{H}=\mathbf{H}_0+\mathbf{H}_{even}+\mathbf{H}_{odd},
\ee
with 
\be
\mathbf{H}_0 =
\left(
\begin{array}{ccccccc}
E_0 & 0  &  0  &  0  &  0  & \ldots & 0 \\
 0  & E_1& 0 &  0  &  0  & \ldots & 0 \\
   0 & 0 & E_2 & 0 &  0  & \ldots & 0 \\
   0 &  0 & 0  & E_3 & 0 & \ldots & 0 \\
\vdots & \vdots & \vdots & \vdots & \vdots & \ddots & \vdots \\
0 &  0  &  0  & 0  & 0 & \ldots & E_{N-1}
\end{array} \right),
\ee
the on-site energies Hamiltonian,
\be
\mathbf{H}_{odd} =
\left(
\begin{array}{cccccc}
  0 & -V  &  0  &  0  &  0  & \ldots  \\
 -V &  0  &  0  &  0  &  0  & \ldots  \\
   0 &  0  &  0  & -V &  0  & \ldots  \\
   0 &  0  & -V &  0  &  0  & \ldots  \\
\vdots & \vdots & \vdots & \vdots & \vdots & \ddots  \\
\end{array} \right),
\ee
the Hamiltonian with the hopping elements between odd-even sites and
\be
\mathbf{H}_{even} =
\left(
\begin{array}{cccccc}
0 & 0  &  0  &  0  &  0  & \ldots  \\
0 & 0  & -V &  0  &  0  & \ldots  \\
0 & -V &  0 &  0  &  0  & \ldots  \\
0 &  0 &  0  &  0  & -V & \ldots  \\
0 &  0 &  0  & -V &   0 & \ldots  \\
\vdots & \vdots & \vdots & \vdots & \vdots & \ddots  \\
\end{array} \right)
\ee
the remaining terms. The decomposition is schematized in Fig. (\ref{fig:Trotter}). 

The full evolution operator writes as $\mathbf{U}_1(\tau) = \exp (-\ii \mathbf{H}_0 \tau)\exp (-\ii \mathbf{H}_{even} \tau)\exp (-\ii \mathbf{H}_{odd} \tau)$. Hence, the evolution is performed by multiplying these matrices in order to the wave function. 
Since $\mathbf{H}_{odd}$ and $\exp (-\ii \mathbf{H}_{even}$ are made up by two by two blocks, the exponentiation is simple. For instance, 
\be
\exp (-\ii \mathbf{H}_{odd} \tau)=
\left(
\begin{array}{cccccc}
  \cos(V\tau) & \ii \sin(V\tau)  &  0  &  0  &  0  & \ldots  \\
 -\ii \sin(V\tau) &  \cos(V\tau)  &  0  &  0  &  0  & \ldots  \\
   0 &  0  &  \cos(V\tau)  &  \ii \sin(V\tau)  &  0  & \ldots  \\
   0 &  0  &  -\ii \sin(V\tau)  &  \cos(V\tau)  &  0  & \ldots  \\
\vdots & \vdots & \vdots & \vdots & \vdots & \ddots  \\
\end{array} \right).
\ee
The evolution of the wave function with one of these operators can be seen as a simple rotation between neighboring elements. The alternate application of the even and odd evolutions resembles a stroboscopic Hamiltonian, which for short $\tau$ quickly converges to an average Hamiltonian with both terms present. Finally, the 
evolution operator $\exp (-\ii \mathbf{H}_0 \tau)$ is diagonal and accordingly is just a phase for each component of the wave function. For higher dimensions, the extension is simply to consider decompositions of two Hamiltonians for each dimension, odds and evens hopping terms, plus the on-site energies Hamiltonian. Additionally, features like the hard walls of the disks in the Lorentz gas (infinite energy potential regions) are easily included in this scheme: a zero hopping term hinders the penetration of the wave function in these areas, and is numerically more stable than using large on--site energies.

Notice that there is no need to store any of these evolution operator matrices in memory when performing the numerical simulation, only the vector $\phi$ and a vector with the energies are needed. This is clearly an improvement in storage resources over direct diagonalization. In addition, the number of operations needed to apply the Trotter evolution operator is proportional to $N^2$, while diagonalization typically requires $~N^3$ steps to perform. Accordingly, for large $N$ one can expect a large improvement in simulation time. 

Regarding the numerical precision, more elaborate schemes exist that take the Trotter evolution operator to higher orders. These are due to Suzuki \cite{Suzuki90,Suzuki93} (see a didactical review in \cite{deRaedt96}), who showed how to construct these operators in such a way that they are always unitary and, at the same time, provided an algorithm to construct any approximation order from the previous one. In particular, a second order Trotter-Suzuki evolution operator is written as
\be
\mathbf{U}_{2}(\tau)=\mathbf{U}_{1}^{T}(\tau/2) \mathbf{U}_{1}(\tau/2),
\ee
where $T$ means transpose. $\mathbf{U}_{2}$ is bounded by $\tau^3$ errors, and therefore is a better approximation to the real evolution. In this thesis the fourth order approximation was used, given by
\be
\mathbf{U}_{4}(\tau)=\mathbf{U}_{2}(p\tau)\mathbf{U}_{2}(p\tau)\mathbf{U}_{2}((1-4p)\tau)\mathbf{U}_{2}(p\tau)\mathbf{U}_{2}(p\tau),
\ee
with $p=(4-4^{1/3})^{-1}$. 

Summarizing, the Trotter-Suzuki evolution operator is a very good approximation to the actual evolution. It is much more efficient than direct diagonalization, it is stable and, furthermore, generalizable to other kinds of Hamiltonians like spin systems \cite{deRaedtSpins}. A particular example of the strengths and advantages of this method over the traditional ones is given by my personal experience with the Lorentz gas: The largest system that one could diagonalize in a 1GB memory computer has $~4 \ \times \ 10^3$ states, and depending on processor power the process can take up to a day of time. In contrast, using the same memory and time with the Trotter-Suzuki algorithm I was able to treat systems with $10^6$ states.

\chapter{The Lorentz Gas: Classical and quantum dynamics}
\label{appe:Lorentz}

This appendix contains some details on the quantum and classical dynamics of the Lorentz gas, used in Chap. \ref{chap:Universality}.

\section{The system}

\begin{figure}[htb]
\begin{center}
\leavevmode
\epsfxsize 3.25in
\epsfbox{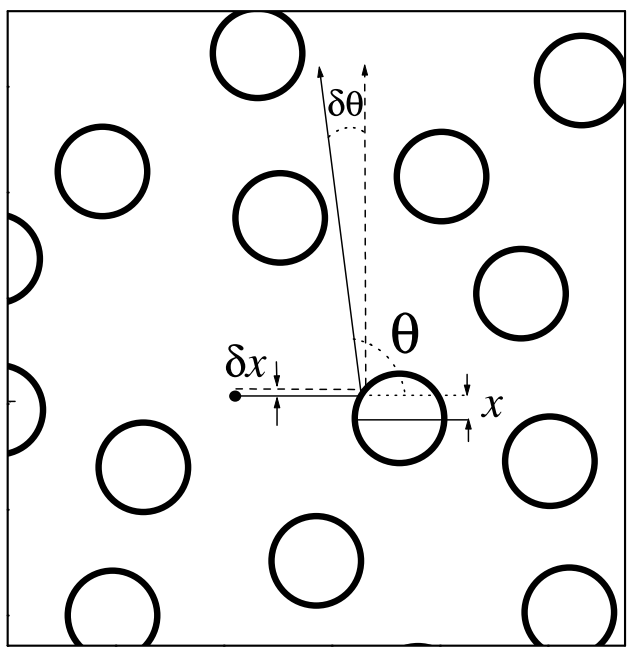}
\caption{Representation of a Lorentz gas. Notice how two initially close trajectories become separated after a collision with an impurity, reflecting the dispersive behavior of the classical dynamics.}
\label{fig:Gas}
\end{center}
\end{figure}

The Lorentz gas is a two dimensional box of sides $L$ where an irregular array of $n$ hard wall disks are fixed [see Fig. (\ref{fig:Gas})]. The classical dynamics of a particle in the system is given by specular reflections against the disks, and against the walls of the box hard wall boundary conditions (associated in quantum mechanics with the Dirichlet boundary conditions $\psi(\br_s)=0$ for points $\br$ on the surface.)

\begin{figure}[htb]
\begin{center}
\leavevmode
\epsfxsize 3.5in
\epsfbox{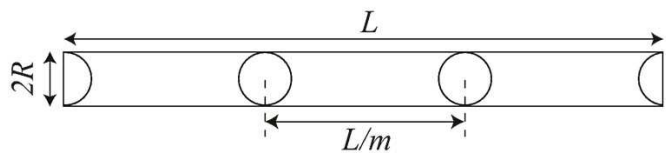}
\caption{Scheme to estimate the mean free path $\ell$ between collisions in the Lorentz gas. $\ell$ is approximately $L/m$ minus the average distance occupied by one disk.}
\label{fig:BoxL}
\end{center}
\end{figure}

\begin{figure}[htb]
\begin{center}
\leavevmode
\epsfxsize 3.25in
\epsfbox{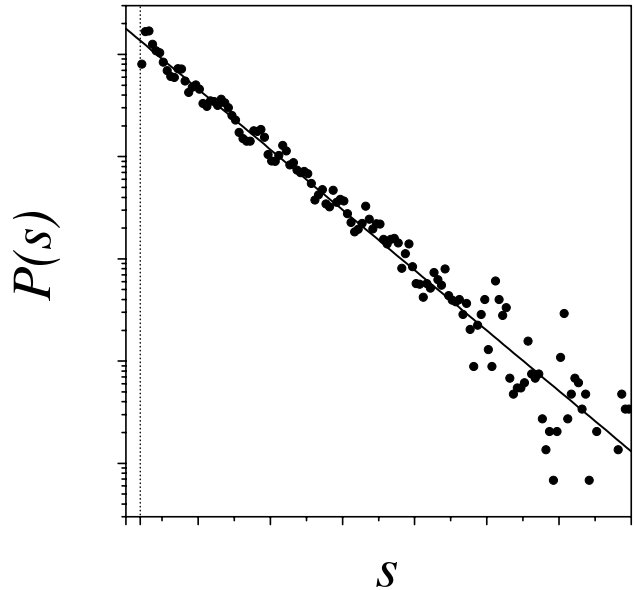}
\caption{Histogram of the distances between collisions with the disks, used in order to obtain numerically the mean free path $\ell$ for the Lorentz gas. The solid line represents Eq. (\ref{LDistribution}) and the dashed vertical line is the cut-off distance $2(R_e - R)$.}
\label{fig:HistogramS}
\end{center}
\end{figure}

If the radius of the disks is $R$, and the concentration (assumed to be uniform) given by the ratio between the area occupied by the disks to the total area of the box is
\be
c=\frac{n \pi R^2}{L^2},
\ee
then the entire system is characterized by the mean free path between collisions $\ell$. A simple argument to estimate this parameter is the following: 
Let us consider a rectangle of sides $L$ and $2R$ representing a typical cross section of the gas [see Fig. (\ref{fig:BoxL})]. In this rectangle we place $m=2cL/\pi R$ disks according to the concentration in the whole box. Locating them equidistant from each other, this leaves a free distance between the disks $L/m - \delta$,  where $\delta$ is the average space occupied by one disk along the short side of the box. In particular, $\delta=\pi R/2$. This free distance between the disks reasonable agrees with numerical computations of $\ell$ [see Fig. (\ref{fig:HistogramS})], and expressed in terms of the other parameters of the system writes
\be
\ell \simeq \frac{\pi R}{2 c}-\frac{\pi R}{2}
\label{MeanL}
\ee

The mean free path between collisions should not be confused with the {\em transport} mean free path which enters in the diffusion equation that describes the classical dynamics of the system. According to this law, the mean square distance traveled by a particle after a time $t$ is given by $<r^2(t)>=2dDt$, where $D$ is the diffusion coefficient and $d$ the dimension (in this case $d=2$). Inserting in the diffusion equation the transport mean free path $\ell_{tr}=v \tau_{tr}$, one obtains $D=v\ell_{tr}/2d$. 

The relationship between the mean free path and the transport mean free path is given by the amount of deflection of the trajectory in each collision. For weak scattering, $\ell_{tr}$  can be much larger than $\ell$. The way to compute their relationship is by weighting every collision with the exit angle $\theta$ after the scattering in the following way,
\be
\ell_{tr}=\int \dd \theta P(\theta) \ell [1-\cos(\theta)]
\ee
Changing variables from $\theta$ to the impact parameter $\rho=R \cos(\theta/2)$, and using $P(\rho)=1/2R$, one obtains that for the Lorentz gas $\ell_{tr}=4 \ell /3$. Numerical simulations of the classical dynamics support this result.

\begin{figure}[htb]
\begin{center}
\leavevmode
\epsfxsize 4in
\epsfbox{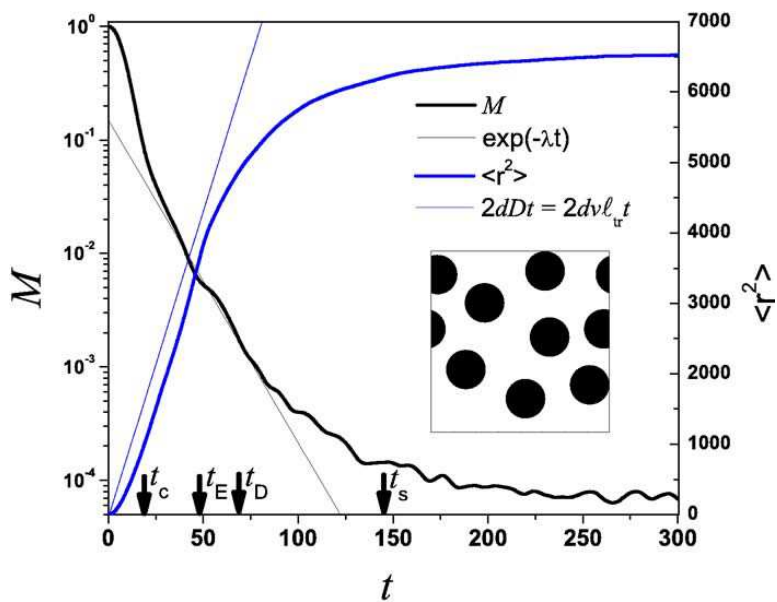}
\caption{In thick blue line, mean dispersion square of a wave packet in a Lorentz gas. For short times it presents ballistic behavior, while for long times it saturates because of the finite size of the sample. In between those regimes, a diffusive behavior is observed that corresponds with the classical one (thin blue line). For comparison, the LE is plotted in the same time scale (thick black line). The time scales marked in the plot are $t_c$ the collision time, $t_E$ the Ehrenfest time, $t_D$ the Thouless time at which diffusion saturates and $t_s$ the time of saturation of the LE.}
\label{fig:Diffusion}
\end{center}
\end{figure}

The quantum dynamics of a localized wave packet follows closely that of the classical distribution of particles (albeit interference effects). One expects for a certain regime to observe diffusive behavior in the propagation of the wave packet, corresponding to the diffusion observed in classical dynamics. Such effect can be observed in Fig. (\ref{fig:Diffusion}), where the average expectation value of $r^2=x^2+y^2$ is plotted as a function of time for a Lorentz gas with $L=200a$, $R=20a$ and $\ell \simeq 100a$.
After an initial ballistic motion, a diffusive behavior sets in that corresponds to the classical case (thin blue line). For long times the wave packet has spread over the whole box and diffusion stops. The finite size effects start to be appreciable at the so called Thouless time $t_D$. 

For comparison, also in Fig. (\ref{fig:Diffusion}) is plotted the decay of the LE for $\alpha=0.07$ (see Chap. \ref{chap:Universality} for details). Notice that the saturation in the diffusive behavior is not correlated to the saturation of the LE.

\section{Classical chaos in the Lorentz gas}

Appart from diffusion, the other relevant feature of the classical dynamics of the Lorentz gas is chaos. In this sense, two particles initially close from each other will separate exponentially fast with the Lyapunov exponent of the system\footnote{The fact that the concentration of impurities is uniform helps to have only one Lyapunov exponent. The general case is that the Lyapunov exponent is a quantity that depends on the location of the trajectories in phase space.} $\lambda$. 
The chaotic character of the dynamics is a consequence of the 
de-focusing nature of the collisions. As illustrated in Fig. (\ref{fig:Gas}), a particle with impact parameter $x$ will be reflected with an angle 
\be
\theta=\pi-2 \arctan{\left[\frac{x}{\sqrt{R^2-x^2}}\right]}.
\ee
Considering a second particle with impact parameter $x+\delta x$, its outgoing angle will be $\theta+\delta \theta$, with
\be
\delta \theta = \frac{2}{\sqrt{R^2-x^2}} \delta x.
\ee
The separation between these two particles when they have travelled a distance $s$ after a collision will grow as 
\be 
\delta d\simeq\delta x+\delta\theta s\simeq\delta x\left(  1+\frac{2s}%
{\sqrt{R^{2}-x^{2}}}\right)  . 
\label{Separation}%
\ee 
The next collision will further amplify the separation, due to the new impact 
parameters and the different incidence angles. 

\begin{figure}[htb]
\begin{center}
\leavevmode
\epsfxsize 3.5in
\epsfbox{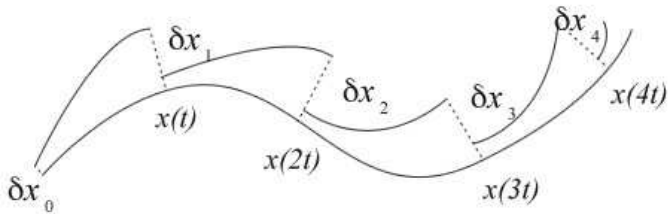}
\caption{Schematics of Benettin's algorithm \cite{Benettin76} to compute the Lyapunov exponent of a chaotic system. Two initially close trajectories are computed up to a time $t$ where the distance is renormalized to the initial one. The Lyapunov exponent is given by the average of the normalization factors.}
\label{fig:Benettin}
\end{center}
\end{figure}

\begin{figure}[htb]
\begin{center}
\leavevmode
\epsfxsize 4.5in
\epsfbox{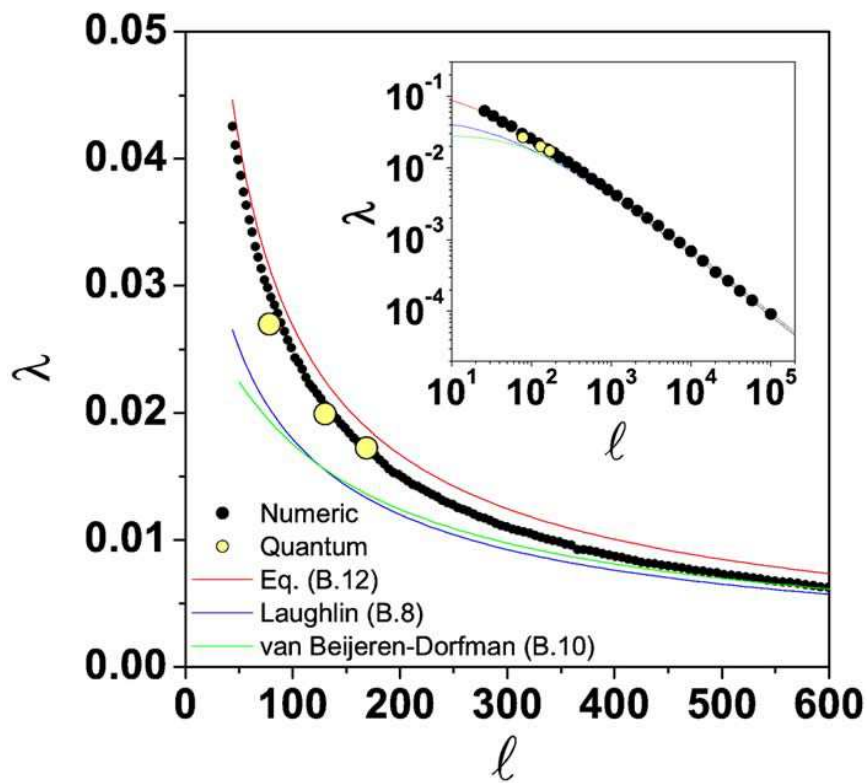}
\caption{Lyapunov exponent $\lambda$ of the Lorentz gas as a function of the mean free path $\ell$. The black dots represent the numerical values obtained through Benettin's method, the red line is the analytical estimate [Eq. (\ref{AnalyticLyapunov})]. The blue and green lines are Laughlin's and van Beijeren-Dorfman's approximations [Eqs. (\ref{Laughlin}) and (\ref{Dorfman})]. The open dots are the quantum values obtained from the decay of the LE [from Chap. (\ref{chap:Universality})]. Inset: the same plot in log-log scale to highlight the agreement between the different approximations in the region of very small concentrations (large $\ell$).}
\label{fig:Lyapunov}
\end{center}
\end{figure}

The usual algorithm for numerical computation of the Lyapunov exponent is that of Benettin et al. \cite{Benettin76}. 
The scheme is the following:
Two nearby 
trajectories are computed, and their separation is periodically scaled down to 
the initial value $\delta x_{0}$. For intermittent chaos like that of the Lorentz gas, the period $t$ should be taken longer than the collision time to avoid computing distances where chaos has not intervened. Also, it should be smaller than the time where the distance enters a diffusive regime, typically given by the moment when the trajectories collide with different impurities. The Lyapunov exponent results from the average over the expanding rates in the different intervals, 
\be 
\lambda=\lim_{n\rightarrow\infty}\frac{v}{n}\ \sum_{j=1}^{n}\frac{1}{s_{j}%
}\ \ln\left[  \frac{\delta x_{j}}{\delta x_{0}}\right]  \ , 
\label{LyapunovAverage}
\ee 
where $s_{j}$ is the length of the $j$-th interval, and $\delta x_{j}$ the 
separation just before the normalization [see Fig. (\ref{fig:Benettin})]. Technically, we should work with distances in phase-space, rather than in configuration space, but the local instability of the Lorentz gas makes this precision unnecessary. The computation of $\lambda$ using this method is presented in Fig.~(\ref{fig:Lyapunov}).

The first estimation of the Lyapunov exponent of the Lorentz gas was given by Laughlin, who considered a periodic Lorentz gas (repeated Sinai billiard) and proposed 
the form \cite{Laughlin87} 
\be 
\lambda=\frac{v}{\ell}\ \ln\left[  1+\frac{\beta\ell}{R}\right]  \ ,
\label{Laughlin}%
\ee
where $\beta$ is a geometrical factor of order $1$. In a similar approach of treating a simpler ordered system, Gaspard and Nicolis \cite{Gaspard90} for the three-disk problem obtained
\be 
\lambda=\frac{v}{2R_{\mathrm{e}}-2R}\ \ln\left[  \frac{2R_{\mathrm{e}%
}-R+\left(  4R_{\mathrm{e}}^{2}-4R_{\mathrm{e}}R\right)  ^{1/2}}{R}\right] 
\ . 
\ee 
A full treatment of the Lorentz gas in the diluted 
limit ($c\ll1$) by van Beijeren and Dorfman \cite{Beijeren95,Beijeren96} showed that 
\be 
\lambda=2\ \frac{N}{L^{2}}\ Rv\left(  1-\ln2-0.577-\ln\left[  \frac{NR^{2}%
}{L^{2}}\right]  \right)  , \label{Dorfman}
\ee 
later confirmed by numerical results \cite{Dellago95}.

A simple approach, presented for the first time in \cite{Cucchietti02Lorentz,Cucchietti03B}, is to use the basis of Benettin's algorithm for an analytical estimate of $\lambda$. For this, we consider the period of renormalization of the distance between trajectories as given by the mean free path. Consequently, we replace $s$ by $\ell$ in Eq. (\ref{Separation}). Using this in Eq. (\ref{LyapunovAverage}), we identify the average over pieces of the trajectory with a geometrical average over impact parameters,
\be 
\lambda=\frac{v}{R \ell} \ \int_{0}^{R} dx \ \ln{\left[  1+\frac{2 \ell}%
{\sqrt{R^{2}-x^{2}}}\right]  } \ . 
\ee 
Performing the integration yields 
\be 
\frac{\lambda}{v} =\frac{1}{\ell} \ \ln{\left[  \frac{\ell}{R}\right]  }+ 
\frac{\pi}{R}+\sqrt{\frac{4}{R^{2}} - \frac{1}{\ell^{2}}} \ \left( 
\arcsin{\left[  \frac{R}{2 \ell}\right]  } - \frac{\pi}{2} \right)  \ . 
\label{AnalyticLyapunov}%
\ee 
As shown in Fig.~(\ref{fig:Lyapunov}), the above expression reproduces 
remarkably well the numerical calculations of the Lyapunov exponent. It agrees 
with the result of van Beijeren and Dorfman and Laughlin in the dilute limit, although 
it appears to have a broader range of validity (for larger concentrations).

\section{The perturbation: distortion of mass tensor}

For a hard wall model, like the one we are considering, one can show that the distortion of the mass tensor [Eq. (\ref{MTPerturbation})] is equivalent to having non-specular reflections. 

\begin{figure}[htb]
\begin{center}
\leavevmode
\epsfxsize 2.5in
\epsfbox{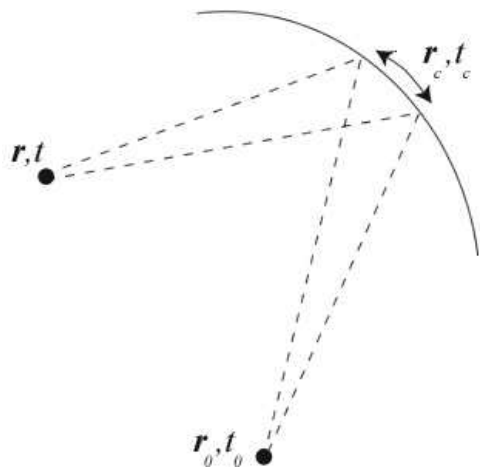}
\caption{A free particle colliding with a hard wall. The fixed points in the trajectory are the initial one $\br_0$ and the final $\br$, the correct trajectory is obtained by minimizing the action with respect the collision point $\br_c$ and time $t_c$.}
\label{fig:Action}
\end{center}
\end{figure}

Let us assume a particle in a free space with mass tensor $\overset 
{\leftrightarrow}{m}$ surrounded by an infinite potential surface (hard wall). 
Suppose that the particle departs from a point $\mathbf{r}_{0}$ at time 
$t_{0}$ and arrives to a final point $\mathbf{r}$ at time $t$. 
The total trajectory is determined by the unknown time $t_{\mathrm{c}}$ and position $\mathbf{r}_{\mathrm{c}}$ along the surface at which the particle collides [see Fig. \ref{fig:Action}]. The action along the trajectory is
 
\be 
S=\frac{(\mathbf{r}_{c} -\mathbf{r}_{0}){\overset{\leftrightarrow}{m}%
}(\mathbf{r}_{c} -\mathbf{r}_{0})}{2(t_{c}-t_{0})}+ \frac{(\mathbf{r} 
-\mathbf{r}_{c}){\overset{\leftrightarrow}{m}}(\mathbf{r} -\mathbf{r}_{c}%
)}{2(t-t_{c})} \ . \label{A-action}%
\ee 
  
We can solve the problem by minimizing the action, taking the derivative of 
Eq.~(\ref{A-action}) along the surface. Introducing the unitary vector 
$\mathbf{n}$ normal to the surface at the point of collision, we can express 
the minimization condition as%
 
\be 
\mathbf{n} \times\nabla_{\mathbf{r}_{c}} S = 0 \ . \label{Min}
\ee 
  
Denoting the initial and final velocities as $\mathbf{v}_{i}=(\mathbf{r}_{c} 
-\mathbf{r}_{0})/(t_{c}-t_{0})$ and $\mathbf{v}_{f}=(\mathbf{r} -\mathbf{r}%
_{c})/(t-t_{c})$, from Eqs. (\ref{A-action}) and (\ref{Min}) we can write%
 
\be 
\mathbf{n} \times{\overset{\leftrightarrow}{m}}(\mathbf{v}_{i}-\mathbf{v}%
_{f})=0 \ . 
\ee 
 
\noindent This, along with the conservation of energy $E=\mathbf{v}%
{\overset{\leftrightarrow}{m}}\mathbf{v}/2$, results in 
a generalized reflection law:%
 
\begin{subequations} 
\label{ReflectionLaw}%
\begin{align} 
v_{xf} &  =\frac{v_{xi}(m_{x}n_{y}^{2} -m_{y}n_{x}^{2})-2v_{yi}%
m_{y}n_{x}n_{y}} {m_{x}n_{y}^{2}+m_{y}n_{x}^{2}} \ , \label{ReflectionLawx}%
\\ 
\displaystyle v_{yf}  &  =\frac{v_{yi}(m_{y}n_{x}^{2}- m_{x}n_{y}%
^{2})-2v_{xi}m_{x}n_{x}n_{y}}{m_{x}n_{y}^{2}+m_{y}n_{x}^{2}} \ . 
\label{ReflectionLawy}%
\end{align} 
\end{subequations}

\begin{figure}[htb]
\begin{center}
\leavevmode
\epsfxsize 4.5in
\epsfbox{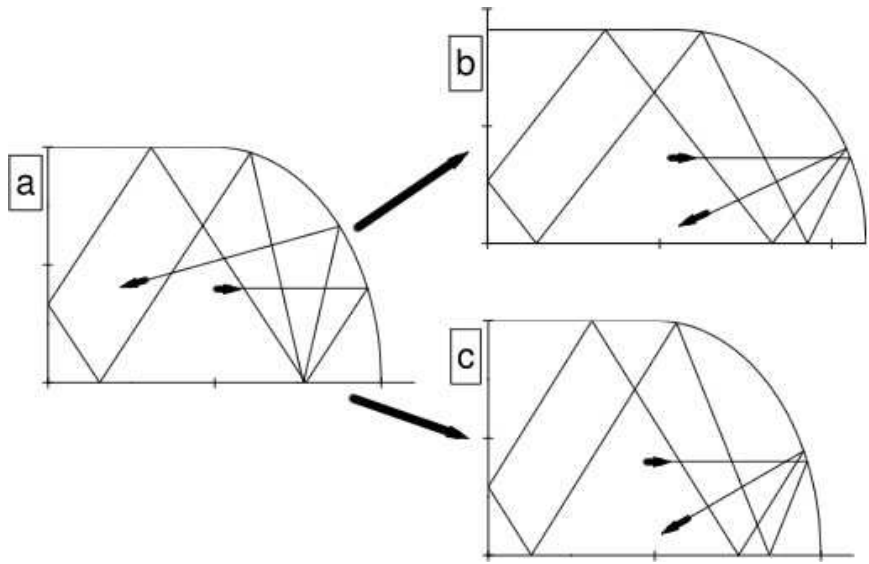}
\caption{(a) A particular trajectory in the Bunimovich stadium. The arrows represent the velocity in the initial and final points. (b) Trajectory with the same initial condition as (a) but in stadium dilated by a factor $\xi$. The change in the boundaries has influenced the trajectory, notice the difference with the final point of (a). (c) Evolution of the same initial point of (a), but now with a distorted mass tensor such that $\xi=\sqrt{1+\alpha}$. The perturbed trajectory is again different from (a), but a simple mapping of the coordinates relates it to (b). Therefore, the distortion of the mass tensor and the dilation are equivalent.}
\label{fig:Stadiums}
\end{center}
\end{figure}

Eqs.~(\ref{ReflectionLaw}) allow to show that the distortion of 
the mass tensor is equivalent to an area conserving deformation of 
the boundaries as $x\rightarrow x(1+\xi)$, $y\rightarrow 
y/(1+\xi),$ as used in other works on the LE\cite{Wisniacki02}, 
where $\xi=\sqrt{1+\alpha}-1$ is the stretching parameter, related to the distortion of ${\overset{\leftrightarrow}{m}}$ as in Eq. (\ref{MTPerturbation}). This equivalence can be observed in Fig. (\ref{fig:Stadiums}), where an unperturbed trajectory in the Bunimovich stadium ($a$) is subjected to both effects, dilation of the stadium ($b$) and distortion of the mass tensor ($c$). As stated above, given the appropriate relation between $\alpha$ and $\xi$, ($b$) and ($c$) are identical.
 
\bibliography{thesis}
\bibliographystyle{alpha}

\end{document}